\documentclass[journal]{IEEEtran}

\usepackage[latin1]{inputenc}
\usepackage[T1]{fontenc}
\usepackage[english]{babel}
\usepackage{amssymb,amsmath,amsfonts}
\usepackage{times}
\usepackage{graphicx}
\usepackage{color}
\usepackage{verbatim}
\usepackage{cite}

\usepackage{subfig}		
\usepackage{pgf}		
\usepackage{tikz}
\usepackage{pgfplots}	
\usetikzlibrary{decorations}
\usetikzlibrary{patterns}
\usetikzlibrary{positioning}
\usetikzlibrary{backgrounds}
\usetikzlibrary{fit}
\usetikzlibrary{shapes.geometric}
\usetikzlibrary{shapes.symbols}
\usetikzlibrary{calc}

\usepackage{stmaryrd}

\usepackage{cancel}
\usepackage[normalem]{ulem}

\usepackage{verbatim}

\newtheorem{definition}	{Definition}
\newtheorem{theorem}	{Theorem}
\newtheorem{proposition}{Proposition}
\newtheorem{lemma}		{Lemma}
\newtheorem{corollary}	{Corollary}
\newtheorem{remark}		{Remark}

\colorlet{vert}{green!70!black}
\definecolor{rouge}{rgb}{.85,0,0.1}

\tikzstyle{sensor}	=[rectangle,draw,text centered]
\tikzstyle{trans}	=[rectangle,draw,solid,text centered,minimum width=1cm,minimum height=.5cm]
\tikzstyle{gentil}	=[blue]	
\tikzstyle{mechant}	=[red]		
\tikzstyle{canal}	=[vert]

\tikzstyle{codeword}=[draw,minimum width=.8cm]
\tikzstyle{cok}		=[fill=gray]

\tikzstyle{oper}	=[inner sep=0pt,minimum width=5mm,draw,circle]

\tikzstyle{mult}	=[fill=lightgray,draw,isosceles triangle,inner sep=1pt]

\tikzstyle{point}	=[circle,fill,inner sep=0pt,minimum size=1.5mm]

\newcommand{\cA}{{\mathcal A}}
\newcommand{\cB}{{\mathcal B}}

\newcommand{\cE}{{\mathcal E}}
\newcommand{\cG}{{\mathcal G}}
\newcommand{\cQ}{{\mathcal Q}}
\newcommand{\cR}{{\mathcal R}}
\newcommand{\cS}{{\mathcal S}}
\newcommand{\cT}{{\mathcal T}}
\newcommand{\cU}{{\mathcal U}}
\newcommand{\cV}{{\mathcal V}}

\newcommand{\cX}{{\mathcal X}}
\newcommand{\cY}{{\mathcal Y}}
\newcommand{\cZ}{{\mathcal Z}}

\newcommand{\bN}{{\mathbb N}}
\newcommand{\bR}{{\mathbb R}}

\newcommand{\interval}[4]{\mathopen{#1}#2\mathclose{},#3\mathclose{#4}}
\newcommand{\intervalLR}[4]{\left #1 #2,#3\right #4}

\newcommand{\intFF}[2]{\interval{[}{#1}{#2}{]}}

\newcommand{\intFO}[2]{\interval{[}{#1}{#2}{)}}

\newcommand{\intFFLR}[2]{\intervalLR{[}{#1}{#2}{]}}

\newcommand{\pr}[1]{\operatorname{Pr}\left\{#1\right\}}

\newcommand{\bE}{{\mathbb E}}

\newcommand{\Var}[1]{\text{Var}\left[#1\right]}

\newcommand{\indpt}{\perp\!\!\!\perp}

\newcommand{\mkv}{-\!\!\!\!\minuso\!\!\!\!-}

\newcommand{\lessnoisy}[1]{\succeq_{\scriptscriptstyle #1}}

\newcommand{\typ}[2]{T_\delta^{#1}(#2)}

\newcommand{\simiid}{\stackrel{\text{i.i.d.}}{\sim}}

\newcommand{\unif}[1]{{\mathcal U}_{\,#1}}
\newcommand{\bern}{{\mathcal B}}

\newcommand{\cN}{{\mathcal N}}

\newcommand{\T}{^\text{\scshape{t}}}

\newcommand{\norm}[1]{\left\lVert#1\right\rVert}
\newcommand{\abs}[1]{\left\lvert#1\right\rvert}

\newcommand{\toas}[1]{\xrightarrow[#1]{}}

\newcommand{\err}{{\mathsf E}}
\newcommand{\Nerr}{\cancel{\mathsf E}}

\usepackage{hyperref}

\title{Secure Transmission of Sources over Noisy Channels with Side Information at the Receivers} 
\author{Joffrey Villard, Pablo Piantanida, and Shlomo Shamai (Shitz)

\thanks{The work of J. Villard is supported by DGA (French Armement Procurement Agency). This research is partially supported by the FP7 Network of Excellence in Wireless COMmunications NEWCOM\#. The material in this paper was presented in part at the IEEE International Symposium on Information Theory, 2011, 2012 and the IEEE Information Theory Workshop (invited paper), 2011.} 
\thanks{J. Villard and P. Piantanida are with the Department of Telecommunications, SUPELEC, 91192 Gif-sur-Yvette, France (e-mail: joffrey.villard@supelec.fr; pablo.piantanida@supelec.fr).}
\thanks{S. Shamai (Shitz) is with the Department of Electrical Engineering, Technion--Israel Institute of Technology, Technion City, Haifa 32000, Israel (e-mail: sshlomo@ee.technion.ac.il).}
}

\begin{document}
\maketitle

\begin{abstract}
This paper investigates the problem of source-channel coding for secure transmission with arbitrarily correlated side informations at both receivers. 
This scenario consists of an encoder (referred to as Alice) that wishes to compress a source and send it through a noisy channel to a legitimate receiver (referred to as Bob). In this context, Alice must simultaneously satisfy the desired requirements on the distortion level at Bob, and the equivocation rate at the eavesdropper (referred to as Eve). This setting can be seen as a generalization of the problems of secure source coding with (uncoded) side information at the decoders, and the wiretap channel. A general outer bound on the rate-distortion-equivocation region, as well as an inner bound based on a pure digital scheme, is derived for arbitrary channels and side informations. In some special cases of interest, it is proved that this digital scheme is optimal and that separation holds. However, it is also shown through a simple counterexample with a binary source that a pure analog scheme can outperform the digital one while being optimal. According to these observations and assuming matched bandwidth, a novel hybrid digital/analog scheme that aims to gather the advantages of both digital and analog ones is then presented. In the quadratic Gaussian setup when side information is only present at the eavesdropper, this strategy is proved to be optimal. Furthermore, it outperforms both digital and analog schemes, and cannot be achieved via time-sharing. By means of an appropriate coding, the presence of any statistical difference among the side informations, the channel noises, and the distortion at Bob can be fully exploited in terms of secrecy. 
\end{abstract}

\section{Introduction}

Consider a system composed of three nodes (or sensors) where each one is measuring an analog source (or random field) as a function of time. 
One of them (referred to as Alice) wishes to transmit a compressed version of its observation to a second node (referred to as Bob) through a noisy (or wireless) channel. In addition, Bob can use his own observation as side information to decode the received message and refine his estimate of Alice's source. The third node (referred to as Eve) is an eavesdropper i.e., a node that can listen to the messages sent by Alice through another noisy channel. 
Considering that Eve is not to be trusted, Alice wishes to leak the smallest amount of information about her source.

Among some major information-theoretic issues, the above scenario involves the notion of secrecy (and its application to source and channel coding), source coding with side information, as well as joint source-channel coding for transmission of sources over noisy channels.
The information-theoretic notion of secrecy, introduced by Shannon~\cite{shannon1949communication}, was first used for secure communication over noisy channels by Wyner~\cite{wyner1975wire}, who studied the so-called wiretap channel.
In particular, allowing the encoder to introduce some independent random noise in the transmitted signal, Wyner showed that it is possible to send information at a positive rate with perfect secrecy as long as the channel of the eavesdropper is a degraded version of the legitimate user's one.
Csisz\`ar and K\"orner~\cite{csiszar1978broadcast} extended this result to the setting of general broadcast channels with any arbitrary level of security, as measured by the equivocation rate --the remaining uncertainty about the message-- at the eavesdropper. 
Extensive research has since been done, yielding several extensions of the original wiretap channel~\cite{liang2008secure,chen2008wiretap,it2008special,liang2009information,liu2010securing}.
 
On the other hand, source coding with side information has been studied by Slepian and Wolf~\cite{slepian1973noiseless}, and Wyner and Ziv~\cite{wyner1976rate}. Security constraints with respect to an additional eavesdropper that must be kept as ignorant as possible of the transmitted source were recently introduced in these source coding problems~\cite{prabhakaran2007secure,6507243,villard2010secure,villard2011securea}. The optimal coding scheme has been characterized in the general case~\cite{villard2010secure,villard2011securea}. It uses standard coding techniques (superposition coding, random binning, etc.) as well as a new evaluation of the equivocation rate at the eavesdropper. As a matter of fact, if the side informations at the decoders are degraded, then Wyner-Ziv coding~\cite{wyner1976rate} is optimal, while it is proved to be insufficient in the general case.

As in the papers cited above, most of the existent work separately considers channel or source coding for secure transmission or compression. 
However, unlike point-to-point communication problems \cite{shannon1948mathematical,merhav2003joint}, there is no general result of separation for multiterminal settings under security constraints.
Recent work~\cite{merhav2008shannon} considered such a setting of source-channel coding for secure transmission by assuming that Eve has a degraded channel with degraded side information with respect to Bob, and shows that separation holds. Along the same line of work, state amplification subject to masking constraints, where Alice wishes to effectively convey --amplify-- the channel state sequence to Bob while masking it from Eve, has been investigated in \cite{Ozan-Soundararajan-Vishwanath-11}. This may indicate that \emph{digital} schemes are well-suited for these multiterminal settings with security constraints. On the other hand, it is well-known that \emph{joint} source-channel coding/decoding is a must for broadcast channels without secrecy constraints~\cite{gastpar2003code,tuncel2006slepian-wolf}, and hybrid digital/analog schemes have been proved useful for point-to-point problems e.g., to handle SNR mismatch (while they can perform as well as digital or analog ones at the true SNR)~\cite{mittal2002hybrid,wilson2010joint}, as well as for some multiterminal settings~\cite{gunduz2008wyner-ziv,lim2010lossy,gao2010new}. By taking advantage of both analog and digital strategies, they may help to solve the considered problem for secure transmission in the more general case without any degradedness condition.

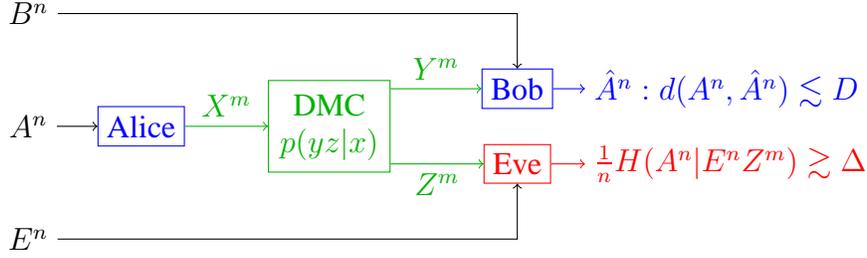
\begin{figure*}[!ht]
\centering
	\begin{tikzpicture}[scale=1,every node/.append style={font=\large}]
	
	\node					(A) 		at (0  ,0  ) {$A^n$};
	\node[sensor,gentil] 	(alice)		at (1.5,0  ) {Alice};
	\node[trans,canal]		(canal)		at (4  ,0  ) {\begin{tabular}{c}DMC\\ $\!\!\!\!\! p(yz|x)\!\!\!\!\!$\end{tabular}};
	\node					(B) 		at (0  ,1.5) {$B^n$};
	\node[sensor,gentil]	(bob) 		at (6.5, .5) {Bob};
	\node[right,gentil]		(hatA) 		at (7.4, .5) {$\hat A^n: d(A^n,\hat A^n)\lesssim D$};
	\node[sensor,mechant]	(eve) 		at (6.5,- .5){Eve};
	\node					(E) 		at (0  ,-1.5){$E^n$};
	\node[right,mechant]	(D) 		at (7.4,- .5){$\frac1n H(A^n|E^n Z^m)\gtrsim\Delta$};
		
	\draw[->]			(A) to (alice);
	\draw[->]			(B) -| (bob);
	\draw[->]			(E)	-| (eve);
	
	\draw[->,canal]		(alice)				to node[above]{$X^m$}	(canal);
	
	\draw[->,canal]		(canal.east|-bob)	to node[above]{$Y^m$}	(bob);
	\draw[->,canal]		(canal.east|-eve)	to node[below]{$Z^m$}	(eve);
		
	\draw[->,gentil]	(bob) to (hatA);
	\draw[->,mechant]	(eve) to (D);
\end{tikzpicture}
\caption{Secure transmission with side information at the receivers.}
\label{fig:sceCh:schema}
\end{figure*}

In this paper, we consider the setup of joint source-channel coding for secure transmission of a source over a noisy channel with an eavesdropper, and in the presence of side information at the receiving terminals, as depicted in Fig.~\ref{fig:sceCh:schema}.
This setting can be seen as the unification of the problems of secure source coding with side information at the decoders~\cite{villard2010secure,villard2011securea}, and the wiretap channel~\cite{wyner1975wire,csiszar1978broadcast}. 
The main goal is to understand how Alice can take simultaneous advantage of the statistical differences among the side informations and the channel noises to reveal the minimum amount of information to Eve, and satisfy the required distortion level at Bob.  
It should be emphasized that the central difficulty of this problem lies in the evaluation of the equivocation at Eve.
As a matter of fact, the presence of side information at the eavesdropper, that can be used together with its channel output to estimate the source, prevents from directly applying secrecy capacity results~\cite{csiszar1978broadcast}.
We derive a general outer bound on the achievable region, referred to as the \emph{rate-distortion-equivocation region}, for arbitrary channels and side informations.
We then propose a pure digital scheme which combines secure source coding~\cite{villard2010secure,villard2011securea} with coding for broadcast channels with confidential messages~\cite{csiszar1978broadcast}, and derive the corresponding single-letter inner bound.
These two bounds do not match in general but we derive two results of optimality when: (\emph{i}) Bob has less noisy side information, and (\emph{ii}) Eve has less noisy channel. 
In these cases, separation holds and the optimal schemes reduce to a \emph{Wyner-Ziv} source encoder~\cite{wyner1976rate} followed by a classical \emph{wiretap} channel encoder~\cite{csiszar1978broadcast}, and a \emph{secure} source encoder~\cite{villard2010secure,villard2011securea, 6283040} followed by a conventional channel encoder~\cite{shannon1948mathematical}, respectively.
However, we show through a simple counterexample with a binary source that a pure analog scheme can outperform the digital one while being optimal.
Then, restricting our attention to the matched-bandwidth case, we propose a novel hybrid digital/analog scheme that aims to gather the advantages of both digital and analog ones, and derive its single-letter inner bound.
In the quadratic Gaussian setup when side information is only present at the eavesdropper, this strategy is proved to be optimal. 
Furthermore, it outperforms both digital and analog schemes and cannot be achieved via time-sharing. 
We also consider secure transmission of a binary source with BEC/BSC side informations over a type-II wiretap channel.
The proposed hybrid digital/analog scheme turns out to be useful also in this setting.

The rest of this paper is organized as follows. 
Section~\ref{sec:general} states definitions along with the general outer bound on the rate-distortion-equivocation region.
Section~\ref{sec:digital} provides a single-letter inner bound based on a digital scheme, as well as special cases where separation holds.
The proof of the inner bound is given in Section~\ref{sec:digital:proof}.
Transmission of a binary source over a type-II wiretap channel is studied in Section~\ref{sec:binary}, providing a counterexample for the optimality of the digital scheme.
A single-letter inner bound based on a hybrid digital/analog scheme is provided in Section~\ref{sec:hybrid}.
The proof is given in Section~\ref{sec:hybrid:proof}.
Section~\ref{sec:binaryCont} (resp. Section~\ref{sec:gaussian}) presents an application example to the transmission of a binary source over a type-II wiretap channel (resp. a Gaussian source over a Gaussian wiretap channel with side informations).
Section~\ref{sec:sceCh:summary} concludes the paper.

\subsection*{Notation}

For any sequence~$(x_i)_{i\in\bN^*}$, notation $x_k^n$ stands for the collection $(x_k,x_{k+1},\dots, x_n)$. $x_1^n$ is simply denoted by $x^n$.
Entropy is denoted by $H(\cdot)$, and mutual information by $I(\cdot;\cdot)$.
We denote typical and conditional typical sets by $\typ{n}{X}$ and $\typ{n}{Y|x^n}$, respectively (see Appendix~\ref{app:typical} for details).
Let $X$, $Y$ and $Z$ be three random variables on some alphabets with probability distribution~$p$. If $p(x|yz)=p(x|y)$ for each $x,y,z$, then they form a Markov chain, which is denoted by $X\mkv Y\mkv Z$. Notation $A \indpt B$ is used to indicate independence between the random variables $A$ and $B$. Random variable $Y$ is said to be less noisy than $Z$ w.r.t.\ $X$ if $I(U;Y) \geq I(U;Z)$ for each random variable $U$ such that $U\mkv X\mkv (Y,Z)$ form a Markov chain. This relation is denoted by $Y\lessnoisy{X}Z$.
The set of nonnegative real numbers is denoted by $\bR_+$.
For each $x\in\bR$, notation $[x]_+$ stands for $\max\{0\,;x\}$.
Logarithms are taken in base $2$ and denoted by $\log(\cdot)$.
The binary entropy function is defined on $\intFF{0}{1}$ as $h_2(x) = -x\log(x) -(1-x)\log(1-x)$.
Its inverse $h_2^{-1}$ is defined on $\intFF{0}{1}$ and takes values in $\intFF{0}{\tfrac12}$.
For each $a,b\in\intFF{0}{1}$, $a\star b = a(1-b) + (1-a)b$.
The Bernoulli distribution of parameter $u$ is denoted by $\bern(u)$.

\section{Problem Definition and General Outer Bound}
\label{sec:general}

\subsection{Problem Definition}
\label{sec:definition}

In this section, we give a more rigorous formulation of the context depicted in Fig.~\ref{fig:sceCh:schema}.
Let $\cA$, $\cB$, $\cE$, $\cX$, $\cY$, and $\cZ$ be six finite sets. 
Alice, Bob, and Eve observe the sequences of random variables 
$(A_i)_{i\in\bN^*}$, $(B_i)_{i\in\bN^*}$, and $(E_i)_{i\in\bN^*}$,
respectively, which take values on $\cA$, $\cB$, and $\cE$, resp.
For each $i\in\bN^*$, the random variables $A_i$, $B_i$, and $E_i$
are distributed according to the joint distribution $p(abe)$ on
$\cA\times\cB\times\cE$.
Moreover, they are independent across time $i$.
Alice can also communicate with Bob and Eve through a discrete memoryless channel with input $X$ on $\cX$, and outputs $Y$, $Z$ on $\cY$, $\cZ$, respectively. This channel is defined by its transition probability $p(yz|x)$.

Let $d \colon \cA\times\cA \to \intFF{0}{d_\text{max}}$ be a finite distortion measure
i.e., such that $0\leq d_\text{max} < \infty$.
We also denote by $d$ the component-wise mean distortion on $\cA^n\times\cA^n$
i.e., for each $a^n,b^n\in\cA^n$, $d(a^n,b^n) = \frac1n\,\sum_{i=1}^n d(a_i,b_i)$.

\begin{definition}[Code]
\label{def:code}
An $(n,m)$-code for source-channel coding is defined by
\begin{itemize}
\item a (stochastic) encoding function at Alice $F \colon \cA^n \to \cX^m$, defined by some transition probability $P_F(x^m|a^n)$,
\item a decoding function at Bob $g \colon \cB^n\times\cY^m \to \cA^n$.
\end{itemize}
The rate of such a code is defined as the number of channel uses per source symbol~$\frac{m}{n}$.
\end{definition}

\begin{definition}[Achievability]
\label{def:achievability}
A tuple $(k,D,\Delta)\in\bR_+^3$ is said to be \emph{achievable} if,
for any $\varepsilon>0$, there exists an $(n,m)$-code $(F,g)$ such that:
\begin{IEEEeqnarray}{rCl}
\frac mn 						&\leq& k+\varepsilon 		\ ,\label{eq:ach:rate}\\[.1em]
\bE\big[ d(A^n,g(B^n,Y^m)) \big]&\leq& D+\varepsilon 		\ ,\label{eq:ach:dist}\\[.1em]
\dfrac1n\,H(A^n|E^n Z^m) 		&\geq& \Delta-\varepsilon 	\ ,\label{eq:ach:equiv}
\end{IEEEeqnarray}
with channel input $X^m$ as the output of the encoder $F(A^n)$.\\
The set of all achievable tuples is denoted by $\cR^*$ and is referred to as the \emph{rate-distortion-equivocation region}.
\end{definition}

\begin{remark}
\label{rem:closedconvex}
Region $\cR^*$ is closed and convex.
\end{remark}

\begin{remark}
\label{rem:marginal}
Quantities involved in Definition~\ref{def:achievability} only depend on the marginal distributions $p(ae)$, $p(ab)$, $p(y|x)$ and $p(z|x)$.
The same applies for subsequent results that provide inner and outer bounds on $\cR^*$.
\end{remark}

\subsection{General Outer Bound}
\label{sec:outer}

The following theorem gives an outer bound on $\cR^*$ i.e., it defines region $\cR_\text{out}\supseteq\cR^*$.

\begin{theorem}[Outer bound]
\label{th:outer}
If $(k,D,\Delta)$ is achievable, then there exist random variables $U$, $V$, $Q$, $T$, $X$ on finite sets $\cU$, $\cV$, $\cQ$, $\cT$, $\cX$, respectively, with joint distribution $p(uvqtabexyz) = p(uv|a)p(abe)\,p(q)p(t|q)p(x|t)p(yz|x)$, and a function $\hat A \colon \cV\times\cB \to \cA$, verifying the following inequalities:
\begin{IEEEeqnarray}{rcl}
I(V;A|B) &\leq& k I(T;Y) 										\ ,\label{eq:outer1}\\[.1em]
D 		& \geq& \bE\big[d(A,\hat A(V,B))\big] 				\ ,\label{eq:outer2}\\[.1em]
\Delta	& \leq & H(A|UE) - \Big[ I(V;A|B) - I(U;A|B)			\nonumber\\[.1em]
 \IEEEeqnarraymulticol{3}{r}{ 	- k \Big( I(T;Y|Q) - I(T;Z|Q) \Big) \Big]_+ \ .\label{eq:outer3} }
\end{IEEEeqnarray}
\end{theorem}
\begin{IEEEproof}
See Appendix~\ref{app:outer:proof}.
\end{IEEEproof}\vspace{1mm}

Let us now give some intuition on Equations \eqref{eq:outer1}-\eqref{eq:outer2} and \eqref{eq:outer3}. Equations \eqref{eq:outer1}-\eqref{eq:outer2} are the conditions for the transmission and distortion of a standard source-channel coding problem. The first term in \eqref{eq:outer3}, i.e., $H(A|UE)$, corresponds to the equivocation rate at Eve when the common message $U$ can be decoded. Let us assume that $Q=T$, in this case the remaining terms capture the equivocation rate at Eve in terms of pure source coding:  
$$
\Delta \leq  H(A|UE) - \Big[ I(A;V|B) - I(A;U|B) \Big]_+ 
$$
where the remaining information rate of Alice, i.e., $I(A;V|B) - I(A;U|B)$, is directly subtracted from the equivocation rate, meaning that it is treated as "raw" bits of $A$. At the same time, if $T\neq Q$ the equivocation rate is increased by the secrecy rate in the wiretap channel, provided that the channel at Bob satisfies $I(T;Y|Q) > I(T;Z|Q)$.

\section{Digital Scheme}
\label{sec:digital}

In this section, we propose a digital coding scheme for secure transmission with side information and derive the corresponding single-letter inner bound $\cR_\text{digital}$ (Theorem~\ref{th:digital}).
This scheme turns out to be optimal under some less-noisy conditions (Propositions~\ref{prop:digital:special:B} and \ref{prop:digital:special:Z}).

\subsection{General Statement}

The following theorem gives an inner bound on $\cR^*$ i.e., it defines region $\cR_\text{digital}\subseteq\cR^*$.
The achievability follows by combining secure source coding of~\cite{villard2010secure,villard2011securea} with coding for broadcast channels with confidential messages~\cite{csiszar1978broadcast}.
This scheme will be referred to as the \emph{digital scheme}.

\begin{theorem}[Digital scheme]
\label{th:digital}
A tuple $(k,D,\Delta)\in\bR_+^3$ is achievable if there exist random variables $U$, $V$, $Q$, $T$, $X$ on finite sets $\cU$, $\cV$, $\cQ$, $\cT$, $\cX$, respectively, with joint distribution $p(uvqtabexyz) = p(u|v)p(v|a)p(abe)\,p(q)p(t|q)p(x|t)p(yz|x)$, and a function $\hat A \colon \cV\times\cB \to \cA$, verifying the following inequalities:
\begin{IEEEeqnarray}{rl}
I(U;A|B) & \leq k I(Q;Y) 										\ ,\label{eq:digital1}\\[.1em]
I(V;A|B) & \leq k I(T;Y) 										\ ,\label{eq:digital2}\\[.1em]
D 		&  \geq \bE\big[d(A,\hat A(V,B))\big] 				\ ,\label{eq:digital3}\\[.1em]
\Delta	  \leq  H(A|UE) &- \Big[ I(V;A|UB)					
			\ ,\label{eq:digital2}\\[.1em]
			&	- k \Big( I(T;Y|Q) - I(T;Z|Q) \Big) \Big]_+ \ .\label{eq:digital4}
\end{IEEEeqnarray}
\end{theorem}

\begin{IEEEproof}
See Section~\ref{sec:digital:proof}.
\end{IEEEproof}

Inequalities~\eqref{eq:digital1}, \eqref{eq:digital2} correspond to sufficient conditions for the transmission of two source layers $U$, $V$ in channel variables $Q$, $T$, respectively.
The first layer $(U,Q)$ can be seen as a \emph{common} message which is considered to be known at Eve, as shown by the term $H(A|UE)$ in \eqref{eq:digital4}.
The second layer $(V,T)$ forms a \emph{private} message which is (partially) protected by adding an independent random noise~\cite{csiszar1978broadcast,liang2009information}. 
The term in square brackets in \eqref{eq:digital4} corresponds to the information that Eve can still obtain on this protected layer.

Notice that the inner and outer bounds $\cR_\text{digital}$ and $\cR_\text{out}$ do not meet in general:
\begin{itemize}
\item Condition~\eqref{eq:digital1} in Theorem~\ref{th:digital}, which is needed in our scheme to characterize the equivocation at Eve, may not be optimal in the general case (see Theorem~\ref{th:outer}).
\item The Markov chain $U\mkv V\mkv A\mkv(B,E)$ is assumed in Theorem~\ref{th:digital} yielding 
$$
I(A;V|B) - I(A;U|B) =  I(A;V|UB)
$$
while only $(U,V)\mkv A\mkv(B,E)$ is proved for arbitrary codes in Theorem~\ref{th:outer}, for which 
$$
I(A;V|B) - I(A;U|B) \leq   I(A;V|UB). 
$$
\end{itemize}
We provide in Section~\ref{sec:digital:special} several cases where $\cR_\text{digital}$ is optimal.

\subsection{Coding Scheme Based On "Operational'' Separation}

In traditional \emph{separated} schemes, two \emph{stand-alone} components successively perform source and channel coding, as depicted in Fig.~\ref{fig:separation}. 
However the scheme that achieves region~$\cR_\text{digital}$ does not satisfy this separation principle: 
The source encoder outputs two layers which are further encoded by using the channel code for a broadcast channel with confidential messages~\cite{csiszar1978broadcast} (see Section~\ref{sec:digital:proof}). 
This results in two independent (but not stand-alone) source and channel components yielding statistically independent source and channel variables (as in~\cite{tuncel2006slepian-wolf} for Slepian-Wolf coding over broadcast channels) i.e., ``operational'' separation holds (see Fig.~\ref{fig:operational}).
As a matter of fact, inequality~\eqref{eq:digital1} in Theorem~\ref{th:digital} prevents from separately choosing variables $U$ and $Q$ which would maximize the equivocation rate at Eve~\eqref{eq:digital4}.

\begin{figure}
\centering
\begin{minipage}[t]{.48\textwidth}
\centering
	\begin{tikzpicture}[scale=.8]
	\useasboundingbox (-.4,-1) rectangle (8.4,1);

	\node					(A) 		at (0,0)	{\small $A^n$};
	\node[rectangle,draw]	(aliceS)	at (2,0)	{\small \begin{tabular}{c}Source\\ encoder\end{tabular}};		
	\node[rectangle,draw]	(aliceC)	at (5.8,0)	{\small \begin{tabular}{c}Channel\\ encoder\end{tabular}};
	\node					(canal)		at (8,0)	{\small $X^m$};
		
	\draw[->]	(A)			to (aliceS);
	\draw[->]	(aliceS)	to	node[above]{\small $r$}	(aliceC);
	\draw[->]	(aliceC)	to  (canal);
\end{tikzpicture}
\caption{Traditional separation.}
\label{fig:separation}
\end{minipage}
\hspace{.02\textwidth}
\begin{minipage}[t]{.48\textwidth}
\centering
	\begin{tikzpicture}[scale=.8]
	\useasboundingbox (-.4,-1) rectangle (8.4,1);
	
	\node					(A) 		at (0,0) 	{\small $A^n$};
	\node[rectangle,draw]	(aliceS)	at (2,0)	{\small \begin{tabular}{c}Source\\ encoder\end{tabular}};		

	\node[above=.1]			(aliceS1)	at (aliceS.east)	{};			
	\node[below=.1]			(aliceS2)	at (aliceS.east)	{};

	\node[rectangle,draw]	(aliceC)	at (5.8,0) 	{\small \begin{tabular}{c}Channel\\ encoder\end{tabular}};
	\node					(canal)		at (8,0)	{\small $X^m$};
			
	\draw[->]	(A)					to (aliceS);
	\draw[->]	(aliceS1.center)	to 	node[above]{\small $r_c$} (aliceC.west|-aliceS1);
	\draw[->]	(aliceS2.center)	to 	node[below]{\small $r_p$} (aliceC.west|-aliceS2);
	\draw[->]	(aliceC)			to  (canal);
\end{tikzpicture}
\caption{Proposed system (``operational'' separation).}
\label{fig:operational}
\end{minipage}
\end{figure}

\subsection{Special Cases}
\label{sec:digital:special}

In this section, we characterize the optimality of the inner bound $\cR_\text{digital}$ for some special cases.

\subsubsection{Bob Has Less Noisy Side Information}

If Bob has less noisy side information than Eve i.e., $B \lessnoisy{A} E$, the optimal coding scheme reduces to a \emph{Wyner-Ziv} source encoder~\cite{wyner1976rate} followed by a classical \emph{wiretap} channel encoder~\cite{csiszar1978broadcast}, and hence separation holds (Fig.~\ref{fig:separation}):

\begin{proposition}
\label{prop:digital:special:B}
If $B \lessnoisy{A} E$, $(k,D,\Delta)\in\bR_+^3$ is achievable if and only if there exist random variables $V$, $Q$, $T$, $X$ on finite sets $\cV$, $\cQ$, $\cT$, $\cX$, respectively, with joint distribution $p(vqtabexyz) = p(v|a)p(abe)\,p(q|t)p(t)p(x|t)p(yz|x)$, and a function $\hat A \colon \cV\times\cB \to \cA$, verifying
\begin{IEEEeqnarray*}{l}
I(V;A|B) \leq k I(T;Y) 										\ ,\\[.1em]
D 		 \geq \bE\big[d(A,\hat A(V,B))\big] 				\ ,\\[.1em]
\Delta	 \leq  H(A|E) - \Big[ I(V;A|B)						
	- k \Big( I(T;Y|Q) - I(T;Z|Q) \Big) \Big]_+ \ .
\end{IEEEeqnarray*}
\end{proposition}

\begin{IEEEproof}
The achievability simply follows from Theorem~\ref{th:digital} by setting the random variable $U$ equal to a constant value. Whereas, the converse follows from Theorem~\ref{th:outer} by noting that the third inequality reads:
\begin{IEEEeqnarray*}{rCl}
\Delta	&\leq&  H(A|UE) 	\,\,\,\textrm{ or}										\\[.1em]
\Delta	&\leq&  H(A|VB) + I(A;B|U) - I(A;E|U)						\\[.1em]
\IEEEeqnarraymulticol{3}{R}{ + k \Big( I(T;Y|Q) - I(T;Z|Q) \Big)	\ .}
\end{IEEEeqnarray*}
Since $B \lessnoisy{A} E$, and $U\mkv A\mkv(B,E)$ form a Markov chain, $I(A;B|U) - I(A;E|U) \leq I(A;B) - I(A;E)$. Moreover $H(A|UE)\leq H(A|E)$. In this case, the outer bound $\cR_\text{out}$ is thus included in (and consequently equal to) $\cR_\text{digital}$.
\end{IEEEproof}

If the informations at Eve (both side information, and channel output) are degraded versions of Bob's ones i.e., if both Markov chains $A\mkv B\mkv E$, and $X\mkv Y\mkv Z$ hold, then Proposition~\ref{prop:digital:special:B} reduces to the results in~\cite{merhav2008shannon}. In this case, variable $Q$ is set to a constant value, and $T=X$.




\subsubsection{Eve Has Less Noisy Channel}

If Eve has less noisy channel than Bob i.e., $Z \lessnoisy{X} Y$, the optimal scheme reduces to a \emph{secure} source encoder~\cite{villard2010secure,villard2011securea} followed by a conventional channel encoder~\cite{shannon1948mathematical}, and hence separation holds (Fig.~\ref{fig:separation}):

\begin{proposition}
\label{prop:digital:special:Z}
If $Z \lessnoisy{X} Y$, $(k,D,\Delta)\in\bR_+^3$ is achievable if and only if there exist random variables $U$, $V$, $X$ on finite sets $\cU$, $\cV$, $\cX$, respectively, with joint distribution $p(uvabexyz) = p(u|v)p(v|a)p(abe)\,p(x)p(yz|x)$, and a function $\hat A \colon \cV\times\cB \to \cA$, verifying
\begin{IEEEeqnarray*}{rcl}
I(V;A|B) &\leq& k I(X;Y) 						\ ,\\[.1em]
D 		&\geq& \bE\big[d(A,\hat A(V,B))\big] 	\ ,\\[.1em]
\Delta	&\leq&  H(A|VB) + I(A;B|U) - I(A;E|U) \ ,
\end{IEEEeqnarray*}
where  it suffices to consider sets~$\cU$ and~$\cV$ such that 
$\norm{\cU}\leq\norm{\cA}+2$ and 
$\norm{\cV}\leq(\norm{\cA}+2)(\norm{\cA}+1)$.
\end{proposition}

\begin{IEEEproof}
The above region is achievable by setting $Q=T=X$ in Theorem~\ref{th:digital}. 
A new proof is needed to obtain the converse part.
Here, auxiliary variables are defined as follows, for each $i\in\{1,\dots,n\}$, and each $j\in\{1,\dots,m\}$:
\begin{IEEEeqnarray*}{rCl}
U_i &=& (\phantom{A^{i-1},B^{i-1}, } B_{i+1}^n,E^{i-1}, Y^m)	\ ,\\
V_i &=& (A^{i-1},B^{i-1},			 B_{i+1}^n,E^{i-1}, Y^m)	\ ,\\[.2em]
Q_j &=& (\phantom{A^n, }	E^n, Y^{j-1}, Z_{j+1}^m) 			\ ,\\
T_j &=& (A^n,				E^n, Y^{j-1}, Z_{j+1}^m) 			\ .
\end{IEEEeqnarray*}
Now, both $U_i\mkv V_i\mkv A_i\mkv (B_i,E_i)$, and $Q_j\mkv T_j\mkv X_j\mkv (Y_j,Z_j)$ form Markov chains (see Fig.~\ref{fig:outer:proof:joint}).
Following the arguments given in Appendix~\ref{app:outer:proof}, we can define new variables $U$, $V$, $Q$, $T$ verifying the above Markov chains and the following inequalities:
\begin{IEEEeqnarray*}{l}
I(V;A|B) \leq k I(T;Y) 										\ ,\\[.1em]
D 		 \geq \bE\big[d(A,\hat A(V,B))\big] 				\ ,\\[.1em]
\Delta	 \leq  H(A|UE) - I(V;A|UB)							
	+ k \Big( I(T;Y|Q) - I(T;Z|Q) \Big) \ .
\end{IEEEeqnarray*}
Since $Z\lessnoisy{X}Y$, and $Q\mkv T\mkv X\mkv(Y,Z)$ form a Markov chain, $I(T;Y|Q) - I(T;Z|Q)\leq0$ and $I(T;Y)\leq I(X;Y)$. This concludes the proof.
\end{IEEEproof}




\subsubsection{Secure Source Coding}

Assuming that all terminals are connected by a free-error link of finite-capacity $R$ and defining this rate by $R\triangleq kI(X;Y)$, Proposition~\ref{prop:digital:special:Z} provides as a special case the single-letter characterization of the \emph{rate-distortion-equivocation} region in the setup of secure source coding with uncoded side information given in ~\cite[Theorem~3]{villard2011securea}.

\begin{figure*}[!ht]
\centering
	\begin{tikzpicture}
	\node[matrix,every node/.style={codeword}] at (-1,-.2){
		\node(uhg){};&	\node{};& 			\node(uhm){};&	\node(uhd){}; 	\\
		\node{};& 		\node{};& 			\node{};& 		\node{}; 		\\
		\node{};& 		\node[cok](uc){};&	\node{};& 		\node{};		\\
		\node{};& 		\node{};& 			\node{};&	 	\node{}; 		\\
		\node{};& 		\node{};& 			\node{};& 	 	\node{}; 		\\
		\node(ubg){};&	\node{};& 			\node{};& 	 	\node(ubd){}; 	\\
	};
	\node at (uhm.north west) [above] {\footnotesize $u^n(s_1)$};

	\node (ubgf) at (ubg.south west) [below=.1] {};
	\node (ubdf) at (ubd.south east) [below=.1] {};
	\draw[<->] (ubgf.center) to node[below]{$2^{nR_1}$} (ubdf.center);
	
	\node (uhgf)  at (uhg.north west) [left=.1] {};
	\node (ubgf2) at (ubg.south west) [left=.1] {};
	\draw[<->] (uhgf.center) to node[left]{$2^{n(S_1-R_1)}$} (ubgf2.center);
		
	\node[matrix,every node/.style={codeword}] at (3.5,.3){
		\node(vhg){};&	\node{};& \node(vhm){};&\node{};& 		\node(vhd){}; 	\\
		\node{};& 		\node{};& \node{};& 	\node{};& 		\node{}; 		\\
		\node{};& 		\node{};& \node{};& 	\node[cok]{};&	\node{};		\\
		\node{};& 		\node{};& \node{};&		\node{};& 		\node{}; 		\\
		\node{};& 		\node{};& \node{};& 	\node{};& 		\node{}; 		\\
		\node{};& 		\node{};& \node{};& 	\node{};& 		\node{}; 		\\
		\node(vbg){};&	\node{};& \node{};& 	\node{};& 		\node(vbd){}; 	\\
	};
	\node at (vhm.north) [above]  {\footnotesize $v^n(s_1,s_2)$};	
	
	\draw[dashed] (uc.north east) to (vhg.north west);
	\draw[dashed] (uc.south east) to (vbg.south west);

	\node (vbgf) at (vbg.south west) [below=.1] {};
	\node (vbdf) at (vbd.south east) [below=.1] {};
	\draw[<->] (vbgf.center) to node[below]{$2^{nR_2}$} (vbdf.center);
	
	\node (vhdf)  at (vhd.north east) [right=.1] {};
	\node (vbdf2) at (vbd.south east) [right=.1] {};
	\draw[<->] (vhdf.center) to node[right]{$2^{n(S_2-R_2)}$} (vbdf2.center);
\end{tikzpicture}
\caption{Digital scheme--Source codebook.}
\label{fig:digital:codebook:source}
\end{figure*}
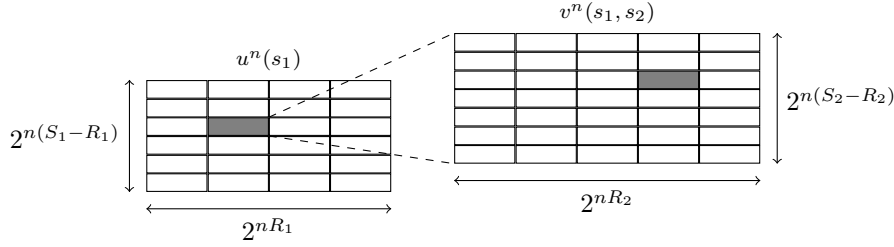

\subsubsection{Wiretap Channel}

Choosing appropriate side informations and auxiliary variables, region $\cR_\text{digital}$ reduces to the achievable region for the wiretap channel~\cite[Eq.~(2.6)]{liang2009information}. To this end, we simply set $V$ to be a degenerated random variable and define $\Delta-H(A|E)\triangleq R_e$ where $R_e$ becomes the conventional secrecy rate in the wiretap channel.

\section{Proof of Theorem~\ref{th:digital} (Digital Scheme)}
\label{sec:digital:proof}

Let $U$, $V$, $Q$, $X$ be four random variables on finite sets $\cU$, $\cV$, $\cQ$, $\cX$, respectively, such that $p(uvqabexyz)=p(u|v)p(v|a)p(abe)\,p(q|x)p(x)p(yz|x)$, a function $\hat A \colon \cV\times\cB \to \cA$, and a tuple $(k,D,\Delta)\in\bR_+^3$.
In this section, we describe a scheme that achieves (under some sufficient conditions) tuple $(k,D,\Delta)$ i.e., for any $\varepsilon>0$, we construct an $(n,m)$-code $(F,g)$ that verifies~\eqref{eq:ach:rate}--\eqref{eq:ach:equiv}.

In this scheme, Alice compresses the source $A$ in descriptions $(U,V)$, with $V$ on the top of $U$.
In view of the side information at Bob $B$, random binning a la Wyner-Ziv~\cite{wyner1976rate} is performed.
The corresponding bin indices $(r_1,r_2)$ are then mapped to indices $(r_c,r_p)$, which are further transmitted to Bob through variables $(Q,X)$ using a code for broadcast channel with confidential messages~\cite{csiszar1978broadcast}, where index $r_p$ is protected by an independent random noise $r_f$.
As in the classical wiretap channel~\cite{csiszar1978broadcast,liang2009information}, its rate $R_f$ satisfies some constraint that allows to characterize the equivocation rate at Eve.

Let $\varepsilon>0$, $R_1, R_2, R_c, R_p, R_f \in\bR_+^*$, $S_1 \geq R_1$, $S_2 \geq R_2$ such that
\begin{equation}
\label{eq:digital:noise}
R_f < (k+\varepsilon)\,I(X;Z|Q) \ ,
\end{equation}
and assume that a local (independent and uniformly distributed) random source with rate $R_f$ is available at Alice.
Define $\gamma = \frac\varepsilon{9\,d_\text{max}}$. 

\subsection{Codebook Generation}

\subsubsection{Source Codewords}

Randomly pick $2^{nS_1}$ sequences $u^n(s_1)$ from $\typ{n}{U}$ and divide them into $2^{nR_1}$ equal size bins $B_1(r_1)$, $r_1\in\{1,\dots,2^{nR_1}\}$.
Then, for each codeword  $u^n(s_1)$, randomly pick $2^{nS_2}$ sequences $v^n(s_1,s_2)$ from $\typ{n}{V|u^n(s_1)}$ and divide them into $2^{nR_2}$ equal size bins $B_2(s_1,r_2)$, $r_2\in\{1,\dots,2^{nR_2}\}$.
See Fig.~\ref{fig:digital:codebook:source}.

\subsubsection{Channel Codewords}

\begin{figure}
\centering
	\begin{tikzpicture}
	\node[matrix,every node/.style={codeword}] at (0,  0){
		\node(qh){};		\\
		\node[cok](qm){};	\\
		\node{}; 			\\
		\node{};			\\
		\node{};			\\
		\node{};			\\
		\node{};			\\
		\node(qb){};		\\
	};

	\node at (qh.north) [above] {\footnotesize $q^m(r_c)$};
	
	\node (qhf) at (qh.north west) [left=.1] {};
	\node (qbf) at (qb.south west) [left=.1] {};
	\draw[<->] (qhf.center) to node[left]{$2^{nR_c}$} (qbf.center);
		
	\node[matrix,every node/.style={codeword}] at (3.5,.3){
		\node(xhg){};&	\node{};& 		\node(xhm){};&\node{};& \node(xhd){}; 	\\
		\node{};& 		\node{};& 		\node{};& 	\node{};& 	\node{}; 		\\
		\node{};& 		\node{};& 		\node{};& 	\node{};&	\node{};		\\
		\node{};& 		\node{};& 		\node{};&	\node{};& 	\node{}; 		\\
		\node{};& 		\node[cok]{};&	\node{};& 	\node{};& 	\node{}; 		\\
		\node{};& 		\node{};& 		\node{};& 	\node{};& 	\node{}; 		\\
		\node(xbg){};&	\node{};& 		\node{};& 	\node{};& 	\node(xbd){}; 	\\
	};

	\node at (xhm.north) [above]  {\footnotesize $x^m(r_c,r_p,r_f)$};	
	
	\draw[dashed] (qm.north east) to (xhg.north west);
	\draw[dashed] (qm.south east) to (xbg.south west);

	\node (xbgf) at (xbg.south west) [below=.1] {};
	\node (xbdf) at (xbd.south east) [below=.1] {};
	\draw[<->] (xbgf.center) to node[below]{$2^{nR_f}$} (xbdf.center);
	
	\node (xhdf)  at (xhd.north east) [right=.1] {};
	\node (xbdf2) at (xbd.south east) [right=.1] {};
	\draw[<->] (xhdf.center) to node[right]{$2^{nR_p}$} (xbdf2.center);
\end{tikzpicture}
\caption{Digital scheme--Channel codebook.}
\label{fig:digital:codebook:channel}
\end{figure}
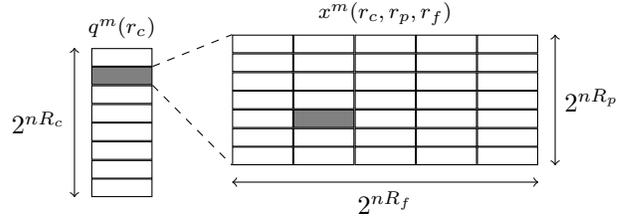

Randomly pick $2^{nR_c}$ sequences $q^m(r_c)$ from $\typ{m}{Q}$.
Then, for each codeword  $q^m(r_c)$,
randomly pick $2^{n(R_p+R_f)}$ sequences $x^m(r_c,r_p,r_f)$ from $\typ{m}{X|q^m(r_c)}$.
See Fig.~\ref{fig:digital:codebook:channel}.

\subsection{Encoding Procedure}

Assume that source sequence $A^n$ and random noise $r_f$ are produced at Alice.

Look for the first codeword $u^n(s_1)$ such that $(u^n(s_1),A^n)\in\typ{n}{U,A}$.
Then look for a codeword $v^n(s_1,s_2)$ such that $(v^n(s_1,s_2),A^n)\in\typ{n}{V,A|u^n(s_1)}$.
Let $B_1(r_1)$ and $B_2(s_1,r_2)$ be the bins of $u^n(s_1)$ and $v^n(s_1,s_2)$, respectively.

Define the mapping $(r_c,r_p) = M(r_1,r_2) \in \{1,\dots,2^{nR_c}\}$ $\times\{1,\dots,2^{nR_p}\}$ where $M$ is an arbitrary fixed one-to-one (invertible) mapping and such that there exists a mapping $M'$ recovering the message $r_1$ from $r_c$ which is $r_1 = M'(r_c)$. 
These two functions can be defined if:
\begin{IEEEeqnarray}{rCl}
R_1 + R_2 	&=& 	R_c + R_p	\ ,\\
R_1 		&\leq& 	R_c 		\ .
\end{IEEEeqnarray}
Notice that the second inequality does not have to necessarily be an equality.  Moreover, Alice then sends $X^m = F(A^n) \triangleq x^m(r_c,r_p,r_f)$.

\subsection{Decoding Procedure}

Assume that Bob observes $B^n$ and receives $Y^m$ from Alice.

Look for the unique codeword $q^m(r_c)$ such that $(q^m(r_c),Y^m)\in\typ{m}{Q,Y}$.
Then look for the unique codeword $x^m(r_c,r_p,r_f)$ such that $(x^m(r_c,r_p,r_f),Y^m)\in\typ{m}{X,Y|q^m(r_c)}$.

Compute $(r_1,r_2) = M^{-1}(r_c,r_p)$.

Look for the unique codeword $u^n(s_1)\in B_1(r_1)$ such that $(u^n(s_1),B^n)\in\typ{n}{U,B}$.
Then look for the unique codeword $v^n(s_1,s_2)\in B_2(s_1,r_2)$ such that $(v^n(s_1,s_2),B^n)\in\typ{n}{V,B|u^n(s_1)}$.

Compute the estimate $g(B^n,Y^m)\in\cA^n$ using the following component-wise relation, for each $i=\{1,\dots,n\}$:
\[
g_i(B^n,Y^m) \triangleq \hat A(v_i(s_1,s_2),B_i) \ .
\]
It is worth mentioning that  given the indices $(r_c,r_p)$ the decoder is able to recover the messages $(r_1,r_2)$. In addition to this, given the common index $r_c$ the decoder is enable to recover at least the message $r_1$.

\subsection{Errors and Constraints}

Denoting by $\err$ the event ``An error occurred during the encoding or decoding steps,'' we expand its probability (averaged over the set of all possible codebooks) as follows: $\pr{\err}\leq P_{t,1}+P_{t,2}+P_{e,1}+P_{e,2}+P_{d,1}+P_{d,2}+P_{d,3}+P_{d,4}$, where each term corresponds to a particular error event, as detailed below.
We derive sufficient conditions on the parameters that make each of these probabilities small for some sufficiently large~$n$.
From now on, let $m=\lfloor n(k+\varepsilon)\rfloor$.\footnote{%
	Note that $m\to\infty$ as $n\to\infty$.}

\subsubsection{Typicality}
From standard properties of typical sequences (see Appendix~\ref{app:typical}), there exists a sequence $\eta_n\toas{n\to\infty}0$ such that
$P_{t,1} \triangleq \pr{(A^n,B^n,E^n)\not\in\typ{n}{A,B,E}} \leq \eta_n$.
Consequently, $P_{t,1} \leq \gamma$ for some sufficiently large~$n$.

Similarly, since the input of the channel $X^m$ is set to some codeword $x^m(r_c,r_p,r_f)\in\typ{m}{X}$, $P_{t,2} \triangleq \pr{(X^m,Y^m,Z^m)\not\in\typ{m}{X,Y,Z}} \leq \gamma$ for some sufficiently large~$n$.

\subsubsection{Encoding}

In the first encoding step, Alice needs to find (at least) one codeword 
$u^n(s_1)$ such that $(u^n(s_1),A^n)\in\typ{n}{U,A}$.
Following standard argument (see e.g.\ \cite[Appendix~B-F-2]{villard2011securea}), we can prove that if $S_1>I(U;A)$, then the probability that this step fails $P_{e,1}$ can be upper bounded by $\gamma$ for some sufficiently large~$n$.

Similarly, the second encoding step succeeds with probability $1-P_{e,2} \geq 1-\gamma$ under condition $S_2>I(V;A|U)$.

\subsubsection{Decoding Indices}

In the first decoding step, Bob looks for the \emph{unique} codeword $q^m(r_c)$ such that $(q^m(r_c),Y^m)\in\typ{m}{Q,Y}$. 
Following standard argument for channel coding, we can prove that if $R_c < (k+\varepsilon) I(Q;Y)$, then the probability that there exists another admissible codeword $P_{d,1}$ can be lowered below $\gamma$ for some sufficiently large~$n$.

Similarly, the second encoding step succeeds with probability $1-P_{d,2} \geq 1-\gamma$ under condition $R_p+R_f < (k+\varepsilon) I(X;Y|Q)$.

\subsubsection{Decoding Source Variables}

In the third decoding step, Bob looks for the \emph{unique} codeword $u^n(s_1)\in B_1(r_1)$ such that $(u^n(s_1),B^n)\in\typ{n}{U,B}$.
Following standard argument for source coding (see e.g.\ \cite[Appendix~B-F-4]{villard2011securea}), we can prove that if $S_1-R_1<I(U;B)$, then the probability that there exists another admissible codeword $P_{d,3}$ can be lowered below $\gamma$ for some sufficiently large~$n$.

Similarly, the fourth decoding step succeeds with probability $1-P_{d,4} \geq 1-\gamma$ under condition $S_2-R_2<I(V;B|U)$.

\subsubsection{Summary}
In this paragraph, we proved that under some sufficient conditions, $\pr{\err} \leq 8\gamma$.

\subsection{Distortion at Bob}

We now check that our code achieves the required distortion level at Bob (averaged over the set of all possible codebooks):
\begin{IEEEeqnarray*}{rCl}
\bE\Big[ d\big(A^n,g(B^n,Y^m)\big) \Big]
	&\leq&		(1-\pr{\err}) \\
	&\times &\bE\left[ d\big(A^n,\hat A\big(v^n(s_1,s_2),B^n\big)\big) \middle| \Nerr \right]\\
	 &+& \pr{\err}d_\text{max}\\
	&\leq&		\bE\big[ d(A,\hat A(V,B)) \big] + \frac\varepsilon9 + \frac{8\varepsilon}9 \ ,
\end{IEEEeqnarray*}
where the last inequality holds for some sufficiently large $n$, and follows from $\pr{\err}\leq 8\gamma$, the definition of $\gamma$, and standard argument in rate-distortion theory from the fact that $(A^n,v^n(s_1,s_2),B^n)\in\typ{n}{A,V,B}$ when no error occured (see e.g.\ \cite[Appendix~B-G]{villard2011securea}).

Condition $D \geq \bE\big[ d(A,\hat A(V,B)) \big]$ is thus sufficient to achieve distortion $D + \varepsilon$ at Bob.

\subsection{Equivocation Rate at Eve}

In the following paragraphs, we prove a lower bound on the equivocation rate at Eve.
We first split up the equivocation into two terms which will be studied separately:
\begin{equation}
\label{eq:digital:proof:Es+Ec}
H(A^n|E^n Z^m)
	= \underbrace{	H(A^n | r_c, r_p E^n Z^m)	}_{E_s}
	+ \underbrace{	I(A^n ; r_c ,r_p | E^n Z^m)	}_{E_c} \ .
\end{equation}

\subsubsection{Study of $E_s$}

The ``source'' term $E_s$ can be written as:
\begin{IEEEeqnarray*}{rCl}
\label{eq:digital:proof:Es}
E_s	&\stackrel{(a)}{=}& 	H(A^n | r_1 ,r_2 E^n) 							\\
	&\stackrel{(b)}{=}&		H(A^n | r_1 E^n) - H(r_2|r_1 E^n)				\\
	&=&						H(A^n | r_1 E^n) - H(r_2) + I(r_2 ; r_1 E^n)	\\
	&\geq & H(A^n | r_1 ,s_1 E^n) - H(r_2) + I(r_2;E^n| r_1) + I(r_2;r_1) \\
	&\geq & H(A^n | s_1 E^n) - H(r_2) + I(r_2;E^n| r_1)  \\
	&\stackrel{(c)}{\geq}&	n H(A | UE) - n\,\tfrac\varepsilon4 - n R_2 + I(r_2 ; E^n | r_1)	\ ,\yesnumber
\end{IEEEeqnarray*}
where
\begin{itemize}
\item step~$(a)$ follows from the Markov chain $A^n\mkv(r_c,r_p,E^n)\mkv Z^m$ and the identity $(r_c,r_p)=M(r_1,r_2)$ where $M$ is a one-to-one mapping,

\item step~$(b)$ from the fact that the bin index $r_2$ is a deterministic function of $A^n$,

\item step~$(c)$ from the fact that the codewords $u^n(s_1)$ are drawn i.i.d., for some sufficiently large~$n$ (see \cite[Lemma~6]{villard2011securea}), the fact that $r_2\in\{1,\dots, 2^{nR_2}\}$, and the non-negativity of mutual information.
\end{itemize}

Note that this term corresponds to the one studied in \cite[Appendix~B-H]{villard2011securea}.
The above lower bound should however be tighter since we do not neglect the remainder term $I(r_2;E^n|r_1)$.

\subsubsection{Study of $E_c$}

The ``channel'' term $E_c$ can be written as:
\begin{IEEEeqnarray*}{rCl}
\label{eq:digital:proof:Ec}
E_c	&=&		H(r_c, r_p | E^n Z^m)		\\[.1em]
	&=&		H(r_p | r_c Z^m) + H(r_c | Z^m) - I(r_c, r_p ; E^n | Z^m) \yesnumber\ ,
\end{IEEEeqnarray*}
where the first step follows from the fact that $(r_c,r_p)$ is a deterministic function of $A^n$.

The first term of the r.h.s.\ of \eqref{eq:digital:proof:Ec} corresponds to the equivocation (of the \emph{private} message, given the \emph{common} message and the output of the channel) in the wiretap channel setting. Following the arguments of~\cite[Section~IV]{csiszar1978broadcast},~\cite[Section~2.3]{liang2009information}, we can easily prove the following lower bound:
\begin{equation}
\label{eq:digital:proof:Ec1}
H(r_p | r_c Z^m) 
	\geq n(R_p + R_f) - m I(X;Z|Q) - 1 - n\,\tfrac\varepsilon2 \ ,
\end{equation}
for some sufficiently large $m$. This proof relies on 
	(\emph{i})		definition $X^m = x^m(r_c,r_p,r_f)$,
	(\emph{ii})	the fact that codewords $x^m(r_c,r_p,r_f)$ are \emph{nearly uniformly} distributed (given $r_c$) over a set of cardinality $2^{n(R_p+R_f)}$,
	(\emph{iii})	the fact that the channel $X\mapsto Z$ is memoryless,
	(\emph{iv})	Fano's inequality together with constraint~\eqref{eq:digital:noise}, which ensures that Eve can decode $x^m(r_c,r_p,r_f)$ from $(r_p,r_f)$ with an arbitrarily small probability of error,
	(\emph{v})		standard properties of typical sequences, and 
	(\emph{vi})	the Markov chain $Q\mkv X\mkv Z$.

\subsubsection{End of Proof}

Gathering \eqref{eq:digital:proof:Es+Ec}--\eqref{eq:digital:proof:Ec1}, we proved that:
\begin{IEEEeqnarray}{rl}
H(A^n|E^n Z^m) & \geq 	n H(A|UE) - n R_2 + n(R_p + R_f)\nonumber\\[.1em]
	&  - m I(X;Z|Q) +	I(r_2 ; E^n | r_1) + H(r_c | Z^m)\nonumber	\\[.1em]
		 &- I(r_c ,r_p ; E^n | Z^m)	-	n\,\tfrac{3\varepsilon}4 - 1 \ .
\label{eq:digital:proof:equiv}
\end{IEEEeqnarray}

We now study the remainder of the r.h.s.\ of the above inequality:
\begin{IEEEeqnarray*}{rCL}
\IEEEeqnarraymulticol{3}{l}{
I(r_2 ; E^n | r_1) + H(r_c | Z^m) - I(r_c, r_p ; E^n | Z^m)}	\\[.1em]\qquad\quad
	&=&					I(r_1 ,r_2 ; E^n) - I(r_1 ; E^n) + H(r_c | Z^m) \\[.1em]
	 \IEEEeqnarraymulticol{3}{R}{- I(Z^m r_c,r_p ; E^n) + I(Z^m ; E^n)}	\\[.1em]
	&\stackrel{(a)}{=}&	- I(r_1 ; E^n) + H(r_c | Z^m) - I(Z^m ; E^n |  r_c, r_p) \\[.1em]
	 \IEEEeqnarraymulticol{3}{R}{+ I(Z^m ; E^n)	}\\[.1em]
	&\stackrel{(b)}{=}&	- I(r_1 ; E^n) + I(r_c Z^m ; E^n) + H(r_c|E^n Z^m)	\\[.1em]
	&\stackrel{(c)}{\geq}&	0	\ ,
\end{IEEEeqnarray*}
where 
\begin{itemize}
\item step~$(a)$ follows from the identity $(r_c,r_p)=M(r_1,r_2)$ where $M$ is a one-to-one mapping,
\item step~$(b)$ follows by noting that $I(Z^m ; E^n |  r_c, r_p)=0$ because $E^n\mkv (r_c,r_p)\mkv Z^m$ form a Markov chain since the output $Z^m$ only depends on $(A^n,E^n,B^n)$ through the messages $(r_c,r_p)$  sent by the channel encoder, 
\item step~$(c)$ from the fact that $r_1=M'(r_c)$ for some mapping $M'$, and the non-negativity of conditional entropy and mutual information.
\end{itemize}

Inequality \eqref{eq:digital:proof:equiv} then yields
\begin{IEEEeqnarray*}{rCL}
\frac1n H(A^n|E^n Z^m) &\geq& H(A|UE) - R_2 + R_p + R_f \\
	&-& \frac{m}{n}\,I(X;Z|Q) - \varepsilon \ ,
\end{IEEEeqnarray*}
for some sufficiently large $n$. 

Condition $\Delta \leq H(A|UE) - R_2 + R_p + R_f - (k+\varepsilon) I(X;Z|Q)$ is thus sufficient to achieve equivocation rate $\Delta - \varepsilon$.

\subsection{Summary of Sufficient Conditions}

Putting all pieces together, we proved that the following inequalities are sufficient conditions for a tuple $(k,D,\Delta)\in\bR_+^3$ to be achievable: For each $\varepsilon > 0$,
\[
\left\{
\begin{IEEEeqnarraybox}[][c]{rCl}
\IEEEeqnarraymulticol{3}{l}{R_1, R_2, R_c, R_p, R_f > 0}				\\
R_1+R_2		&=&		R_c+R_p												\\
R_1		 	&\leq& 	R_c 												\\
S_1			&\geq&	R_1													\\
S_2			&\geq&	R_2													\\
S_1			&>&		I(U;A)												\\
S_2			&>&		I(V;A|U)											\\
R_c 		&<&		(k+\varepsilon) I(Q;Y)								\\
R_p + R_f	&<&		(k+\varepsilon) I(X;Y|Q)							\\
S_1-R_1		&<&		I(U;B)												\\
S_2-R_2		&<&		I(V;B|U)											\\
R_f 		&<&		(k+\varepsilon) I(X;Z|Q)							\\
D 			&\geq&	\bE\big[d(A,\hat A(V,B))\big]						\\
\Delta		&\leq&	H(A|UE) - R_2 + R_p  + R_f  \\
 \IEEEeqnarraymulticol{3}{R}{- (k+\varepsilon) I(X;Z|Q)}
\end{IEEEeqnarraybox}
\right.
\]

Using Fourier-Motzkin elimination~\cite{villard2011fourier-motzkin},
it is straightforward to prove that this system of inequalities is equivalent to:
\[
\left\{
\begin{IEEEeqnarraybox}[][c]{rCl}
I(U;A|B)		&<&	(k+\varepsilon) I(Q;Y)									\\
I(V;A|B)		&<&	(k+\varepsilon) I(X;Y) 									\\
D 				&\geq&	\bE\big[d(A,\hat A(V,B))\big]						\\
\Delta 			&<&	H(A|UE) 												\\
\Delta			&<&	H(A|UE) - I(V;A|UB) 
							\\
							&+& (k+\varepsilon) \Big( I(X;Y|Q) - I(X;Z|Q) \Big)
\end{IEEEeqnarraybox}
\right.
\]

\subsection{Channel Prefixing}

For each random variable $T$ on some finite set $\cT$ such that $T\mkv X\mkv (Y,Z)$ form a Markov chain, we can use the above scheme considering the DMC $T\mapsto(Y,Z)$ instead of $X\mapsto(Y,Z)$.
In this case, the above sufficient conditions write
\[
\left\{
\begin{IEEEeqnarraybox}[][c]{rCl}
I(U;A|B)		&<&	(k+\varepsilon) I(Q;Y) 									\\
I(V;A|B)		&<&	(k+\varepsilon) I(T;Y)									\\
D 				&\geq&	\bE\big[d(A,\hat A(V,B))\big]						\\
\Delta 			&<&	H(A|UE)													\\
\Delta			&<&	H(A|UE) - I(V;A|UB) \\
							&+& (k+\varepsilon) \Big( I(T;Y|Q) - I(T;Z|Q) \Big)
\end{IEEEeqnarraybox}
\right.
\]
Since region $\cR^*$ is closed, this proves Theorem~\ref{th:digital}.
\endIEEEproof

\section{Secure Transmission of a Binary Source With BEC/BSC Side Informations Over a Type-II Wiretap Channel}
\label{sec:binary}

\subsection{System Model}

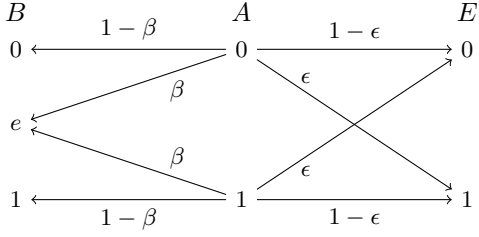
\begin{figure}
\centering
	\begin{tikzpicture}
	\node	(A) 	at (0,.5) 	{$A$};
	\node	(A0) 	at (0,0) 	{\small $0$};
	\node	(A1)	at (0,-2)	{\small $1$};
	
	\node	(B) 	at (-3,.5) 	{$B$};
	\node	(B0) 	at (-3,0) 	{\small $0$};
	\node	(Be)	at (-3,-1)	{\small $e$};
	\node	(B1)	at (-3,-2)	{\small $1$};
	
	\node	(E) 	at (3,.5) 	{$E$};
	\node	(E0) 	at (3,0) 	{\small $0$};
	\node	(E1)	at (3,-2)	{\small $1$};
	
	\draw[->] (A0) to node[midway,above]		{\small $1-\epsilon$}	(E0);
	\draw[->] (A0) to node[near start,above]	{\small $\epsilon$}		(E1);
	\draw[->] (A1) to node[near start,below]	{\small $\epsilon$}		(E0);
	\draw[->] (A1) to node[midway,below]		{\small $1-\epsilon$}	(E1);
	\draw[->] (A0) to node[midway,above]		{\small $1-\beta$}		(B0);
	\draw[->] (A0) to node[near start,below]	{\small $\beta$}		(Be);
	\draw[->] (A1) to node[near start,above]	{\small $\beta$}		(Be);
	\draw[->] (A1) to node[midway,below]		{\small $1-\beta$}		(B1);
\end{tikzpicture}
\caption{Binary source with BEC/BSC side informations.}
\label{fig:binary:source}
\end{figure}

\begin{figure}
\centering
	\begin{tikzpicture}[scale=0.92]
	\node	(ci)	at (-.4,0)	{};
	\node	(c0) 	at (0,0) 	{};
	\node	(c1)	at (2.5,0)	{};
	\node	(c2)	at (5,0)	{};
	\node	(c3)	at (8,0)	{};
	\node	(e) 	at (9,0) 	{};
	
	\draw[->] (ci) to (e);
	\draw (c0) to node[below] {\small $A\mkv B\mkv E$}		(c1);
	\draw (c1) to node[below] {\small $B\lessnoisy{A}E$}	(c2);
	\draw (c2) to node[below] {\small $I(A;B)\geq I(A;E)$}	(c3);
	
	\draw (c0)+(0,-2pt) -- +(0,2pt)	node[above]	{$0$};
	\draw (c1)+(0,-2pt) -- +(0,2pt)	node[above]	{$2\epsilon$};
	\draw (c2)+(0,-2pt) -- +(0,2pt) node[above] {$4\epsilon(1-\epsilon)$};
	\draw (c3)+(0,-2pt) -- +(0,2pt) node[above] {$h_2(\epsilon)$};
	\draw (e) 						node[above] {$\beta$};
\end{tikzpicture}
\caption{Relative properties of the side informations as a function of $(\beta,\epsilon)$.}
\label{fig:binary:cas}
\end{figure}
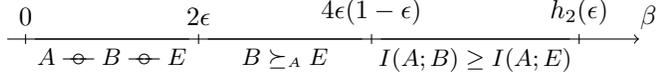

Consider the source model depicted in Fig.~\ref{fig:binary:source}, where the source is binary uniformly distributed ($A\sim\cB\left(\tfrac12\right)$). The side information at Bob is the output of a binary erasure channel (BEC) with erasure probability $\beta\in(0,1]$ and  input $A$. The side information at Eve is the output of a  binary symmetric channel (BSC) with crossover probability $\epsilon\in\intFF{0}{\tfrac12}$  and input $A$. Recall that according to the values of the parameters $(\beta,\epsilon)$, the side informations satisfy the properties summarized in Fig.~\ref{fig:binary:cas} \cite{nair2009capacitya}.

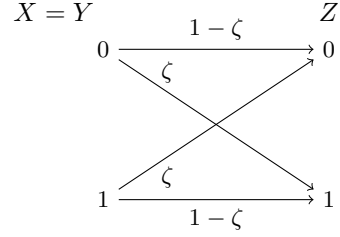
\begin{figure}
\centering
	\begin{tikzpicture}
	\node[left]	(X) 	at (0,.5) 	{$X=Y$};
	\node		(X0) 	at (0,0) 	{\small $0$};
	\node		(X1)	at (0,-2)	{\small $1$};
	
	\node		(Z) 	at (3,.5) 	{$Z$};
	\node		(Z0) 	at (3,0) 	{\small $0$};
	\node		(Z1)	at (3,-2)	{\small $1$};
	
	\draw[->] (X0) to node[midway,above]		{\small $1-\zeta$}	(Z0);
	\draw[->] (X0) to node[near start,above]	{\small $\zeta$}	(Z1);
	\draw[->] (X1) to node[near start,below]	{\small $\zeta$}	(Z0);
	\draw[->] (X1) to node[midway,below]		{\small $1-\zeta$}	(Z1);
\end{tikzpicture}
\caption{Type-II wiretap channel.}
\label{fig:binary:channel}
\end{figure}

The communication channel is similar to the type-II wiretap channel of~\cite{wyner1975wire}: It consists of a noiseless channel from Alice to Bob, and a BSC with crossover probability $\zeta\in\intFF{0}{\tfrac12}$, from Alice to Eve (see Fig.~\ref{fig:binary:channel}).

In this section, we focus on \emph{lossless} reconstruction at Bob ($d$ is the Hamming distance and $D=0$) and \emph{matched bandwidth} ($k=1$).

\subsection{Performance of Coding Schemes}

From the general outer bound of Theorem~\ref{th:outer}, we can easily derive the following result.
\begin{proposition}[Outer bound]
\label{prop:binary:outer}
If $(k=1,D=0,\Delta)$ is achievable, then there exist $u,q\in\intFF{0}{\tfrac12}$ such that
\begin{IEEEeqnarray*}{lcl}
\Delta	 		&\leq & h_2(\epsilon)+h_2(u)-h_2(\epsilon\star u)	\\
		&-& \Big[ \beta h_2(u)				
		- \Big( h_2(\zeta) + h_2(q) - h_2(\zeta\star q) \Big) \Big]_+	\ .
\end{IEEEeqnarray*}
\end{proposition}

\begin{IEEEproof}
The proof is similar to the one of the converse part of Proposition~\ref{prop:binary:digital} below, given in Appendix~\ref{app:binary:digital}.
Details are omitted.
\end{IEEEproof}

The following proposition provides a simple expression of region $\cR_\text{digital}$.

\begin{proposition}[Digital scheme]
\label{prop:binary:digital}
$(k=1,D=0,\Delta)\in\cR_\text{digital}$ if and only if there exist $u,q\in\intFF{0}{\tfrac12}$ such that
\begin{IEEEeqnarray*}{rcl}
\beta(1-h_2(u))	&\leq& 1-h_2(q)	 										\ ,\\ [.1cm]
\Delta	 		&\leq&  h_2(\epsilon)+h_2(u)-h_2(\epsilon\star u)	\\ [.1cm]
		&-& \Big[ \beta h_2(u)				
		- \Big( h_2(\zeta) + h_2(q) - h_2(\zeta\star q) \Big) \Big]_+	\ .
\end{IEEEeqnarray*}
\end{proposition}

\begin{IEEEproof}
The proof of the converse part is given in Appendix~\ref{app:binary:digital}.

The direct part follows from Theorem~\ref{th:digital} after some straightforward manipulations choosing auxiliary variables as follows (details are omitted):
$V=A$;
$X\sim\bern\left(\tfrac12\right)$;
$U$ (resp. $Q$) is the output of a BSC with crossover probability $u\in\intFF{0}{\tfrac12}$ (resp. $q\in\intFF{0}{\tfrac12}$) and input $A$ (resp. $X$).
\end{IEEEproof}

Notice that if $\beta\leq 4\epsilon(1-\epsilon)$, then $B\lessnoisy{A}E$ (see Fig.~\ref{fig:binary:cas}), and hence Proposition~\ref{prop:digital:special:B} holds i.e., the above inner bound is optimal and separation holds.

In the following, we will compare the above digital scheme with a pure analog one, consisting in directly sending the source over the channel.
Its performance is given by the following proposition.

\begin{proposition}[Analog scheme]
\label{prop:binary:analog}
A tuple $(k=1,D=0,\Delta)\in\bR_+^3$ is achievable through an analog scheme if
\[
\Delta \leq h_2(\epsilon) + h_2(\zeta) - h_2(\zeta\star\epsilon) \ .
\]
\end{proposition}

\begin{IEEEproof}
Letting $X=A$ yields zero distortion at Bob (since $Y=X$) and equivocation rate $H(A|EZ)$ at Eve.
The above expression follows after some straightforward manipulations.
Details are omitted.
\end{IEEEproof}

\subsection{Counterexample for the Optimality of Theorem~\ref{th:digital}}\label{sec:binary:numericalr}

Let now assume that Bob does not have any side information i.e., $B=\emptyset$ or equivalently $\beta=1$, and let $\epsilon=\zeta=0.1$, so that $A\mkv E\mkv B$ and $X\mkv Y\mkv Z$ form Markov chains, and neither Proposition~\ref{prop:digital:special:B}, nor Proposition~\ref{prop:digital:special:Z} applies. 

This setting provides a  counterexample for the general optimality of the inner bound in Theorem~\ref{th:digital}: 
Numerical optimization over $u$ and $q$ in Proposition~\ref{prop:binary:digital} indicates that the proposed digital scheme achieves an equivocation rate $\Delta= 0.056$, while the naive analog scheme of Proposition~\ref{prop:binary:analog} achieves $\Delta = 0.258$.
Furthermore, the latter coincides with the outer bound of Proposition~\ref{prop:binary:outer}.
This shows that a \emph{joint} source-channel scheme may achieve better performance in some cases.

\section{Hybrid Coding}
\label{sec:hybrid}

Based on the observations of the previous section about the usefulness of \emph{analog schemes}, we propose in this section a \emph{hybrid digital/analog scheme} that yields a new single-letter inner bound $\cR_\text{hybrid}$ in the matched-bandwidth case (Theorem~\ref{th:hybrid}).

\subsection{General Statement}

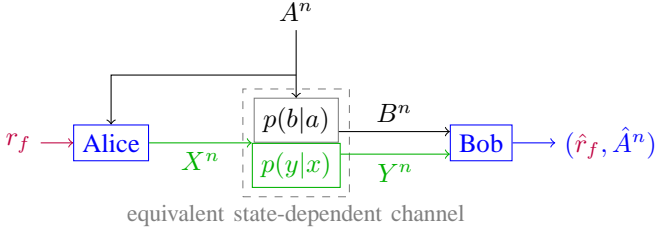
\begin{figure}
\centering
	\begin{tikzpicture}[scale=0.82]	
	\node[purple]			(noise)		at (  0,  0)		{$r_f$};
	\node[sensor,gentil]	(alice)		at (1.5,  0)	 	{Alice};
	
	\node[matrix,draw=gray,dashed] (transitions) at (4.5,  0){
		\node[trans]		(probB)		{$p(b|a)$}; \\
		\node[trans,canal]	(canal)		{$p(y|x)$}; \\
	};	
	
	\node[below=.7,gray]	at (transitions)	
		{\small equivalent state-dependent channel};
	
	\node[above=1.5]		(A) 		at (transitions)	{$A^n$};
	\node[sensor,gentil]	(bob) 		at (7.5,  0)	 	{Bob};
	\node[right,gentil]		(hatA) 		at (8.7,  0)		{$({\color{purple}\hat r_f},\hat A^n)$};
		
	\draw[->,purple]	(noise) to (alice);
	
	\draw[->]			(A)			to (probB);
	\draw[->]			(A|-0,1.1)	-| (alice);
	
	\node[above=.15]	(bg) 			at (probB.south east) {};
	\draw[->]			(bg.south)	to node[above]{$B^n$} (bob.west|-bg.south);
	
	\draw[->,canal]		(alice)		to node[below]{$X^n$} (canal.north west);
	
	\node[below=.15] 	(yg) at (canal.north east)  {};
	\draw[->,canal]		(yg.north)	to node[below]{$Y^n$} (bob.west|-yg.north);
	
	\draw[->,gentil]	(bob)	to (hatA);
\end{tikzpicture}
\caption{Alice and Bob as a system with state-dependent channel and CSIT.}
\label{fig:gelfand}
\end{figure}

Channels $A\mapsto B$ and $X\mapsto Y$ can be viewed together as a state-dependent channel with input $X$, state $A$ and output $(B,Y)$.
In this perspective, Alice and Bob form a communication system with channel state information non-causally known at the transmitter (CSIT), as depicted in Fig.~\ref{fig:gelfand}.
Roughly speaking, the proposed scheme consists in sending independent digital random noise $r_f$ using a Gelfand-Pinsker code~\cite{gel'fand1980coding} for this equivalent state-dependent channel.

\begin{theorem}[Hybrid scheme]
\label{th:hybrid}
A tuple $(k=1,D,\Delta)\in\bR_+^3$ is achievable if there exist random variables $U$, $V$, $X$ on finite sets $\cU$, $\cV$, $\cX$, with joint distribution $p(uvabexyz) = p(u|v) p(vx|a) p(abe) (yz|x)$, $x=x(v,a)$, and a function $\hat A \colon \cV\times\cB\times\cY \to \cA$, verifying
\begin{IEEEeqnarray}{rcl}
I(U;A)		&\leq& I(U;BY) 								\ ,	\label{eq:hybrid1}\\[.1cm]
I(V;A|U)	&\leq& I(V;BY|U)								\ ,	\label{eq:hybrid2}\\[.1cm]
D			&\geq& \bE\big[d(A,\hat A(V,B,Y))\big]		\ ,	\label{eq:hybrid3}\\[.1cm]
\Delta		&\leq&  H(A|UE) - I(V;A|U) - I(X;Z|UE) \nonumber\\
			&+& \min\Big\{ I(V;BY|U) \,; I(V;AZ|U) \Big\}	\ .\qquad\label{eq:hybrid4}
\end{IEEEeqnarray}
\end{theorem}

\begin{IEEEproof}
See Section~\ref{sec:hybrid:proof}.
\end{IEEEproof}

Inequalities~\eqref{eq:hybrid1}, \eqref{eq:hybrid2} correspond to sufficient conditions for the transmission of descriptions $U$, $V$ of $A$.
The first layer $U$ can be seen as a \emph{common} message which is considered to be known at Eve, as shown by the term $H(A|UE)$ in~\eqref{eq:hybrid4}.
Digital random noise $r_f$ helps to secure the second layer $V$ against Eve.

\subsection{Special Cases}

\subsubsection{Analog Schemes}
\label{sec:hybrid:special:analog}
The proposed scheme can reduce to a pure analog one (as the simple one of Proposition~\ref{prop:binary:analog}).
Hence, $\cR_\text{hybrid}$ contains tuples that may not be in $\cR_\text{digital}$: $\cR_\text{hybrid}\not\subset\cR_\text{digital}$.

\subsubsection{Digital Schemes}\label{sec:hybrid:special:digital}

By defining the variables in Theorem~\ref{th:hybrid} as pairs of independent source and channel components, we can obtain the structure of those in Theorem~\ref{th:digital}, but such variables do not verify all inequalities and thus $\cR_\text{digital}\not\subset\cR_\text{hybrid}$.

\subsubsection{Wiretap Channel}
Choosing independent source and channel variables with appropriate rates, region $\cR_\text{hybrid}$ reduces to the achievable region for the wiretap channel~\cite[Eq.~(2.6)]{liang2009information}.

\section{Proof of Theorem~\ref{th:hybrid} (Hybrid Scheme)}
\label{sec:hybrid:proof}

Let $U$, $V$, $X$ be three random variables on finite sets $\cU$, $\cV$, $\cX$, respectively, such that $p(uvabexyz)=p(u|v)p(vx|a)p(abe)p(yz|x)$, $x=x(v,a)$, a function $\hat A \colon \cV\times\cB\times\cY \to \cA$, and a tuple $(D,\Delta)\in\bR_+^2$.
In this section, we describe a scheme that achieves (under some sufficient conditions) tuple $(k=1,D,\Delta)$ i.e., for any $\varepsilon>0$, we construct an $(n,n)$-code $(F,g)$ such that:
\begin{IEEEeqnarray*}{rCl}
\bE\big[ d\big(A^n,g(B^n,Y^n)\big) \big]	&\leq& D+\varepsilon 		\ ,\\
\dfrac1n\,H(A^n|E^n Z^n)	 				&\geq& \Delta-\varepsilon 	\ .
\end{IEEEeqnarray*}

In this scheme, Alice compresses the source $A$ in descriptions $(U,V)$, with $V$ on the top of $U$.
Digital random noise $r_f$ is also transmitted on $V$ (a la Gelfand-Pinsker) to take advantage of the possibly better quality of Bob's channel, and prevent Eve from decoding the whole message.
As in the classical wiretap channel~\cite{csiszar1978broadcast,liang2009information}, its rate $R_f$ satisfies some constraint that allows to characterize the equivocation rate at Eve.
Alice finally sends some deterministic function $x(V,A)$ of $V$ and $A$.

Let $\varepsilon>0$, $R_1,R_2,R_f\in\bR_+^*$ such that
\begin{equation}
\label{eq:hybrid:noise}
R_2 + R_f < I(V;AZ|U) \ ,
\end{equation}
and assume that a local (independent and uniformly distributed) random source with rate $R_f$ is available at Alice.
Define $\gamma = \frac\varepsilon{7\,d_\text{max}}$.

\subsection{Codebook Generation}

\begin{figure}[!t]
\centering
	\begin{tikzpicture}
	\node[matrix,every node/.style={codeword}] at (0,  0){
		\node(uh){};		\\
		\node{};			\\
		\node{}; 			\\
		\node[cok](um){};	\\
		\node{};			\\
		\node{};			\\
		\node{};			\\
		\node(ub){};		\\
	};

	\node at (uh.north) [above] {\footnotesize $u^n(r_1)$};
	
	\node (uhf) at (uh.north west) [left=.1] {};
	\node (ubf) at (ub.south west) [left=.1] {};
	\draw[<->] (uhf.center) to node[left]{$2^{nR_1}$} (ubf.center);
		
	\node[matrix,every node/.style={codeword}] at (3.5,.3){
		\node(vhg){};&	\node{};& \node(vhm){};&\node{};& 		\node(vhd){}; 	\\
		\node{};& 		\node{};& \node{};& 	\node{};& 		\node{}; 		\\
		\node{};& 		\node{};& \node{};& 	\node[cok]{};&	\node{};		\\
		\node{};& 		\node{};& \node{};&		\node{};& 		\node{}; 		\\
		\node{};& 		\node{};& \node{};& 	\node{};& 		\node{}; 		\\
		\node{};& 		\node{};& \node{};& 	\node{};& 		\node{}; 		\\
		\node(vbg){};&	\node{};& \node{};& 	\node{};& 		\node(vbd){}; 	\\
	};

	\node at (vhm.north) [above]  {\footnotesize $v^n(r_1,r_2,r_f)$};	
	
	\draw[dashed] (um.north east) to (vhg.north west);
	\draw[dashed] (um.south east) to (vbg.south west);

	\node (vbgf) at (vbg.south west) [below=.1] {};
	\node (vbdf) at (vbd.south east) [below=.1] {};
	\draw[<->] (vbgf.center) to node[below]{$2^{nR_f}$} (vbdf.center);
	
	\node (vhdf)  at (vhd.north east) [right=.1] {};
	\node (vbdf2) at (vbd.south east) [right=.1] {};
	\draw[<->] (vhdf.center) to node[right]{$2^{nR_2}$} (vbdf2.center);
\end{tikzpicture}
\caption{Hybrid scheme--Codebook.}
\label{fig:hybrid:codebook}
\end{figure}
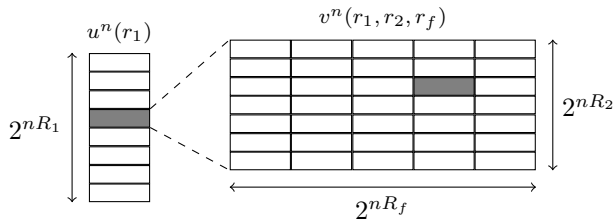

Randomly pick $2^{nR_1}$ sequences $u^n(r_1)$ from $\typ{n}{U}$.
Then, for each codeword  $u^n(r_1)$, randomly pick $2^{n(R_2+R_f)}$ sequences $v^n(r_1,r_2,r_f)$ from $\typ{n}{V|u^n(r_1)}$.
See Fig.~\ref{fig:hybrid:codebook}.

\subsection{Encoding Procedure}
\label{sec:hybrid:proof:encoding}

Assume that source sequence $A^n$ and random noise $r_f$ are produced at Alice.

Look for the first codeword $u^n(r_1)$ such that $(u^n(r_1),A^n)\in\typ{n}{U,A}$.
Then look for the first codeword $v^n(r_1,r_2,r_f)$ such that $(v^n(r_1,r_2,r_f),A^n)\in\typ{n}{V,A|u^n(r_1)}$.

Alice then sends $X^n = F(A^n)$ defined by the following component-wise relation, for each $i=\{1,\dots,n\}$:
\[
X_i \triangleq x(v_i(r_1,r_2,r_f),A_i)\ .
\]
It is worth mentioning here that this scheme, as opposed to the pure digital scheme in Theorem~2, does not use binning. The reason for this is that it is not needed since decoding at the destination (Bob)  is jointly performed  over both the side information and the channel output creating the same effect than binning (similar conclusions are drawn in~\cite{tuncel2006slepian-wolf}).

\subsection{Decoding Procedure}
\label{sec:hybrid:proof:decoding}

Assume that Bob observes $B^n$ and receives $Y^n$ from Alice.

Look for the unique codeword $u^n(r_1)$ such that $(u^n(r_1),B^n,Y^n) \in \typ{n}{U,B,Y}$.
Then look for the unique $v^n(r_1,r_2,r_f)$ such that $(v^n(r_1,r_2,r_f),B^n,Y^n) \in \typ{n}{V,B,Y|u^n(r_1)}$.

Compute the estimate $g(B^n,Y^n)\in\cA^n$ using the following component-wise relation, for each $i=\{1,\dots,n\}$:
\[
g_i(B^n,Y^n) \triangleq \hat A(v_i(r_1,r_2,r_f),B_i,Y_i) \ .
\]

\subsection{Errors and Constraints}
\label{sec:hybrid:proof:errors}

Denoting by $\err$ the event ``An error occurred during the encoding or decoding steps,'' we expand its probability (averaged over the set of all possible codebooks) as follows:\linebreak
$\pr{\err}\leq P_{t,1}+P_{t,2}+P_{e,1}+P_{e,2}+P_{d,1}+P_{d,2}$, where each term corresponds to a particular error event, as detailed below.
We derive sufficient conditions on the parameters that make each of these probabilities small.

\subsubsection{Typicality}

From standard properties of typical sequences (see Appendix~\ref{app:typical}), there exists a sequence $\eta_n\toas{n\to\infty}0$ such that $P_{t,1} \triangleq \pr{(A^n,B^n,E^n)\not\in\typ{n}{A,B,E}} \leq \eta_n$.
Consequently, $P_{t,1} \leq \gamma$ for some sufficiently large $n$.

Similarly, since the input of the channel $X^n$ is typical (when no error occurs during the encoding steps), $P_{t,2} \triangleq \pr{(X^n,Y^n,Z^n)\not\in\typ{n}{X,Y,Z}} \leq \gamma$ for some sufficiently large $n$.

\subsubsection{Encoding}

In the first encoding step, Alice needs to find (at least) one codeword $u^n(r_1)$ such that $(u^n(r_1),A^n)\in\typ{n}{U,A}$.
Following standard argument (see e.g.\ \cite[Appendix~B-F-2]{villard2011securea}, we can prove that if $R_1>I(U;A)$, then the probability that this step fails $P_{e,1}$ can be upper bounded by $\gamma$ for some sufficiently large $n$.

Similarly, the second encoding step succeeds with probability $1-P_{e,2} \geq 1-\gamma$ under condition $R_2>I(V;A|U)$.

\subsubsection{Decoding}

In the first decoding step, Alice should find the unique codeword $u^n(r_1)$ such that $(u^n(r_1),B^n,Y^n) \in \typ{n}{U,B,Y}$.
The corresponding error probability $P_{d,1}$ must be carefully handled.

As previously noted in~\cite{lapidoth2010sending} and~\cite{lim2010lossy}, the conventional random coding proof technique does not apply here.
In the proposed joint coding scheme, a single codebook plays both roles of source and channel codebooks.
For a given source sequence $a^n$, the indices $(r_1,r_2)$ thus depend on the entire codebooks, and the averaging over the set of all possible codebooks cannot be performed in the usual way.
Similarly to~\cite{lim2010lossy}, we can prove the following lemma.
\begin{lemma}
\label{lem:hybrid:proof:decoding}
There exists $\kappa<1$ and a sequence $\eta_n\toas{n\to\infty}0$ such that
\[
P_{d,1} \leq \frac{2^{n( R_1 - I(U;BY) + \eta_n )}}{(1 - \kappa)^2}	\ ,
\]
for some sufficiently large $n$.
\end{lemma}

\begin{IEEEproof}
See Appendix~\ref{app:hybrid:proof:decoding}.
\end{IEEEproof}

From Lemma~\ref{lem:hybrid:proof:decoding}, if $R_1<I(U;BY)$, then probability $P_{d,1}$ vanishes as $n$ tends to infinity, and hence can be upper bounded by $\gamma$ for some sufficiently large $n$.
		
Using similar arguments, we can prove that the second decoding step succeeds with probability $1-P_{d,2} \geq 1-\gamma$ under condition $R_2+R_f<I(V;BY|U)$.

\subsubsection{Summary}
In this paragraph, we proved that under some sufficient conditions, $\pr{\err} \leq 6\gamma$.

\subsection{Distortion at Bob}

We now check that our code achieves the required distortion level at Bob (averaged over the set of all possible codebooks):
\begin{IEEEeqnarray*}{ll}
\bE\Big[ d\big(A^n,\, &g(B^n,Y^n)\big) \Big]\\[.1cm]
	&\leq		(1-\pr{\err}) \\[.1cm]
	&\times \bE\left[ d\big(A^n,\hat A\big(v^n(r_1,r_2,r_f),B^n,Y^n\big)\big) \middle| \Nerr \right] \\[.1cm]
	&+ \pr{\err}d_\text{max}\\[.1cm]
	&\leq		\bE\big[ d(A,\hat A(V,B,Y)) \big] + \frac\varepsilon7 + \frac{6\varepsilon}7 \ ,
\end{IEEEeqnarray*}
where the last inequality holds for some sufficiently large $n$, and follows from $\pr{\err}\leq 6\gamma$, the definition of $\gamma$, and standard argument in rate-distortion theory from the fact that $(A^n,v^n(r_1,r_2,r_f),B^n,Y^n)\in\typ{n}{A,V,B,Y}$ when no error occured (see e.g.\ \cite[Appendix~B-G]{villard2011securea}).

Condition $D \geq \bE\big[ d(A,\hat A(V,B,Y)) \big]$ is thus sufficient to achieve distortion $D + \varepsilon$.

\subsection{Equivocation Rate at Eve}

\providecommand{\resteEps}{\tfrac\varepsilon4}

The equivocation at Eve can be divided in ``source'' and ``channel'' terms. 
Each one is studied using standard properties of typical sequences, and following the arguments of \cite[Appendix~B-H]{villard2011securea} and~\cite[Section~2.3.\emph{Step 3}]{liang2009information}:
\begin{IEEEeqnarray*}{ll}
\label{eq:equivocation}
&H(A^n | E^n Z^n)\\[.1cm]
	&= 					H(r_1 r_2 r_f A^n X^n | E^n Z^n) 
								- H(r_1 r_2 r_f X^n | A^n E^n Z^n)			\\ [.1cm]
	&\stackrel{(a)}{=}		H(r_1 r_2 r_f A^n X^n | E^n Z^n) 
								- H(r_2 r_f | r_1 A^n Z^n)					\\[.1cm]
	&\stackrel{(b)}{\geq}	H(r_1 r_2 r_f A^n X^n | E^n Z^n) - n\,\resteEps	\\[.1cm]
	&\stackrel{(c)}{\geq}	H(r_2 r_f A^n X^n | r_1 E^n Z^n) - n\,\resteEps	\\[.1cm]
	&=						H(r_2 r_f A^n X^n | r_1) 
								+ H(E^n Z^n | r_1 r_2 r_f A^n X^n) \\[.1cm]
								&- H(E^n Z^n | r_1) - n\,\resteEps			\\[.1cm]
	&\stackrel{(d)}{=}		H(r_f A^n | r_1) + H(E^n Z^n | A^n X^n)	
								- H(E^n Z^n | r_1) - n\,\resteEps			\\[.1cm]
	&\stackrel{(e)}{=}		H(A^n | r_1) + H(r_f) + H(E^n | A^n)  + H(Z^n | X^n)\\[.1cm]
 &- H(E^n Z^n | r_1) - n\,\resteEps	\ ,\quad\yesnumber
\end{IEEEeqnarray*}
where
\begin{itemize}
\item step~$(a)$ follows from the fact that $X^n$ (resp. $r_1$) is a deterministic function of $(r_1,r_2,r_f,A^n)$ (resp. $A^n$), and the Markov chain $(r_2,r_f)\mkv (A^n,Z^n)\mkv E^n$,

\item step~$(b)$ from condition~\eqref{eq:hybrid:noise} (which ensures that Eve can decode $v^n(r_1,r_2,r_f)$ from $(u^n(r_1),\linebreak A^n,Z^n)$ with an arbitrarily small probability of error, using a decoder similar to Bob's one--see Sections~\ref{sec:hybrid:proof:decoding}, \ref{sec:hybrid:proof:errors}), and Fano's inequality (for some sufficiently large $n$),

\item step~$(c)$ from the fact that conditioning reduces the entropy,

\item step~$(d)$ from the fact that $(r_2,X^n)$ is a deterministic function of $(r_1,r_f,A^n)$, and the Markov chain $(E^n,Z^n)\mkv(A^n,X^n) \mkv(r_1,r_2,r_f)$,

\item step~$(e)$ from the fact that $r_f$ is independent of $(r_1,A^n)$, and the Markov chains $E^n\mkv A^n \mkv(X^n,Z^n)$, $Z^n\mkv X^n \mkv(A^n,E^n)$.
\end{itemize}

We now separately study each term of the r.h.s.\ of~\eqref{eq:equivocation}:
\begin{itemize}
\item From the fact that the codewords $u^n(r_1)$ are drawn i.i.d., and following the argument of \cite[Lemma~6]{villard2011securea}, we can prove that
\[
H(A^n | r_1) \geq n \left( H(A|U) - \resteEps \right) \ ,
\]
for some sufficiently large $n$.

\item Since the random source $r_f$ is uniformly distributed with rate $R_f$:
\[
H(r_f) = n R_f \ .
\]

\item Since the sources are i.i.d.:
\[
H(E^n | A^n) = n H(E|A) \ .
\]

\item Since the channel is memoryless and the input $X^n$ is typical (see~\cite[Eq.~(2.46)]{liang2009information}),
\[
H(Z^n | X^n) \geq n \left( H(Z|X) - \resteEps \right) \ .
\]

\item From the fact that $(u^n(r_1),E^n,Z^n)$ are jointly typical, and following the arguments of~\cite[Eq.~(2.54)]{liang2009information}, we can prove that
\[
H(E^n Z^n | r_1)  \leq n \left( H(EZ|U) + \resteEps \right) \ .
\]
\end{itemize}
\begin{figure*}[!ht]
\centering


\begin{tikzpicture}

\tikzstyle{outer}	=[			color=black]
\tikzstyle{hybrid}	=[			color=red,	thick]
\tikzstyle{analog}	=[dotted,	color=blue,	thick]
\tikzstyle{digital}	=[dashed,	color=vert,thick]
\tikzstyle{limites}	=[dashed,	color=black]

\newcommand{\xmin}{0}
\newcommand{\xmax}{1}
\newcommand{\ymin}{.1}
\newcommand{\ymax}{.5}

\newcommand{\xlabel}{$\beta$}
\newcommand{\ylabel}{$\Delta$}

\begin{axis}[xlabel={\xlabel}, ylabel={\ylabel}, 
			xmin=\xmin, xmax=\xmax, ymin=\ymin, ymax=\ymax, 
			grid=both]

\pgfplotsset{every axis grid/.style={dotted}} 
\pgfplotsset{every axis legend/.append style={
	cells={anchor=west},
	at={(1.03,0)},
	anchor=south west
}}

\newcommand{\valP} {1  };
\newcommand{\valPY}{0.5};
\newcommand{\valPZ}{  1};
\newcommand{\valPE}{  1};

\addplot[outer] coordinates{
	(0,0.469)(0.1,0.469)(0.2,0.469)(0.3,0.469)(0.36,0.469)(0.36,0.469)(0.41,0.469)(0.46,0.469)(0.469,0.469)(0.469,0.469)(0.479,0.46146)(0.489,0.45418)(0.499,0.44715)(0.509,0.44036)(0.519,0.4338)(0.529,0.42744)(0.539,0.42129)(0.549,0.41533)(0.559,0.40955)(0.569,0.40394)(0.579,0.3985)(0.589,0.39322)(0.599,0.38808)(0.609,0.3831)(0.619,0.37825)(0.629,0.37353)(0.639,0.36893)(0.649,0.36446)(0.659,0.36011)(0.669,0.35586)(0.679,0.35173)(0.689,0.34769)(0.699,0.34376)(0.709,0.33992)(0.719,0.33617)(0.729,0.33251)(0.739,0.32893)(0.749,0.32544)(0.759,0.32202)(0.769,0.31869)(0.779,0.31542)(0.789,0.31223)(0.799,0.3091)(0.809,0.30604)(0.819,0.30305)(0.829,0.30012)(0.839,0.29724)(0.849,0.29443)(0.859,0.29167)(0.869,0.28897)(0.879,0.28632)(0.889,0.28372)(0.899,0.28117)(0.909,0.27867)(0.919,0.27622)(0.929,0.27381)(0.939,0.27144)(0.949,0.26912)(0.959,0.26684)(0.969,0.26461)(0.979,0.26241)(0.989,0.26025)(0.999,0.25813)
};
\addlegendentry{Outer bound (Prop.~\ref{prop:binary:outer})};

\addplot[hybrid] coordinates{
	(0,0.469)(0.1,0.469)(0.2,0.469)(0.3,0.469)(0.36,0.469)(0.36,0.469)(0.41,0.469)(0.46,0.469)(0.469,0.469)(0.469,0.469)(0.479,0.459)(0.489,0.449)(0.499,0.439)(0.509,0.429)(0.519,0.419)(0.529,0.409)(0.539,0.399)(0.549,0.389)(0.559,0.379)(0.569,0.369)(0.579,0.359)(0.589,0.349)(0.599,0.33932)(0.609,0.33046)(0.619,0.32238)(0.629,0.315)(0.639,0.30828)(0.649,0.30217)(0.659,0.29663)(0.669,0.29161)(0.679,0.28708)(0.689,0.28301)(0.699,0.27936)(0.709,0.2761)(0.719,0.27321)(0.729,0.27067)(0.739,0.26844)(0.749,0.2665)(0.759,0.26482)(0.769,0.2634)(0.779,0.26219)(0.789,0.26119)(0.799,0.26037)(0.809,0.2597)(0.819,0.25918)(0.829,0.25878)(0.839,0.25849)(0.849,0.25827)(0.859,0.25812)(0.869,0.25803)(0.879,0.25797)(0.889,0.25794)(0.899,0.25792)(0.909,0.25792)(0.919,0.25791)(0.929,0.25791)(0.939,0.25791)(0.949,0.25791)(0.959,0.25791)(0.969,0.25791)(0.979,0.25791)(0.989,0.25791)(0.999,0.25791)	
};
\addlegendentry{Hybrid (Th.~\ref{th:hybrid})};

\addplot[digital] coordinates{
	(0,0.469)(0.1,0.469)(0.2,0.469)(0.3,0.469)(0.36,0.469)(0.36,0.469)(0.41,0.469)(0.46,0.469)(0.469,0.469)(0.469,0.469)(0.479,0.459)(0.489,0.449)(0.499,0.439)(0.509,0.429)(0.519,0.419)(0.529,0.409)(0.539,0.399)(0.549,0.389)(0.559,0.379)(0.569,0.36906)(0.579,0.35938)(0.589,0.34995)(0.599,0.34077)(0.609,0.3318)(0.619,0.32304)(0.629,0.31447)(0.639,0.30607)(0.649,0.29784)(0.659,0.28976)(0.669,0.28182)(0.679,0.27401)(0.689,0.26633)(0.699,0.25876)(0.709,0.25129)(0.719,0.24393)(0.729,0.23665)(0.739,0.22946)(0.749,0.22235)(0.759,0.21531)(0.769,0.20833)(0.779,0.20142)(0.789,0.19456)(0.799,0.18775)(0.809,0.18099)(0.819,0.17428)(0.829,0.1676)(0.839,0.16095)(0.849,0.15434)(0.859,0.14775)(0.869,0.14119)(0.879,0.13464)(0.889,0.12812)(0.899,0.12161)(0.909,0.1151)(0.919,0.10861)(0.929,0.10213)(0.939,0.095642)(0.949,0.089159)(0.959,0.082675)(0.969,0.076187)(0.979,0.069693)(0.989,0.063192)(0.999,0.056681)
};
\addlegendentry{Digital (Prop.~\ref{prop:binary:digital})};

\addplot[analog] coordinates{
	(0,0.25791)(0.1,0.25791)(0.2,0.25791)(0.3,0.25791)(0.36,0.25791)(0.36,0.25791)(0.41,0.25791)(0.46,0.25791)(0.469,0.25791)(0.469,0.25791)(0.479,0.25791)(0.489,0.25791)(0.499,0.25791)(0.509,0.25791)(0.519,0.25791)(0.529,0.25791)(0.539,0.25791)(0.549,0.25791)(0.559,0.25791)(0.569,0.25791)(0.579,0.25791)(0.589,0.25791)(0.599,0.25791)(0.609,0.25791)(0.619,0.25791)(0.629,0.25791)(0.639,0.25791)(0.649,0.25791)(0.659,0.25791)(0.669,0.25791)(0.679,0.25791)(0.689,0.25791)(0.699,0.25791)(0.709,0.25791)(0.719,0.25791)(0.729,0.25791)(0.739,0.25791)(0.749,0.25791)(0.759,0.25791)(0.769,0.25791)(0.779,0.25791)(0.789,0.25791)(0.799,0.25791)(0.809,0.25791)(0.819,0.25791)(0.829,0.25791)(0.839,0.25791)(0.849,0.25791)(0.859,0.25791)(0.869,0.25791)(0.879,0.25791)(0.889,0.25791)(0.899,0.25791)(0.909,0.25791)(0.919,0.25791)(0.929,0.25791)(0.939,0.25791)(0.949,0.25791)(0.959,0.25791)(0.969,0.25791)(0.979,0.25791)(0.989,0.25791)(0.999,0.25791)
};
\addlegendentry{Analog (Prop.~\ref{prop:binary:analog})};

\addplot[limites] coordinates{
	(0.36,0)(0.36,1)
};
\addplot[limites] coordinates{
	(0.469,0)(0.469,1)
};

\end{axis}
\end{tikzpicture}
\caption{Equivocation rate $\Delta$ as a function of erasure probability $\beta$ ($\epsilon=0.1$, $\zeta=0.1$).}
\label{fig:binary:Delta=fbeta}
\end{figure*}
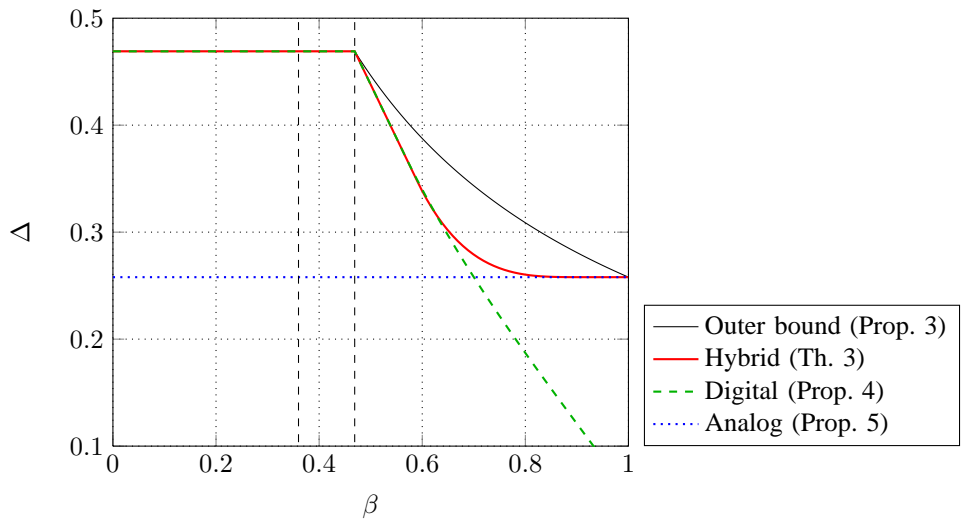
Gathering all the above equations, we proved that
\begin{IEEEeqnarray}{rCl}
H(A^n | E^n Z^n)
	&\geq& n\Big( H(A|U) + R_f + H(E|A)\nonumber  \\
	&+& H(Z|X) - H(EZ|U) - \varepsilon \Big) \ ,
\end{IEEEeqnarray}
for some sufficiently large $n$.

After some algebraic manipulations using the Markov chains $U\mkv A\mkv E$ and $(U,E)\mkv X\mkv Z$, we proved that the following condition is sufficient to achieve equivocation rate  $\Delta - \varepsilon$ at Eve:
\[
\Delta \leq H(A|UE) - I(X;Z|UE) + R_f	\ .
\]

\subsection{End of Proof}

In this section, we proved that sufficient conditions for the achievability of a tuple $(k=1,D,\Delta)$ are given by the following system of inequalities:
\[
\left\{
\begin{IEEEeqnarraybox}[][c]{rCl}
R_1			&>&		I(U;A)								\\
R_2			&>&		I(V;A|U)							\\
R_f			&>&		0									\\
R_1 		&<&		I(U;BY)								\\
R_2 + R_f	&<&		I(V;BY|U)							\\
R_2 + R_f	&<&		I(V;AZ|U)							\\
D			&\geq&	\bE\big[ d(A,\hat A(V,B,Y)) \big]	\\
\Delta		&\leq&	H(A|UE) - I(X;Z|UE) + R_f
\end{IEEEeqnarraybox}
\right.
\]
Fourier-Motzkin elimination~\cite{villard2011fourier-motzkin} then yields:
\[
\left\{
\begin{IEEEeqnarraybox}[][c]{rCl}
I(U;A)		&<&		I(U;BY) 									\\
I(V;A|U)	&<&		I(V;BY|U) 									\\
D			&\geq&	\bE\big[ d(A,\hat A(V,B,Y)) \big]			\\
\Delta		&<&		H(A|UE) - I(X;Z|UE) \\
&+& I(V;BY|U) - I(V;A|U) 	\\	
\Delta		&<&		H(A|UE) - I(X;Z|UE) \\
&+& I(V;AZ|U) - I(V;A|U)
\end{IEEEeqnarraybox}
\right.
\]
This proves Theorem~\ref{th:hybrid}.
\endIEEEproof

\section{Secure Transmission of a Binary Source With BEC/BSC Side Informations Over a Type-II Wiretap Channel (Continued)}
\label{sec:binaryCont}

In this section, we go back on the binary example introduced in Section~\ref{sec:binary} and compare a hybrid coding scheme based on Theorem~\ref{th:hybrid} with the ones analyzed is Section~\ref{sec:binary}, namely the digital scheme of Section~\ref{sec:digital} (see Proposition~\ref{prop:binary:digital}) and a pure analog one consisting in directly sending the source over the channel (see Proposition~\ref{prop:binary:analog}).

\subsection{Hybrid Coding}

We consider the hybrid scheme of Theorem~\ref{th:hybrid} choosing variables $U$, $V$ and $X$ as follows: 
\begin{IEEEeqnarray}{rCl}
U &=& V \oplus W							\ ,\label{eq:binaryDefU}\\
V &\stackrel{\indpt A}{\sim}& \bern(\tfrac12) \ ,\label{eq:binaryDefV}\\
X &=& V \oplus A							\ ,\label{eq:binaryDefX}
\end{IEEEeqnarray}
where $\oplus$ stands for the binary exclusive-or operator, $W$ is independent of $A$ and $V$, and $W\sim\bern(u)$ for some crossover probability $u\in\intFF{0}{\tfrac12}$.

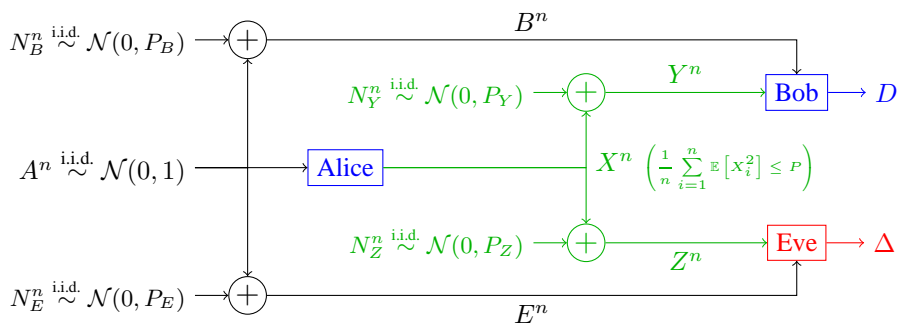
\begin{figure*}[!ht]
\centering
	\begin{tikzpicture}	
	\node[left]				(A) 		at (0,0) 		{$A^n\simiid\cN(0,1)$};
	\node[sensor,gentil]	(alice)		at (2,0)		{Alice};
	
	\node[left]				(NB)		at (0,1.7)		{\small $N_B^n\simiid\cN(0,P_B)$};
	\node[oper]				(addB)		at (.7,1.7)		{\large $+$};
	
	\node[left]				(NE)		at (0,-1.7)		{\small $N_E^n\simiid\cN(0,P_E)$};
	\node[oper]				(addE)		at (.7,-1.7)	{\large $+$};
	
	\node[sensor,gentil]	(bob) 		at (8,1)		{Bob};
	\node[right=.5,gentil]	(D) 		at (bob.east)	{$D$};
	
	\node[sensor,mechant]	(eve) 		at (8,-1)		{Eve};
	\node[right=.5,mechant]	(Delta) 	at (eve.east)	{$\Delta$};
	
	\draw[->] (A) 		to (alice);
	
	\draw[->] (A) 		-| (addB);
	\draw[->] (NB) 		to (addB);	
	\draw[->] (addB) 	to node[above]{$B^n$} (addB-|bob) to (bob);	
	
	\draw[->] (A)		-| (addE);	
	\draw[->] (NE) 		to (addE);	
	\draw[->] (addE) 	to node[below]{$E^n$} (addE-|eve) to (eve);

	\node[canal,oper]	(addY)		at (5.2,1)	{\large $+$};
	\node[canal,left=.7](NY)		at (addY)	{\small $N_Y^n\simiid\cN(0,P_Y)$};
	
	\node[canal,oper]	(addZ)		at (5.2,-1)	{\large $+$};
	\node[canal,left=.7](NZ)		at (addZ)	{\small $N_Z^n\simiid\cN(0,P_Z)$};
	
	\node[right,canal] at (addY|-alice)			{$X^n$ \tiny$\displaystyle{\left(\frac1n \sum_{i=1}^n \bE\left[ X_i^2 \right]\leq P\right)}$};
	
	\draw[->,canal]		(alice) 	-| (addY);
	\draw[->,canal]		(NY) 		to (addY);	
	\draw[->,canal]		(addY) 		to node[above]{$Y^n$} (bob);	
	\draw[->,canal]		(alice)		to (alice-|addZ) to (addZ);
	\draw[->,canal]		(NZ) 		to (addZ);	
	\draw[->,canal]		(addZ) 		to node[below]{$Z^n$} (eve);

	\draw[->,gentil]	(bob)		to (D);
	\draw[->,mechant]	(eve)		to (Delta);
\end{tikzpicture}
\caption{Transmission of a Gaussian source over a Gaussian wiretap channel with side information.}
\label{fig:sceCh:gaussian}
\end{figure*}

\subsection{Numerical Results}

Fig.~\ref{fig:binary:Delta=fbeta} represents the largest equivocation rate at Eve $\Delta$ as a function of the erasure probability~$\beta$ for 
\begin{enumerate}
\item the outer bound of Proposition~\ref{prop:binary:outer},
\item the hybrid digital/analog scheme of Theorem~\ref{th:hybrid} with variables~\eqref{eq:binaryDefU}--\eqref{eq:binaryDefX} (and optimizing over $u$),
\item the digital scheme of~Proposition~\ref{prop:binary:digital} (optimizing over $u$ and $q$),
\item the analog scheme of Proposition~\ref{prop:binary:analog},
\end{enumerate}
for parameter values $\epsilon=0.1$, $\zeta=0.1$.

If $\beta\leq 4\epsilon(1-\epsilon)$, $B$ is less noisy than $E$ (see Fig.~\ref{fig:binary:cas}), and the digital scheme is optimal (as stated by Proposition~\ref{prop:digital:special:B}), as well as the proposed hybrid one. Here, this result also holds when $B$ is only more capable than $E$ i.e., for $\beta\leq h_2(\epsilon)$. On the other hand, for $\beta=1$, as already noted in Section~\ref{sec:binary:numericalr}, the naive pure analog scheme outperforms the digital one. However,  depending on the parameters $\epsilon$, $\zeta$, this is not the case for all values $\beta$ in $\left[ h_2(\epsilon), 1 \right)$ as it is shown in Fig.~\ref{fig:binary:Delta=fbeta}.  In particular, for $\beta\in \left( h_2(\epsilon), 1 \right)$ the side information at Bob and Eve cannot be ordered (see Fig.~\ref{fig:binary:cas}), and hence it appears that the hybrid digital/analog scheme performs strictly better than both: pure analogue and the digital schemes. Furthermore, this region cannot be achieved via time sharing between the digital and analogue schemes. This can be explained by noting that in this regime successive refinement coding, via the two layers used in the digital scheme, does not improve on the hybrid digital/analog scheme with a single digital layer.

\section{Secure Transmission of a Gaussian Source Over a Gaussian Wiretap Channel}
\label{sec:gaussian}

\subsection{System Model}

In this section, we consider the transmission of a Gaussian source over a Gaussian wiretap channel with matched bandwidth.
More precisely, as depicted in Fig.~\ref{fig:sceCh:gaussian}, the source at Alice $A$ is standard Gaussian, and observations at Bob and Eve are the outputs of independent additive white Gaussian noise (AWGN) channels with input $A$ and respective noise powers $P_B$ and $P_E$.
Communication channels from Alice to Bob and Charlie are AWGN channels with respective noise powers $P_Y$ and $P_Z$. 
The average input power of this channel is limited to~$P$. 
One channel use is allowed per source symbol.

Euclidean distance on $\bR$ is used to measure distortion at Bob ($d(a,b) = (a-b)^2$, for each $a,b\in\bR$). 
Differential entropy $h(\cdot)$ measures uncertainty yielding equivocation rates $\Delta\in\bR$. 
We also introduce quantity $D_E = 2^{2\Delta}/(2\pi e)$, which provides a lower bound on the minimum mean-square error of any estimator of $A$ at Eve \cite[Theorem~8.6.6]{cover2006elements}. It is worth mentioning that the information leakage reads: 
\begin{IEEEeqnarray*}{rCl}
\dfrac1n I(A^n;E^n Z^n)&=& H(A) - \dfrac1n H(A^n | E^n Z^n)\\
&\leq & H(A) - \Delta \\
& =& H(A) - \dfrac12 \log_2 (2\pi e) D_E\, .
\end{IEEEeqnarray*}

\begin{definition}[Achievability]
In this section, a tuple $(D,D_E)\in{\bR_+^*}^2$ is said to be \emph{achievable} if,
for any $\varepsilon>0$, there exists an $(n,n)$-code $(F,g)$ such that:
\begin{IEEEeqnarray*}{rCl}
\bE\big[ \norm{A^n - g(B^n,Y^n)}^2 \big]	&\leq& D+\varepsilon 		\ ,\\
\dfrac1n\,h(A^n|E^n Z^n) 					&\geq& \frac12\log\left( 2\pi e\,D_E \right)-\varepsilon	\ ,\\
\frac1n \sum_{i=1}^n \bE\big[ X_i^2 \big]	&\leq& P+\varepsilon 		\ ,
\end{IEEEeqnarray*}
with channel input $X^n$ as the output of the encoder $F(A^n)$.
\end{definition}

\subsection{First Results}

Although Theorems~\ref{th:outer}, \ref{th:digital}, \ref{th:hybrid} are stated and proved for finite alphabets, we take the liberty to use their statements as inner/outer regions also for this quadratic Gaussian case. The involved probability distributions should now also verify condition
\begin{equation}
\label{eq:power}
\Var{X} \leq P \ .
\end{equation}
The corresponding regions will be denoted with an additional $\cdot^P$ i.e., $\cR_\text{out}^P$, $\cR_\text{digital}^P$ and $\cR_\text{hybrid}^P$. 

Notice that due to the Gaussian additive noises, and depending on the relative values of $P_B$, $P_E$ (resp. $P_Y$, $P_Z$), one side information (resp. one channel) is a stochastically degraded version of the other. 
There exist four different cases and, from the results of Section~\ref{sec:digital:special}, separation holds for three of them, as summarized in Table~\ref{tab:gaussian:combin}.
In these cases, we can moreover prove closed-from expressions for the achievable region, as stated by Propositions~\ref{prop:gaussian:B} and \ref{prop:gaussian:EZ} below. 

\begin{proposition}
\label{prop:gaussian:B}
Assume that $P_B\leq P_E$.
A tuple $(D,D_E)$ is achievable if and only if
\begin{IEEEeqnarray*}{rCl}
D	&\geq& \frac1{1+\frac1{P_B}}\cdot\frac1{1+\frac{P}{P_Y}}					\ ,\\[.1cm]
D_E	&\leq& \frac1{1+\frac1{P_E}}												\ ,\\[.1cm]
D_E	&\leq& \frac{1+\frac1{P_B}}{1+\frac1{P_E}}\cdot D\cdot
			\max\left\{ 1 \,; \frac{1+\frac{P}{P_Y}}{1+\frac{P}{P_Z}} \right\}	\ .
\end{IEEEeqnarray*}
\end{proposition}

\begin{IEEEproof}
The proof of the converse part is given in Appendix~\ref{app:gaussian:B}.

The direct part follows from Proposition~\ref{prop:digital:special:B} by choosing appropriate Gaussian auxiliary random variables: $U$ constant, $X=T=Q$ to a Gaussian random variable of variance $P$ and $V$ is the usual random variable for the Gaussian Wyner-Ziv problem. Details are omitted.
\end{IEEEproof}

\begin{proposition}
\label{prop:gaussian:EZ}
Assume that $P_B > P_E$ and $P_Y \geq P_Z$.
A tuple $(D,D_E)$ is achievable if and only if
\begin{IEEEeqnarray*}{rCl}
D	&\geq& \frac1{1+\frac1{P_B}}\cdot\frac1{1+\frac{P}{P_Y}}					\ ,\\
D_E	&\leq& \frac1{1+\frac1{P_E}}												\ ,\\
D_E	&\leq& \frac1{\frac1D + \frac1{P_E}-\frac1{P_B} }
\end{IEEEeqnarray*}
\end{proposition}

\begin{IEEEproof}
The proof of the converse part is given in Appendix~\ref{app:gaussian:EZ}.

The direct part follows from Proposition~\ref{prop:digital:special:Z} by choosing appropriate Gaussian auxiliary random variables. Details are omitted.
\end{IEEEproof}

\begin{figure*}[!ht]
\centering
\begin{tikzpicture}
	\node[matrix,draw,every node/.style={codeword}] (codebook) at (0,0){
		\node{};&		\node{};&		\node{};& 		\node{}; 	\\
		\node{};& 		\node{};& 		\node{};& 		\node{}; 	\\
		\node{};& 		\node{};& 		\node{};& 		\node{}; 	\\
		\node{};& 		\node{};& 		\node[cok]{};&	\node{}; 	\\
		\node{};&		\node{};& 		\node{};& 		\node{}; 	\\
	};
	\node[left=.7]			(A)		at (codebook.west) 	{$A^n\simiid\cN(0,1)$};
	\node[above=.4]			(r) 	at (codebook.north) {$r_f\sim\unif{\{1,\dots,2^{nR_f}\}}$};
	
	\node[mult,below=1.2]	(mult)	at (codebook)	{$\times$};
	\node[below]					at (mult.south)	{$(\alpha+\beta)$};
	
	\node[oper,right=.8]	(add)	at (codebook.east)	{\large $+$};
	\node[left=.05]					at (add.south west) {\footnotesize $-$};
	\node[below=.05]				at (add.south west) {\footnotesize $+$};
	
	\node[mult,right=.5]	(multP)	at (add.east)	{$\times$};
	\node[below]					at (multP.south){$\sqrt{P}$};
	
	\node[right=1]	(X) at (multP) {$X^n$};
	
	\draw[->]		(A)	to (codebook);
	\draw[->]		(codebook.west) ++ (-.4,0) |- (mult);
	
	\draw[->]		(r)	to (codebook);
	\draw[->]		(codebook)	to node[above]{$v^n$} (add);
	\draw[->]		(mult)	-| (add);
	
	\draw[->]		(add)	to (multP);
	\draw[->]		(multP)	to (X);
\end{tikzpicture}
\caption{Hybrid digital/analog scheme for secure transmission of a Gaussian source over a Gaussian wiretap channel.}
\label{fig:gaussian:hybrid}
\end{figure*}
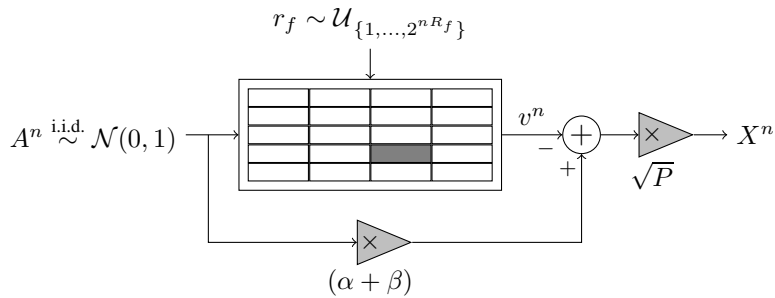

\begin{table}
\renewcommand{\arraystretch}{1.4}
\centering
\begin{tabular}{c||c|c}
				&	$P_B \leq P_E$								
				&	$P_B >    P_E$				\\\hline\hline
$P_Y <    P_Z$	&	{\color{vert}\checkmark}
				& 	{\color{red} \bfseries ?}	\\\hline
$P_Y \geq P_Z$	&	{\color{vert}\checkmark}
				&	{\color{vert}\checkmark}
\end{tabular}
\caption{Cases where $\cR_\text{digital}$ is tight and separation holds.}
\label{tab:gaussian:combin}
\end{table}

As a matter of fact, the case when Bob has ``better'' channel ($P_Y < P_Z$) and ``worse'' side information ($P_B > P_E$) than Eve is still open. 
In this quadratic Gaussian example, the outer bound $\cR_\text{out}^P$ reduces to the one given by Proposition~\ref{prop:gaussian:EY} below.

\begin{proposition}
\label{prop:gaussian:EY}
Assume that $P_B > P_E$ and $P_Y < P_Z$.
Any achievable tuple $(D,D_E)$ verifies the following inequalities:
\begin{IEEEeqnarray*}{rCl}
D	&\geq& \frac1{1+\frac1{P_B}}\cdot\frac1{1+\frac{P}{P_Y}}					\ ,\\
D_E	&\leq& \frac1{1+\frac1{P_E}}												\ ,\\
D_E	&\leq& \frac1{\frac1D\cdot\frac{1+\frac{P}{P_Z}}{1+\frac{P}{P_Y}} + \frac1{P_E}-\frac1{P_B} }
\end{IEEEeqnarray*}
\end{proposition}

\begin{IEEEproof}
See Appendix~\ref{app:gaussian:EY}.
\end{IEEEproof}

We next propose a hybrid digital/analog scheme based on Theorem~\ref{th:hybrid} that turns out to be optimal i.e., that achieves the region of Proposition~\ref{prop:gaussian:EY}, when $P_Y < P_Z$ and $P_B\to\infty$.

\subsection{Hybrid Coding}
\label{sec:gaussian:hybrid}

Proposition~\ref{prop:gaussian:hybrid} below follows from Theorem~\ref{th:hybrid} choosing variables $U$, $V$ and $X$ as follows:
\begin{IEEEeqnarray}{rCl}
U &=& \emptyset										\ ,\label{eq:hybrid:defU}\\
V &=& \alpha A + \gamma N							\ ,\label{eq:hybrid:defV}\\
X &=& \big( \beta A - \gamma N \big)\sqrt{P} 		\ ,\label{eq:hybrid:defX}
\end{IEEEeqnarray}
where $\alpha\in\bR$, $\beta\in\intFO{0}{1}$, $\gamma = \sqrt{1 - \beta^2}$
and $N\sim\cN(0,1)$ is a standard Gaussian random variable independent of $A$.
Note that $X\sim\cN(0,P)$ can be written as a deterministic function of $A$ and $V$:
\begin{equation}
\label{eq:gaussX(A,V)}
X = \big( (\alpha+\beta) A - V \big)\sqrt{P} 		\ .
\end{equation}
Function $\hat A$ is defined as the MMSE estimator of $A$ from $(V,B,Y)$.

The hybrid digital/analog scheme of Section~\ref{sec:hybrid} with variables \eqref{eq:hybrid:defU}--\eqref{eq:hybrid:defX} reduces to the one depicted in Fig.~\ref{fig:gaussian:hybrid}.

\begin{proposition}[Hybrid scheme]
\label{prop:gaussian:hybrid}
A tuple $(D,D_E)\in{\bR_+^*}^2$ is achievable if
\begin{IEEEeqnarray}{rCl}
D	&\geq& \frac1{1 + \frac1{P_B} + \frac{\alpha^2}{\gamma^2}
										+ \frac{P}{P_Y}(\alpha+\beta)^2}	
	\ ,\label{eq:GaussD}	\\
D_E	&\leq& \frac1{1 + \frac1{P_E} + \frac{P}{P_Z}\left( 1 + \frac{\gamma^2}{P_E} \right)}\\
		&\cdot& \min\left\{ 
			\frac	{1 + \frac1{P_B} + \frac{P}{P_Y}\left( 1 + \frac{\gamma^2}{P_B} \right)}
					{1 + \frac1{P_B} + \frac{\alpha^2}{\gamma^2} + \frac{P}{P_Y}(\alpha+\beta)^2}
			\,;
			1 + \gamma^2\,\frac{P}{P_Z}
		\right\} 															
	\ ,\qquad\label{eq:GaussDE}
\end{IEEEeqnarray}
for some $\alpha\in\bR$, $\beta\in\intFO{0}{1}$ such that
\begin{equation}
\label{eq:GaussRate}
\frac{\alpha^2}{\gamma^2} + \frac{P}{P_Y} (\alpha+\beta)^2
	\leq \frac{P}{P_Y} \left( 1 + \frac{\gamma^2}{P_B} \right) \ ,
\end{equation}
where
\begin{equation}
\label{eq:defgamma}
\gamma = \sqrt{1-\beta^2} \ .
\end{equation}
\end{proposition}

\begin{IEEEproof}
See Appendix~\ref{app:gaussian:hybrid}.
\end{IEEEproof}

\begin{remark}
In the proposed scheme, unlike dirty-paper coding for point-to-point communication without secrecy constraint~\cite{costa1983writing}, the source $A$ (that can be viewed as the state of some channel, known at the encoder --see Fig.~\ref{fig:gelfand}) and the channel input $X$ are not independent.
\end{remark}

\subsection{Special Case: $P_Y < P_Z$, $P_B\to\infty$}
\label{sec:gaussian:noB}

\begin{figure*}[!ht]
\centering


\begin{tikzpicture}

\tikzstyle{optimal}	=[			color=red,	thick]
\tikzstyle{analog}	=[dotted,	color=blue,	thick]
\tikzstyle{digital}	=[dashed,	color=vert,thick]
\tikzstyle{limites}	=[dashed,	color=black]

\newcommand{\xmin}{.3}
\newcommand{\xmax}{1}
\newcommand{\ymin}{.25}
\newcommand{\ymax}{.52}

\newcommand{\xlabel}{$D$}
\newcommand{\ylabel}{$D_E$}

\begin{axis}[xlabel={\xlabel}, ylabel={\ylabel}, 
			xmin=\xmin, xmax=\xmax, ymin=\ymin, ymax=\ymax, 
			grid=both]

\pgfplotsset{every axis grid/.style={dotted}} 
\pgfplotsset{every axis legend/.append style={
	cells={anchor=west},
	at={(1.03,0)},
	anchor=south west
}}

\newcommand{\valP} {1  };
\newcommand{\valPY}{0.5};
\newcommand{\valPZ}{  1};
\newcommand{\valPE}{  1};

\addplot[optimal,domain=1/(1+\valP/\valPY):1] {
	1/( 1/\valPE + max(1 , 1/x * (1+\valP/\valPZ)/(1+\valP/\valPY)) )
};
\addlegendentry{Optimal (Th.~\ref{th:gaussian:noB})};

\addplot[digital] coordinates{
	(0.33333,0.26795)(0.34722,0.27911)(0.36111,0.29028)(0.375,0.30144)(0.38889,0.31261)(0.40278,0.32377)(0.41667,0.33494)(0.43056,0.3461)(0.44444,0.35727)(0.45833,0.36842)(0.47222,0.37931)(0.48611,0.38983)(0.5,0.4)(0.51389,0.40984)(0.52778,0.41935)(0.54167,0.42857)(0.55556,0.4375)(0.56944,0.44615)(0.58333,0.45455)(0.59722,0.46269)(0.61111,0.47059)(0.625,0.47826)(0.63889,0.48571)(0.65278,0.49296)(0.66667,0.5)(0.66667,0.5)(0.68056,0.5)(0.69444,0.5)(0.70833,0.5)(0.72222,0.5)(0.73611,0.5)(0.75,0.5)(0.76389,0.5)(0.77778,0.5)(0.79167,0.5)(0.80556,0.5)(0.81944,0.5)(0.83333,0.5)(0.84722,0.5)(0.86111,0.5)(0.875,0.5)(0.88889,0.5)(0.90278,0.5)(0.91667,0.5)(0.93056,0.5)(0.94444,0.5)(0.95833,0.5)(0.97222,0.5)(0.98611,0.5)(1,0.5)
};
\addlegendentry{Digital (Prop.~\ref{prop:gaussian:noB:digital})};

\addplot[analog,domain=1/(1+\valP/\valPY):1] {
	1/( 1 + 1/\valPE + (1/x - 1)*\valPY/\valPZ)
};
\addlegendentry{Analog (Prop.~\ref{prop:gaussian:noB:analog})};

\addplot[limites] coordinates{
	(1/3,0)(1/3,1)
};

\end{axis}
\end{tikzpicture}
\caption{Quantity $D_E$ as a function of the distortion at Bob $D$ ($P=1$, $P_Y=0.5$, $P_Z=1$, $P_E=1$).}
\label{fig:gaussian:DE=fD}
\end{figure*}
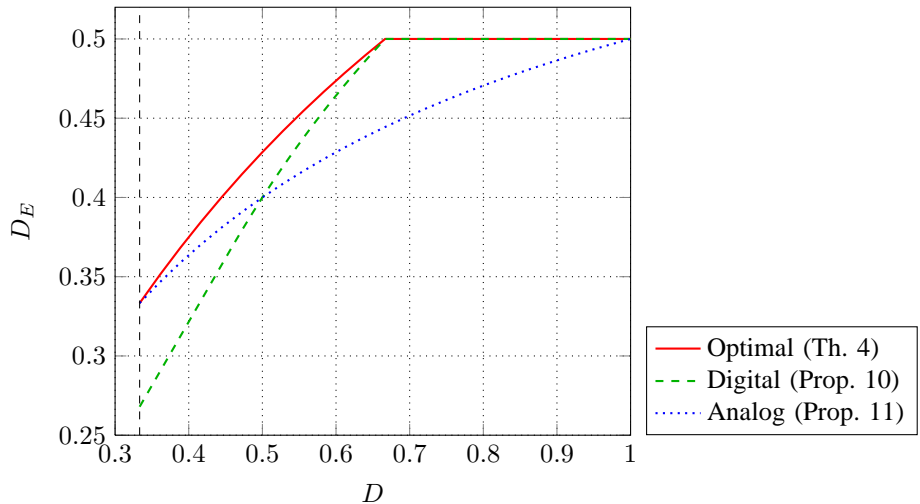

From now on, we focus on the unsolved case (represented by ``{\color{red}\bfseries ?}'' in Table~\ref{tab:gaussian:combin}).
In particular, we assume that $P_Y < P_Z$.
Then, if Bob does not have any side information i.e., $B=\emptyset$ or equivalently $P_B\to\infty$:

\begin{itemize}
\item The hybrid digital/analog scheme of Proposition~\ref{prop:gaussian:hybrid} is optimal and yields Theorem~\ref{th:gaussian:noB}.
\item The digital scheme of Theorem~\ref{th:digital} is strictly sub-optimal, as shown by Proposition~\ref{prop:gaussian:noB:digital} and Fig.~\ref{fig:gaussian:DE=fD}.
\end{itemize}

\begin{theorem}[Closed-form characterization]
\label{th:gaussian:noB}
If $P_Y < P_Z$ and $B=\emptyset$, $(D,D_E)\in{\bR_+^*}^2$ is achievable if and only if
\begin{IEEEeqnarray}{rCl}
D 	&\geq& \frac1{1+\frac{P}{P_Y}}	\ ,\label{eq:Gauss0D}\\
D_E	&\leq& \frac1{\max\left\{ 1 \,; \frac1D \cdot \frac{1+\frac{P}{P_Z}}{1+\frac{P}{P_Y}} \right\} + \frac1{P_E}}			\ .\label{eq:Gauss0DE}
\end{IEEEeqnarray}
\end{theorem}

\begin{IEEEproof}
The converse part directly follows from Proposition~\ref{prop:gaussian:EY} by letting $P_B$ tend to infinity.

The direct part follows after Proposition~\ref{prop:gaussian:hybrid} by letting $P_B$ tend to infinity and choosing, for any distortion level: 
$$
D\in\left[	\frac1{1+\frac{P}{P_Y}}\,, \frac{1+\frac{P}{P_Z}}{1+\frac{P}{P_Y}} \right)\,,
$$
\begin{IEEEeqnarray}{rCl}
\alpha	&=&  \frac	
			{ \beta	+ \gamma^2 \sqrt{\frac1{D}\left( \frac{P}{P_Y}-\frac{P}{P_Z} \right)} }
			{ 1 + \gamma^2 \frac{P}{P_Y}}	  
			- \beta									\ ,\label{eq:defalpha}\\
\beta	&=& \sqrt{\frac{P_Z}{P}}\sqrt{ 1 + \frac{P}{P_Z} - D \left(1+\frac{P}{P_Y}\right) } \label{eq:defbeta}\ .
\end{IEEEeqnarray}
Details are provided in Appendix~\ref{app:gaussian:noB:direct}.
\end{IEEEproof}

The following proposition provides a simple expression of region $\cR_\text{digital}^P$ i.e., the set of all tuples achievable by the digital scheme of Section~\ref{sec:digital}.

\begin{proposition}[Digital scheme]
\label{prop:gaussian:noB:digital}
If $P_Y < P_Z$ and $B=\emptyset$, $(D,D_E)\in\cR_\text{digital}^P$ if and only if
\begin{IEEEeqnarray}{l}
D	\geq \frac1{1+\frac{P}{P_Y}}										\ ,\label{eq:gaussian:noB:digital:D}\\
D_E \leq \frac1{\frac1\mu + \frac1{P_E}}\cdot	\min\left\{ 1 \,;
		\frac	{D\left(1+\frac{P}{P_Y}\right)}
				{1 + \mu\frac{P}{P_Z} - (1-\mu)\frac{P_Y}{P_Z}} \right\}\ ,\label{eq:gaussian:noB:digital:DE}
\end{IEEEeqnarray}
for some $\mu\in\intFFLR{\frac1{1+\frac{P}{P_Y}}}{1}$.
\end{proposition}

\begin{IEEEproof}
The proof of the converse part is given in Appendix~\ref{app:gaussian:noB:digital}.

The direct part follows from Theorem~\ref{th:digital} with Gaussian variables $U$, $V$, $Q$ and $T=X$ after some straightforward derivations. 
Details are omitted.
\end{IEEEproof}\vspace{.2cm}

\begin{remark}
\label{rem:digitalOpt}
If $D \geq \frac{1+\frac{P}{P_Z}}{1+\frac{P}{P_Y}}$, then $\mu=1$ is optimal in Proposition~\ref{prop:gaussian:noB:digital}, yielding inequalities \eqref{eq:Gauss0D}, \eqref{eq:Gauss0DE} in Theorem~\ref{th:gaussian:noB}.
This implies that the digital scheme of Section~\ref{sec:digital} is optimal in this region.
For such distortion levels, the quantity $D_E = \frac1{1+\frac1{P_E}} = \Var{A|E}$ is achievable, meaning that Eve cannot retrieve additional information from the communication between Alice and Bob.
\end{remark}

In the following, we will also compare the two above schemes with a pure analog one, consisting in directly sending a scaled version of the source over the channel.
Its performance is given by the following proposition.

\begin{proposition}[Analog scheme]
\label{prop:gaussian:noB:analog}
If $B=\emptyset$, $(D,D_E)\in{\bR_+^*}^2$ is achievable through an analog scheme if
\begin{IEEEeqnarray*}{rCl}
D 	&\geq& \frac1{1+\frac{P}{P_Y}}																\ ,\\
D_E	&\leq& \frac1{1 + \frac1{P_E} + \left[ \left(\frac1D - 1\right)\frac{P_Y}{P_Z} \right]_+}	\ .
\end{IEEEeqnarray*}
\end{proposition}

\begin{IEEEproof}
See Appendix~\ref{app:gaussian:noB:analog}.
\end{IEEEproof}

\begin{remark}
\label{rem:analogOpt}
If $D = D_{\min} \triangleq \frac1{1+\frac{P}{P_Y}}$, then Proposition~\ref{prop:gaussian:noB:analog} yields inequalities \eqref{eq:Gauss0D}, \eqref{eq:Gauss0DE} in Theorem~\ref{th:gaussian:noB} i.e., the above analog scheme is optimal.
\end{remark}

\begin{remark}
When there is no secrecy requirement i.e., $D_E = 0$, all the above schemes can achieve distortion $D_{\min}$, as stated in~\cite{goblick1965theoretical,gastpar2003code}.
\end{remark}

\subsubsection*{Numerical Results}

Fig.~\ref{fig:gaussian:DE=fD} represents the largest achievable $D_E$ as a function of the distortion level at Bob $D$ for 
\begin{enumerate}
\item the optimal hybrid digital/analog scheme of Theorem~\ref{th:gaussian:noB},
\item the digital scheme of Proposition~\ref{prop:gaussian:noB:digital} (optimizing over $\mu$),
\item the analog scheme of Proposition~\ref{prop:gaussian:noB:analog},
\end{enumerate}
for parameter values $P=1$, $P_Y=0.5$, $P_Z=1$, $P_E=1$.

As a matter of fact, the proposed hybrid digital/analog scheme outperforms both pure analog and digital schemes.
Furthermore, while the digital scheme is optimal for $D \geq \frac{{1+\frac{P}{P_Z}}}{1+\frac{P}{P_Y}}$ (Remark~\ref{rem:digitalOpt}) and the analog one for $D = D_{\min}$ (Remark~\ref{rem:analogOpt}), a time-sharing combination of these falls short to achieve the entire region, as shown by Fig.~\ref{fig:gaussian:DE=fD} and Theorem~\ref{th:gaussian:noB}.

\begin{remark}
While the hybrid digital/analog scheme of Section~\ref{sec:gaussian:hybrid} can be used regardless of the values of the parameters, we did not manage to prove a result of optimality in the more general case where $P_B < \infty$.
However, numerical optimization over $\alpha$ and $\beta$ in Proposition~\ref{prop:gaussian:hybrid} tends to show that this scheme does not achieve the outer bound of Proposition~\ref{prop:gaussian:EY}.
\end{remark}

\section{Summary}
\label{sec:sceCh:summary}

In this paper, we have investigated the general problem of source-channel coding for secure transmission of sources over noisy channels with side information at the receivers.
This setting can be seen as a generalization of the problems of secure source coding with side information at the decoders, and the wiretap channel. 
A general outer bound on the corresponding achievable region has been derived, as well as two inner bounds based on (\emph{i}) a pure digital scheme which combines secure source coding of~\cite{villard2010secure,villard2011securea} with coding for broadcast channels with confidential messages~\cite{csiszar1978broadcast}, and (\emph{ii}) a novel hybrid digital/analog scheme (in the matched-bandwidth case).

The proposed bounds do not match in general, but the digital scheme turns out to be optimal under some less noisy conditions.
However, a simple counterexample shows that a \emph{joint} source-channel scheme may achieve better performance in some other cases. 
At first look, this is not surprising since it is well-known that \emph{joint} source-channel coding/decoding are well-suited for broadcast channels without secrecy constraints~\cite{tuncel2006slepian-wolf}, when all decoders must perfectly reconstruct the source. 
But the \emph{secure} setting is rather different because Alice only wants to help one receiver (Bob), while she wants to blur the other one (Eve). 
Therefore, the intuition indicates that the optimal strategy would be the opposite i.e., separation between source and channel encoders, as in Propositions~\ref{prop:digital:special:B} and~\ref{prop:digital:special:Z}. 
On the other hand, the proposed hybrid digital/analog scheme can be useful in terms of secrecy.
In a quadratic Gaussian setup when side information is only present at the eavesdropper, this strategy turns out to be optimal. 
However, in a more general case where both receivers have side information, a new scheme seems to be needed.

\appendices

\section{Strongly Typical Sequences and Delta-Convention}
\label{app:typical}

Following~\cite{csiszar1982information}, we use in this paper \emph{strongly typical sets} and the so-called \emph{Delta-Convention}. 
Some useful facts are recalled here.
Let $X$ and $Y$ be random variables on some finite sets $\cX$ and $\cY$, respectively. We denote by $p_{X,Y}$ (resp. $p_{Y|X}$, and $p_X$) the joint probability distribution of $(X,Y)$ (resp. conditional distribution of $Y$ given $X$, and marginal distribution of $X$). 

\begin{definition}[Number of occurrences]
For any sequence $x^n\in\cX^n$ and any symbol $a\in\cX$, notation $N(a|x^n)$ stands for the number of occurrences of $a$ in $x^n$.
\end{definition}

\begin{definition}[Typical sequence]
A sequence $x^n\in\cX^n$ is called \emph{(strongly) $\delta$-typical} w.r.t.\ $X$ (or simply \emph{typical} if the context is clear) if
\[
\abs{\frac1n N(a|x^n) - p_X(a)} \leq \delta \ \text{ for each } a\in\cX \ ,
\]
and $N(a|x^n)=0$ for each $a\in\cX$ such that $p_X(a)=0$.
The set of all such sequences is denoted by $\typ{n}{X}$.
\end{definition}

\begin{definition}[Conditionally typical sequence]
Let $x^n\in\cX^n$.
A sequence $y^n\in\cY^n$ is called \emph{(strongly) $\delta$-typical (w.r.t.\ $Y$) given $x^n$}  if,  for each  $a\in\cX, b\in\cY$
\[
\abs{\frac1n N(a,b|x^n,y^n) - \frac1n N(a|x^n)p_{Y|X}(b|a)} \leq \delta  \ ,
\]
and, $N(a,b|x^n,y^n)=0$ for each $a\in\cX$, $b\in\cY$ such that $p_{Y|X}(b|a)=0$.
The set of all such sequences is denoted by $\typ{n}{Y|x^n}$.
\end{definition}

\newtheorem{DeltaConvention}{Delta-Convention~\cite{csiszar1982information}}
\begin{DeltaConvention}
For any sets $\cX$, $\cY$, there exists a sequence $\{\delta_n\}_{n\in\bN^*}$ such that the lemmas stated below hold.\footnote{%
	As a matter of fact, $\delta_n\to0$ and $\sqrt{n}\,\delta_n\to\infty$ as $n\to\infty$.}
From now on, typical sequences are understood with $\delta=\delta_n$. 
Typical sets are still denoted by $\typ{n}{\cdot}$.
\end{DeltaConvention}

\begin{lemma}[\!{\cite[Lemma~1.2.12]{csiszar1982information}}]
There exists a sequence $\eta_n\toas{n\to\infty}0$ such that
\[
p_X(\typ{n}{X}) \geq 1 - \eta_n \ .
\]
\end{lemma}

\begin{lemma}[\!{\cite[Lemma~1.2.13]{csiszar1982information}}]
\label{lem:cardTyp}
There exists a sequence $\eta_n\toas{n\to\infty}0$ such that, for each $x^n\in\typ{n}{X}$,
\begin{IEEEeqnarray*}{c}
\abs{\frac1n \log \norm{\typ{n}{X}} - H(X)} \leq \eta_n 		\ ,\\
\abs{\frac1n \log \norm{\typ{n}{Y|x^n}} - H(Y|X)} \leq \eta_n	\ .
\end{IEEEeqnarray*}
\end{lemma}

\begin{lemma}[Asymptotic equipartition property]
\label{lem:AEP}
There exists a sequence $\eta_n\toas{n\to\infty}0$ such that, for each $x^n\in\typ{n}{X}$ and each $y^n\in\typ{n}{Y|x^n}$,
\begin{IEEEeqnarray*}{c}
\abs{-\frac1n \log p_X(x^n) - H(X)} \leq \eta_n 				\ ,\\
\abs{-\frac1n \log p_{Y|X}(y^n|x^n) - H(Y|X)} \leq \eta_n		\ .
\end{IEEEeqnarray*}
\end{lemma}

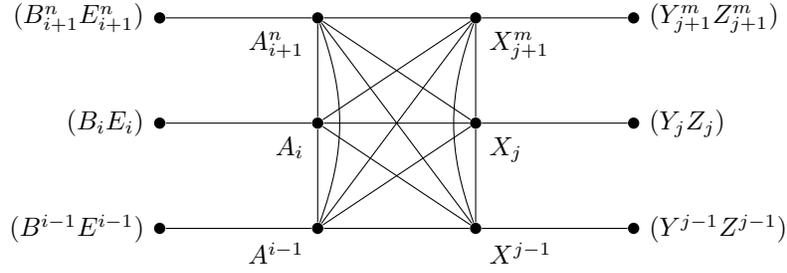
\begin{figure*}[!ht]
\centering
	\begin{tikzpicture}[scale=1.4]					
	\node[point,label=left:$(B_{i+1}^n E_{i+1}^n)$]	(BE+)at (0,1)	{};
	\node[point,label=left:$(B_i E_i)$]				(BEi)at (0,0)	{};
	\node[point,label=left:$(B^{i-1} E^{i-1})$]		(BE-)at (0,-1)	{};
	
	\node[point,label=below left:$A_{i+1}^n$]		(A+)at 	(1.5,1)	{};
	\node[point,label=below left:$A_i$]				(Ai)at 	(1.5,0)	{};
	\node[point,label=below left:$A^{i-1}$]			(A-)at 	(1.5,-1){};

	\node[point,label=below right:$X_{j+1}^m$]		(X+)at 	(3,1)	{};
	\node[point,label=below right:$X_j$]			(Xj)at 	(3,0)	{};
	\node[point,label=below right:$X^{j-1}$]		(X-)at 	(3,-1) 	{};
	
	\node[point,label=right:$(Y_{j+1}^m Z_{j+1}^m)$](YZ+)at (4.5,1)	{};
	\node[point,label=right:$(Y_j Z_j)$]			(YZj)at (4.5,0)	{};
	\node[point,label=right:$(Y^{j-1} Z^{j-1})$]	(YZ-)at (4.5,-1){};
				
	\draw	(BE+)	to (A+);
	\draw	(BEi)	to (Ai);
	\draw	(BE-)	to (A-);
	
	\draw	(A+)to (Ai);
	\draw	(Ai)to (A-);
	\draw	(A+)to [out=-70,in=70] (A-);
	
	\draw	(A+) to (X+);
	\draw	(A+) to (Xj);
	\draw	(A+) to (X-);
	\draw	(Ai) to (X+);
	\draw	(Ai) to (Xj);
	\draw	(Ai) to (X-);
	\draw	(A-) to (X+);
	\draw	(A-) to (Xj);
	\draw	(A-) to (X-);
						
	\draw	(X+)to (Xj);
	\draw	(Xj)to (X-);
	\draw	(X+)to [out=-110,in=110] (X-);
	
	\draw	(X+)	to (YZ+);
	\draw	(Xj)	to (YZj);
	\draw	(X-)	to (YZ-);
\end{tikzpicture}
\caption{Outer bound--Graphical representation of distribution $p(a^n b^n e^n x^m y^m z^m)$.}
\label{fig:outer:proof:joint}
\end{figure*}
\begin{lemma}[Joint typicality lemma~\cite{elgamal2010lecture}]
\label{lem:jointTypicality}
There exists a sequence $\eta_n\toas{n\to\infty}0$ such that
\[
\abs{-\frac1n \log p_Y(\typ{n}{Y|x^n}) - I(X;Y)} \leq \eta_n \
\text{ for each }x^n\in\typ{n}{X}\ .
\]
\end{lemma}

\begin{IEEEproof}
\begin{IEEEeqnarray*}{rCl}
p_Y(\typ{n}{Y|x^n})
	&=&						\sum_{y^n\in\typ{n}{Y|x^n}} p_Y(y^n)				\\
	&\stackrel{(a)}{\leq}&	\norm{\typ{n}{Y|x^n}}\,2^{-n[H(Y)-\alpha_n]}		\\
	&\stackrel{(b)}{\leq}&	2^{n[H(Y|X)+\beta_n]}\,2^{-n[H(Y)-\alpha_n]}	\\
	&=&						2^{-n[I(X;Y)-\beta_n-\alpha_n]}					\ ,
\end{IEEEeqnarray*}
where
\begin{itemize}
\item step~$(a)$ follows from the fact that $\typ{n}{Y|x^n}\subset\typ{n}{Y}$ and Lemma~\ref{lem:AEP}, for some sequence $\alpha_n\toas{n\to\infty}0$,
\item step~$(b)$ from Lemma~\ref{lem:cardTyp}, for some sequence $\beta_n\toas{n\to\infty}0$.
\end{itemize}
The reverse inequality $p_Y(\typ{n}{Y|x^n})\geq 2^{-n[I(X;Y)+\beta_n+\alpha_n]}$ can be proved following similar argument. 
\end{IEEEproof}

\section{Graphical Representation of Probability Distributions}
\label{app:graphical}

Following~\cite[Section~II]{permuter2010two-way}, we use in this paper a technique based on undirected graphs, that provides a sufficient condition for establishing Markov chains from a joint distribution. 
Such a technique for establishing conditional independence was introduced in~\cite{pearl1986fusion} for Bayesian networks, and further generalized to various types of graphs~\cite{kramer2003capacity}.
This paragraph recalls the main points of this technique.

Assume that a sequence of random variables $X^n$ has joint distribution with the following form:
\[
p(x^n) = f_1(x_{\cS_1}) f_2(x_{\cS_2}) \cdots f_k(x_{\cS_k})	\ ,
\]
where, for each $i\in\{1,\dots,k\}$, $\cS_i$ is a subset of $\{1,\dots,n\}$,  notation $x_{\cS_i}$ stands for collection $(x_j)_{j\in\cS_i}$, and $f_i$ is some nonnegative function.

\subsection{Drawing the Graph}
Draw an undirected graph where all involved random variables e.g., $(X_j)_{j\in\{1,\dots,n\}}$, are nodes.
For each $i\in\{1,\dots,k\}$, draw edges between all the nodes in $X_{\cS_i}$.

\subsection{Checking Markov Relations}
Let $\cG_1$, $\cG_2$, and $\cG_3$ be three disjoint subsets of $\{1,\dots,n\}$.
If all paths in the graph from a node in $X_{\cG_1}$ to a node in $X_{\cG_3}$ pass through a node in $X_{\cG_2}$, then $X_{\cG_1}\mkv X_{\cG_2}\mkv X_{\cG_3}$ form a Markov chain. 
The proof of this result can be found in~\cite{permuter2010two-way} and is omitted here.

\section{Proof of Theorem~\ref{th:outer} (Outer Bound)}
\label{app:outer:proof}

Let $(k,D,\Delta)$ be an achievable tuple, and $\varepsilon>0$.
There exists an $(n,m)$-code $(F,g)$ s.t.
\begin{IEEEeqnarray}{rCl}
\frac mn 							&\leq& k+\varepsilon 		
											\ ,\label{eq:outer:code:rate}\\
\bE\left[ d(A^n,g(B^n,Y^m)) \right]	&\leq& D+\varepsilon 
											\ ,\label{eq:outer:code:D}\\
\dfrac1n\,H(A^n|E^n Z^m) 			&\geq& \Delta-\varepsilon 	
											\ ,\label{eq:outer:code:Delta}
\end{IEEEeqnarray}
with channel input $X^m$ as the output of the encoder $F(A^n)$.

From the fact that random variables $A_i$, $B_i$, $E_i$ are independent across time and the channel $X\mapsto(Y,Z)$ is memoryless, the joint distribution of $(A^n,B^n,E^n,X^m,Y^m,Z^m)$ can be written as follows, for each $i\in\{1,\dots,n\}$ and each $j\in\{1,\dots,m\}$:
\begin{multline*}
p(a^n,b^n,e^n,x^m,y^m,z^m)
	=	p(a^{i-1},b^{i-1},e^{i-1})\\ 
	\times  p(a_i,b_i,e_i)\,p(a_{i+1}^n,b_{i+1}^n,e_{i+1}^n)\,P_F(x^m|a^n)\,\\
		\times 
		p(y^{j-1},z^{j-1}|x^{j-1})\,p(y_j,z_j|x_j)\,p(y_{j+1}^m,z_{j+1}^m|x_{j+1}^m)\ .
\end{multline*}
Following the technique described in Appendix~\ref{app:graphical} and using the above expression, we can obtain the graph of Fig.~\ref{fig:outer:proof:joint} that will be used to establish Markov chains.

For each $i\in\{1,\dots,n\}$ (resp. each $j\in\{1,\dots,m\}$), define the source (resp. channel) auxiliary random variables $U_i$, $V_i$ (resp. $Q_j$, $T_j$) as 
\begin{IEEEeqnarray}{rCl}
U_i &=& (\phantom{A^{i-1},B^{i-1}, } B_{i+1}^n,E^{i-1}, Z^m)
													\ ,\label{eq:outer:defU}\\
V_i &=& (A^{i-1},B^{i-1},			 B_{i+1}^n,E^{i-1}, Y^m)
													\ ,\label{eq:outer:defV}\\[.15cm]
Q_j &=& (\phantom{A^n, }	B^n, Y^{j-1}, Z_{j+1}^m)\ ,\label{eq:outer:defQ}\\
T_j &=& (A^n,				B^n, Y^{j-1}, Z_{j+1}^m)\ .\label{eq:outer:defT}
\end{IEEEeqnarray}
Note that $(U_i,V_i)\mkv A_i\mkv (B_i,E_i)$, and $Q_j\mkv T_j\mkv X_j\mkv (Y_j,Z_j)$ form Markov chains (see Fig.~\ref{fig:outer:proof:joint}).

Following the usual technique, we introduce independent random variables $K$ and $J$, uniformly distributed over the sets $\{1,\dots,n\}$ and $\{1,\dots,m\}$, respectively.
We also define random variables $A=A_K$, $B=B_K$, $E=E_K$, $U=(K,U_K)$, $V=(K,V_K)$, $X=X_J$, $Y=Y_J$, $Z=Z_J$, $Q=(J,Q_j)$, and $T=(J,T_j)$.
$(U,V)\mkv A\mkv (B,E)$ and $Q\mkv T\mkv X\mkv (Y,Z)$ still form Markov chains. 
$(A,B,E)$, resp. $(X,Y,Z)$, is distributed according to $p(abe)$, resp. $p(xyz)$, i.e., the original distribution of $(A_i,B_i,E_i)$, resp. $(X_j,Y_j,Z_j)$.

\subsection{Rate}

We first prove the rate inequality~\eqref{eq:outer1} in Theorem~\ref{th:outer}:
\begin{IEEEeqnarray*}{rCl}
I(A^n;Y^m|B^n)
	&\stackrel{(a)}{=}&	\sum_{i=1}^n I(A_i;Y^m|A^{i-1} B^n) 		\\
	&\stackrel{(b)}{=}&	\sum_{i=1}^n I(A_i;Y^m|A^{i-1} B^n E^{i-1}) \\
	&=&					\sum_{i=1}^n \Big[ I(A_i;A^{i-1} B^{i-1} B_{i+1}^n E^{i-1} Y^m | B_i)\\
	 &-& I(A_i;A^{i-1} B^{i-1} B_{i+1}^n E^{i-1} | B_i) \Big] \\
	&\stackrel{(c)}{=}&	\sum_{i=1}^n I(A_i;A^{i-1} B^{i-1} B_{i+1}^n E^{i-1} Y^m | B_i) \\
	&\stackrel{(d)}{=}&	\sum_{i=1}^n I(A_i;V_i | B_i) \ ,
\end{IEEEeqnarray*}
where
\begin{itemize}
\item step~$(a)$ follows from the chain rule for conditional mutual information,
\item step~$(b)$ from the Markov chain $(A_i,Y^m) \mkv (A^{i-1},B^n) \mkv E^{i-1}$ (see Fig.~\ref{fig:outer:proof:joint}),
\item step~$(c)$ from the independence of random variables $A_i$, $B_i$, and $E_i$ across time,
\item step~$(d)$ from definition~\eqref{eq:outer:defV}.
\end{itemize}

We now find an upper bound for the latter quantity:
\begin{IEEEeqnarray*}{rCl}
I(A^n;Y^m|B^n)
	&\stackrel{(a)}{=}&		\sum_{j=1}^m I(A^n ; Y_j | B^n Y^{j-1} ) 			\\
	&\stackrel{(b)}{\leq}&	\sum_{j=1}^m I(A^n B^n Y^{j-1} Z_{j+1}^m ; Y_j )	\\
	&\stackrel{(c)}{\leq}&	\sum_{j=1}^m I(T_j ; Y_j )							\ ,
\end{IEEEeqnarray*}
where
\begin{itemize}
\item step~$(a)$ follows from the chain rule for conditional mutual information,
\item step~$(b)$ from the non-negativity of mutual information,
\item step~$(c)$ from definition~\eqref{eq:outer:defT}.
\end{itemize}

Putting all pieces together, we proved that
$\displaystyle{
\sum_{i=1}^n I(A_i;V_i | B_i) \leq \sum_{j=1}^m I(T_j;Y_j) \ .
}$
Using random variables $K$ and $J$, this inequality can be written as
\[
\sum_{i=1}^n I(A_K ; V_K | B_K, K=i) \leq \sum_{j=1}^m I(T_J ; Y_J | J=j) \ ,
\]
i.e.,
\begin{equation}
\label{eq:outer:proof:rate}
I(A ; V | B) \leq \frac mn\, I(T ; Y) \ .
\end{equation}

\subsection{Distortion at Bob}

Bob reconstructs $g(B^n,Y^m)$. 
The $i$-th coordinate of this estimate is
\[
g_i(Y^m,B^{i-1},B_i,B_{i+1}^n) \triangleq	\hat A_i(V_i,B_i) \ .
\]
The component-wise mean distortion at Bob thus can be written as:
\begin{IEEEeqnarray*}{rl}
\label{eq:outer:proof:distortion}
\bE\big[ d(A^n,&g(B^n,Y^m)) \big]\\
	&=	\frac1n \sum_{i=1}^n \bE\left[ d(A_i,\hat A_i(V_i,B_i)) \right]	\\
	&=	\frac1n \sum_{i=1}^n \bE\left[ d(A_K,\hat A_K(V_K,B_K)) \ \middle|\ K=i \right]	\\
	&= \bE\left[ d(A,\hat A(V,B)) \right] \ ,\yesnumber
\end{IEEEeqnarray*}
where we defined function $\hat A$ on $\cV\times\cB$ by 
$\hat A(V,B) = \hat A(K,V_K,B_K) \triangleq \hat A_K(V_K,B_K)$.

\subsection{Equivocation Rate at Eve}

We expand the equivocation at Eve $H(A^n | E^n Z^m)$ in two ways.

\subsubsection{}

First, 
\begin{IEEEeqnarray*}{rCl}
H(A^n | E^n Z^m)
	&\stackrel{(a)}{=}&		\sum_{i=1}^n H(A_i | A_{i+1}^n E^n Z^m) 			\\
	&\stackrel{(b)}{=}&		\sum_{i=1}^n H(A_i | A_{i+1}^n B_{i+1}^n E^i Z^m) 	\\
	&\stackrel{(c)}{\leq}&	\sum_{i=1}^n H(A_i | B_{i+1}^n E^i Z^m)			 	\\
	&\stackrel{(d)}{=}&		\sum_{i=1}^n H(A_i | U_i E_i) 						\ ,
\end{IEEEeqnarray*}
where 
\begin{itemize}
\item step~$(a)$ follows from the chain rule for conditional entropy,
\item step~$(b)$ from the Markov chain $A_i \mkv (A_{i+1}^n,E^i,Z^m)\mkv (B_{i+1}^n,E_{i+1}^n)$ (see Fig.~\ref{fig:outer:proof:joint}),
\item step~$(c)$ from the fact that conditioning reduces the entropy,
\item step~$(d)$ from definition~\eqref{eq:outer:defU}.
\end{itemize}
Using random variable $K$ defined above, the equivocation rate at Eve can be bounded as follows:
\begin{IEEEeqnarray*}{rCl}
\label{eq:outer:proof:equiv1}
\frac1n H(A^n | E^n Z^m)	
	&\leq&	\frac1n\sum_{i=1}^n H(A_K | U_K E_K,K=i)	\\
	&=&		H(A   | U E) 								\yesnumber\ .
\end{IEEEeqnarray*}

\subsubsection{}
Second, from standard properties of conditional entropy and mutual information:
\begin{IEEEeqnarray*}{rCl}
\label{eq:outer:decompositionDelta}
&&H(A^n | E^n Z^m)=\\[.1em]
	&&	H(A^n | B^n Y^m) + I(A^n ; B^n Y^m)- I(A^n ; E^n Z^m)	\\[.1em]
	&&=	H(A^n | B^n Y^m) + I(A^n ; Y^m |  B^n) + I(A^n ; B^n) 		\\[.1em]
	&& - I(A^n ; E^n | Z^m) - I(A^n ; Z^m) 				\\[.1em]
	&&\stackrel{(a)}{=}	H(A^n | B^n Y^m) + I(A^n ; Y^m |  B^n) + I(A^n Z^m ; B^n)	\\[.1em]
	 &&- I(A^n ; E^n | Z^m) - I(B^n A^n ; Z^m) 			\\[.1em]
	&&=	\underbrace{	H(A^n | B^n Y^m) + I(A^n ; B^n | Z^m) - I(A^n ; E^n | Z^m) }_{\Delta_s}	\\[.1em]
	\IEEEeqnarraymulticol{3}{R}{
	+ 	\underbrace{ I(A^n ; Y^m | B^n) - I(A^n ; Z^m | B^n) }_{\Delta_c} \ ,}\yesnumber%
\end{IEEEeqnarray*}
where step~$(a)$ follows from the Markov chain $B^n\mkv A^n\mkv Z^m$.

We now separately study the ``source'' term $\Delta_s$ and the ``channel'' term $\Delta_c$.

\paragraph{Study of $\Delta_s$}

The "source'' term $\Delta_s$ can be studied following the argument for the converse part of~\cite[Theorem~3]{villard2011securea} (see~\cite[Appendix~E-C]{villard2011securea}):
\begin{IEEEeqnarray*}{rCl}
\label{eq:outer:DeltaS}
\Delta_s
	&=&	H(A^n|B^n Y^m) + I(A^n;B^n|Z^m) - I(A^n;E^n|Z^m) \\[.1em]
	&\stackrel{(a)}{=}& H(A^n|B^n Y^m) + I(A^n;B^n) - I(Z^m;B^n)\\[.1em]
	& -& I(A^n;E^n) + I(Z^m;E^n) \\[.1em]	
         &=&	\sum_{i=1}^n \Big[ H(A_i|A^{i-1} B^n Y^m) + I(A_i;B_i) \\[.1em]
	&-& I(Z^m;B_i| B_{i+1}^n ) - I(A_i;E_i) + I( Z^m; E_i|E^{i-1} ) \Big] \\[.1em]	
	&\stackrel{(b)}{=}&	\sum_{i=1}^n \Big[ H(A_i|A^{i-1} B^n Y^m) + I(A_i;B_i) \\[.1em]
	&-& I(B_{i+1}^n Z^m;B_i) - I(A_i;E_i) + I(E^{i-1} Z^m; E_i) \Big] \\[.1em]
	&\stackrel{(c)}{=}&	\sum_{i=1}^n \Big[ H(A_i|A^{i-1} B^n E^{i-1} Y^m) + I(A_i;B_i)\\ [.1em]
	&-& I(B_{i+1}^n Z^m;B_i) - I(A_i;E_i)+ I(E^{i-1} Z^m; E_i)  \\[.1em]
	&+& I(E_i;B_{i+1}^n|E^{i-1} Z^m) - I(B_i;E^{i-1}|B_{i+1}^n Z^m) \Big]	\\[.1em]
	&=&					\sum_{i=1}^n \Big[ H(A_i|A^{i-1} B^n E^{i-1} Y^m) + I(A_i;B_i) - I(A_i;E_i) \\[.1em]
	&+& I(E_i;B_{i+1}^n E^{i-1} Z^m) - I(B_i;B_{i+1}^n E^{i-1} Z^m) \Big]	\\[.1em]
	&\stackrel{(d)}{=}&	\sum_{i=1}^n \Big[ H(A_i|V_i B_i) + I(A_i;B_i) - I(A_i;E_i) \\[.1em]
	&+& I(E_i;U_i) - I(B_i;U_i) \Big] \\[.1em]
	&=&	\sum_{i=1}^n \Big[ H(A_i|V_i B_i) + H(A_i|E_i) -H(A_i|B_i)  \\[.1em]
	&+& H(U_i| B_i) - H(U_i| E_i) \Big] \\[.1em]
        &=&	\sum_{i=1}^n \Big[ H(A_i|U_i E_i) -I(A_i;V_i| B_i) + I(U_i;A_i|B_i) \\ [.1em]
	&-& \Big( I(A_i;U_i|E_i) + H(U_i| A_i) - H(U_i| E_i)  \Big) \Big]  \\ [.1em]
	&\stackrel{(e)}{=}&	\sum_{i=1}^n \Big[ H(A_i|U_i E_i) \\ [.1em]
	&-& \Big( I(A_i;V_i|B_i) - I(A_i;U_i|B_i) \Big) \Big] \ ,\yesnumber
\end{IEEEeqnarray*}
where 
\begin{itemize}
\item step~$(a)$ follows from the Markov chain $(B^n,E^n)\mkv A^n\mkv Z^m$,
\item step~$(b)$ from the chain rules for conditional entropy and mutual information, and the fact that random variables $A_i$, $B_i$ and $E_i$ are independent across time,
\item step~$(c)$ from the Markov chain $A_i\mkv (A^{i-1},B^n,Y^m)\mkv E^{i-1}$ (see Fig.~\ref{fig:outer:proof:joint}) and Csisz\'ar and K\"orner's equality~\cite[Lemma~7]{csiszar1978broadcast},
 $$
 \sum_{i=1}^n  I(E_i;B_{i+1}^n|E^{i-1} Z^m)  = \sum_{i=1}^n I(B_i;E^{i-1}|B_{i+1}^n Z^m) \,,
 $$
\item step~$(d)$ from definitions~\eqref{eq:outer:defU} and~\eqref{eq:outer:defV},
\item step~$(e)$ from the Markov chain $U_i\mkv A_i\mkv (B_i,E_i)$.
\end{itemize}

\paragraph{Study of $\Delta_c$}

The ``channel'' term $\Delta_c$ can be studied following the argument for the converse part of~\cite[Theorem~1]{csiszar1978broadcast} (see \cite[Section~V]{csiszar1978broadcast}, \cite[Section~2.4]{liang2009information}):
\begin{IEEEeqnarray*}{rl}
\label{eq:outer:DeltaC}
&\Delta_c
	=				I(A^n ; Y^m | B^n) - I(A^n ; Z^m | B^n) \\[.1em]
	&\stackrel{(a)}{=} \sum_{j=1}^m \Big[ I(A^n;Y_j|B^n Y^{j-1}) - I(A^n;Z_j|B^n Z_{j+1}^m) \Big] \\[.1em]
	&= 				\sum_{j=1}^m \Big[ I(A^n Z_{j+1}^m;Y_j|B^n Y^{j-1}) \\[.1em]
	 &- I(Z_{j+1}^m;Y_j|A^n B^n Y^{j-1}) - I(A^n Y^{j-1};Z_j|B^n Z_{j+1}^m) \\[.1em]
	 &+ I(Y^{j-1};Z_j|A^n B^n Z_{j+1}^m) \Big] \\[.1em]	
	&\stackrel{(b)}{=}	\sum_{j=1}^m \Big[ I(A^n Z_{j+1}^m;Y_j|B^n Y^{j-1}) - I(A^n Y^{j-1};Z_j|B^n Z_{j+1}^m) \Big] \\[.1em]
	&=					\sum_{j=1}^m \Big[ I(A^n;Y_j|B^n Y^{j-1} Z_{j+1}^m) + I(Z_{j+1}^m;Y_j|B^n Y^{j-1}) \\[.1em]
	&- I(A^n;Z_j|B^n Y^{j-1} Z_{j+1}^m) - I(Y^{j-1};Z_j|B^n Z_{j+1}^m) \Big] \\[.1em]
	&\stackrel{(c)}{=}   \sum_{j=1}^m \Big[ I(A^n;Y_j|B^n Y^{j-1} Z_{j+1}^m) - I(A^n;Z_j|B^n Y^{j-1} Z_{j+1}^m) \Big] \hspace{3cm}\\[.1em]
	&\stackrel{(d)}{=}	\sum_{j=1}^m \Big[ I(T_j;Y_j|Q_j) - I(T_j;Z_j|Q_j) \Big] \ ,\yesnumber
\end{IEEEeqnarray*}
where 
\begin{itemize}
\item step~$(a)$ follows from the chain rule for conditional mutual information,
\item steps~$(b)$ and $(c)$ from Csisz\'ar and K\"orner's equality~\cite[Lemma~7]{csiszar1978broadcast},
\item step~$(d)$ from definitions~\eqref{eq:outer:defQ} and \eqref{eq:outer:defT}.
\end{itemize}

\paragraph{}
Gathering \eqref{eq:outer:decompositionDelta}--\eqref{eq:outer:DeltaC}, using variables $J$, $K$, and new source-channel variables, the equivocation rate at Eve can be written as:
\begin{multline}
\label{eq:outer:proof:equiv2}
\frac1n H(A^n|E^n Z^m)
	=	H(A|U E) - \Big[ I(A;V|B) - I(A;U|B) \\[.1em]
				 - \frac mn \Big( I(T;Y|Q) - I(T;Z|Q) \Big)	\Big] \ .
\end{multline}

\subsection{End of Proof}

Inequalities~\eqref{eq:outer:proof:rate}--\eqref{eq:outer:proof:equiv1}, and \eqref{eq:outer:proof:equiv2} only involve \emph{marginal} distributions of auxiliary variables $p(uv|a)$ and $p(qtx)$. 
Consequently, we can define new variables $\tilde U$, $\tilde V$, $\tilde Q$, $\tilde T$, $\tilde X$, with identical marginal distributions $p(uv|a)$ and $p(qtx)$ (and hence verifying \eqref{eq:outer:proof:rate}--\eqref{eq:outer:proof:equiv1}, and \eqref{eq:outer:proof:equiv2}) such that the (global) joint distribution can be written as 
$p(uvqtabexyz) = p(uv|a)p(abe)\,p(q|t)p(t) p(x|t)p(yz|x)$ i.e., such that source and channel variables are independent.

Gathering inequalities \eqref{eq:outer:proof:rate}--\eqref{eq:outer:proof:equiv1}, \eqref{eq:outer:proof:equiv2}, and \eqref{eq:outer:code:rate}--\eqref{eq:outer:code:Delta}, we then proved that, for each achievable tuple $(k,D,\Delta)$ and each $\varepsilon>0$, there exist random variables $U$, $V$, $Q$, $T$, $X$ with joint distribution $p(uvqtabexyz) = p(uv|a)p(abe)p(q|t)p(t)p(x|t)p(yz|x)$, and a function $\hat A$ such that
\begin{IEEEeqnarray*}{rCl}
I(A;V|B)				&\leq&	(k+\varepsilon) I(T;Y) 						\ ,\\[.1em]
D + \varepsilon			&\geq&	\bE\big[d(A,\hat A(V,B))\big] 				\ ,\\[.1em]
\Delta - \varepsilon	&\leq&	H(A|UE) - \Big[ I(V;A|B) - I(U;A|B) \ ,\\[.1em]
				&-& (k+\varepsilon) \Big( I(T;Y|Q) - I(T;Z|Q) \Big) \Big]_+	\ ,
\end{IEEEeqnarray*}
i.e., $(k+\varepsilon,D+\varepsilon,\Delta-\varepsilon)\in\cR_\text{out}$.
Letting $\varepsilon$ tend to zero then proves Theorem~\ref{th:outer}.
\endIEEEproof

\section{Proof of the Converse Part of Proposition~\ref{prop:binary:digital}}
\label{app:binary:digital}

Let $(k=1,D=0,\Delta)\in\cR_\text{digital}$ i.e., such that there exist random variables $U$, $V$, $Q$, $T$, $X$ with joint distribution $p(uvqtaexyz) = p(u|v)p(v|a)p(ae)p(q|t)p(tx)p(yz|x)$, and a function $\hat A$, verifying

\begin{IEEEeqnarray*}{rcl}
I(U;A|B) & \leq  & I(Q;Y) 										\ ,\\[.1em]
I(V;A|B) & \leq  & I(T;Y) 										\ ,\\[.1em]
0 		 & \geq & \bE\big[d(A,\hat A(V,B))\big]					\ ,\\[.1em]
\Delta	& \leq & H(A|UE) - \Big[ I(V;A|UB) -  \Big( I(T;Y|Q) \\[.1em]
				\IEEEeqnarraymulticol{3}{R}{	- I(T;Z|Q) \Big) \Big]_+ \ .}
\end{IEEEeqnarray*}

From the assumptions of Section~\ref{sec:binary}, we can easily prove the following inequalities:
\begin{IEEEeqnarray}{rcl}
\beta (1 - H(A|U) )	&\leq& 1 - H(X|Q)	\ ,\label{eq:binary:digital1}\\[.1em]
\Delta				&\leq& H(A|U) + h_2(\epsilon) - H(E|U)	-\Big[ \beta H(A|U)	 \nonumber\\[.1em]
	&-&  \Big( H(X|Q) - H(Z|Q) +h_2(\zeta) \Big) \Big]_+	\ .
\end{IEEEeqnarray}

Since $0\leq H(A|U)\leq H(A)=1$ and $0\leq H(X|Q)\leq H(X)\leq 1$, we now introduce two parameters: $u=h_2^{-1}\big(H(A|U)\big)$, $q=h_2^{-1}\big(H(X|Q)\big)$.

Then, from the fact that $E$ is the output of a BSC with crossover probability $\epsilon$ and input $A$, Mrs.\ Gerber's lemma~\cite{wyner1973theorem} yields:
\begin{equation}
\label{eq:binary:gerber1}
H(E|U) \geq h_2(\epsilon \star u)\ .
\end{equation}
Similarly, since $Z$ is the output of a BSC with crossover probability $\zeta$ and input $X$:
\begin{equation}
\label{eq:binary:gerber2}
H(Z|Q) \geq h_2( \zeta \star q)\ .
\end{equation}

Gathering \eqref{eq:binary:digital1}--\eqref{eq:binary:gerber2}, we obtain
\begin{IEEEeqnarray*}{rcl}
\beta (1 - h_2(u) )	&\leq &1 - h_2(q)	\ ,\\
\Delta				&\leq& h_2(u) + h_2(\epsilon) - h_2(\epsilon \star u) \\ 
&-& \Big[ \beta h_2(u) - \Big( h_2(q) - h_2( \zeta \star q) + h_2(\zeta) \Big) \Big]_+	\ .
\end{IEEEeqnarray*}
This proves the converse part of Proposition~\ref{prop:binary:digital}.
\endIEEEproof

\section{Proof of Lemma~\ref{lem:hybrid:proof:decoding}}
\label{app:hybrid:proof:decoding}

In this section, we prove Lemma~\ref{lem:hybrid:proof:decoding} following the argument of~\cite{lim2010lossy}.
In the decoding procedure described in Section~\ref{sec:hybrid:proof:decoding}, an error occurs in the first step if there exists another admissible codeword $u^n(r_1')$ with $r_1'\neq r_1$.
The probability of this event can be written as

\begin{IEEEeqnarray*}{rCl}
\label{eq:hybrid:proof:Pd1}
P_{d,1}
	&\triangleq&	\pr{ \exists\ r_1'\neq r_1 : (u^n(r_1'),B^n,Y^n) \in \typ{n}{U,B,Y} }									\\
	&\leq&			\sum_{r_1'=1}^{2^{nR_1}} \pr{ r_1\neq r_1', (u^n(r_1'),B^n,Y^n) \in \typ{n}{U,B,Y} }									\\
	&=&			\sum_{r_1'=1}^{2^{nR_1}}\sum_{a^n} p(a^n) \Pr \Big\{ r_1\neq r_1', (u^n(r_1'),B^n,Y^n) \\
	\IEEEeqnarraymulticol{3}{R}{\in \typ{n}{U,B,Y}\ \Big|\ A^n=a^n \Big\}}	\ .\quad\yesnumber
\end{IEEEeqnarray*}

We now study each term of the above summation: For each $r_1'$, and each $a^n$,
\begin{IEEEeqnarray*}{rCl}
\IEEEeqnarraymulticol{3}{l}{ \pr{ r_1\neq r_1', (u^n(r_1'),B^n,Y^n) \in \typ{n}{U,B,Y}\ \Big|\ A^n=a^n } }	\\
	&\leq&	 \Pr \Big\{ (u^n(r_1'),B^n,Y^n)    \in \typ{n}{U,B,Y}\ \Big|A^n=a^n, r_1\neq r_1' \Big\} \\
	&=&	 \!\!	\sum_{(u^n,b^n,y^n)\in\typ{n}{U,B,Y}}\Pr\Big\{ u^n(r_1')=u^n, B^n=b^n, Y^n=y^n \\
	\IEEEeqnarraymulticol{3}{R}{  \Big| A^n=a^n, r_1\neq r_1' \Big\}} \\
	&=&	 \!\!		\sum_{(u^n,b^n,y^n)\in\typ{n}{U,B,Y}}
				\pr{ u^n(r_1')=u^n \ \Big|\ A^n=a^n, r_1\neq r_1'   }		\\
			&	& \times\pr{ B^n=b^n, Y^n=y^n\ \Big|\ A^n=a^n, r_1\neq r_1'}. \quad 
			\yesnumber
\label{eq:hybrid:proof:Pd1_2}
\end{IEEEeqnarray*}

For each $r_1'$, and each $a^n$, according to the encoding procedure described in Section~\ref{sec:hybrid:proof:encoding},
\begin{IEEEeqnarray*}{rl}
& \pr{ r_1 = r_1' \ \Big|\ A^n=a^n } =\\[.1em]
	&	\Pr\Big\{ \left( \bigcap_{k=1}^{r_1'-1} \big\{ u^n(k)\notin\typ{n}{U|a^n} \big\} \right) \cap \big\{ u^n(r_1')\in\typ{n}{U|a^n} \big\} \Big\}	\\[.1em]
	&\leq	\pr{ u^n(r_1')\in\typ{n}{U|a^n} }	\\[.1em]
	&\leq	2^{-n(I(U;A)-\eta_n)}	\ ,
\end{IEEEeqnarray*}
where the last inequality holds for some sequence $\eta_n\toas{n\to\infty}0$ (see Lemma~\ref{lem:jointTypicality} in Appendix~\ref{app:typical}) from the fact that the codewords are uniformly distributed over $\typ{n}{U}$, independent of the source, and $a^n\in\typ{n}{A}$.
From this inequality, there exists $\kappa<1$ such that, for some sufficiently large $n$,
\begin{equation}
\label{eq:hybrid:proof:kappa}
\pr{ r_1 = r_1' \ \Big|\ A^n=a^n } \leq \kappa \ .
\end{equation}

The above upper bound yields the following inequality, for each $u^n$, $r_1'$, and $a^n$:
\begin{IEEEeqnarray*}{rCl}
\IEEEeqnarraymulticol{3}{l}{ \pr{ u^n(r_1')=u^n \ \Big|\ A^n=a^n, r_1\neq r_1' } } \\\hspace{1.5cm}
	&=&  \pr{ u^n(r_1')=u^n \ \Big|\ A^n=a^n }\\[.1em]
	&\times& 	\frac	{ \pr{ r_1\neq r_1' \ \Big|\ A^n=a^n, u^n(r_1')=u^n } }
				{ 1 - \pr{ r_1 = r_1' \ \Big|\ A^n=a^n } }	\\[.1em]
	&\leq&  \frac{ \pr{ u^n(r_1')=u^n } }{ 1 - \kappa }	\ ,\yesnumber
\label{eq:hybrid:proof:kappa1}
\end{IEEEeqnarray*}
where the above equation follows from \eqref{eq:hybrid:proof:kappa} and the fact that $u^n(r_1')$ and $A^n$ are independent.

Plugging \eqref{eq:hybrid:proof:kappa1} into \eqref{eq:hybrid:proof:Pd1_2} yields, for each $r_1'$, and each $a^n$, 
\begin{multline}
\label{eq:hybrid:proof:Pd1_3}
\pr{ r_1\neq r_1', (u^n(r_1'),B^n,Y^n) \in \typ{n}{U,B,Y}\ \Big|\ A^n=a^n } \\[.1em]
	\leq	\sum_{(u^n,b^n,y^n)\in\typ{n}{U,B,Y}}
				\frac{ \pr{ u^n(r_1')=u^n } }{ 1 - \kappa } \\[.1em]
				\times \pr{ B^n=b^n, Y^n=y^n\ \Big|\ A^n=a^n, r_1\neq r_1' } \ .
\end{multline}

From \eqref{eq:hybrid:proof:kappa} once again, the last term in the r.h.s.\ of the above equation can be upper bounded as follows, for each $r_1'$, each $a^n$, and each $(b^n,y^n)\in\typ{n}{B,Y}$:
\begin{IEEEeqnarray*}{ll}
& \pr{ B^n=b^n, Y^n=y^n\ \Big|\ A^n=a^n, r_1\neq r_1'} \\
	&= \frac	{ \pr{ B^n=b^n, Y^n=y^n, r_1\neq r_1' \ \Big|\ A^n=a^n } }
				{ \pr{ r_1\neq r_1' \ \Big|\ A^n=a^n } } 	\\
	&\leq  \frac{ \pr{ B^n=b^n, Y^n=y^n \ \Big|\ A^n=a^n } }{ 1 - \kappa } \ .\yesnumber
\label{eq:hybrid:proof:kappa2}
\end{IEEEeqnarray*}

Gathering~\eqref{eq:hybrid:proof:Pd1}, \eqref{eq:hybrid:proof:Pd1_3} and \eqref{eq:hybrid:proof:kappa2}, we obtain, from the fact that the codewords are identically distributed,
\begin{IEEEeqnarray*}{ll}
P_{d,1} 
	\leq& \frac{2^{nR_1}}{(1 - \kappa)^2} \sum_{(u^n,b^n,y^n)\in\typ{n}{U,B,Y}} \pr{ u^n(1)=u^n }\\
				\IEEEeqnarraymulticol{2}{R}{   \times \pr{ B^n=b^n, Y^n=y^n }}	\ .
\end{IEEEeqnarray*}
Finally, from the joint typicality lemma (Lemma~\ref{lem:jointTypicality} in Appendix~\ref{app:typical}), there exists a sequence $\eta_n'\toas{n\to\infty}0$ such that
\[
P_{d,1} \leq \frac{2^{nR_1}}{(1 - \kappa)^2}\,2^{-n(I(U;BY) - \eta_n')}	\ .
\]
This proves Lemma~\ref{lem:hybrid:proof:decoding}.
\endIEEEproof

\section{Proof of the Converse Part of Proposition~\ref{prop:gaussian:B}}
\label{app:gaussian:B}

Assume that $P_B \leq P_E$, and let $(D,D_E)$ be an achievable tuple.
From Proposition~\ref{prop:digital:special:B}, there exist random variables $V$, $Q$, $T$, $X$ with joint distribution $p(vqtabexyz) = p(v|a)p(abe)\,p(q|t)p(t)\linebreak p(x|t)p(yz|x)$,
and a function $\hat A$, verifying
\begin{IEEEeqnarray*}{l}
I(V;A|B)\leq I(T;Y) 												\ ,\\[.1em]
D 		\geq \bE\big[d(A,\hat A(V,B))\big] 							\ ,\\[.1em]
\Delta	\leq  h(A|E) - \Big[ I(V;A|B) - I(T;Y|Q) + I(T;Z|Q) \Big]_+ \ ,\\[.1em]
\Var{X}  \leq P 													\ ,
\end{IEEEeqnarray*}
where $\Delta=\frac12\log(2\pi e\,D_E)$.

Then, from standard properties of differential entropy, and from the fact that distortion measure~$d$ is the Euclidean distance on $\bR$, the following sequence of inequalities holds true:
\[
2^{2h(A|VB)}/(2\pi e) \leq \Var{A|VB} \leq \bE\big[d(A,\hat A(V,B))\big] \ ,
\]Ä
and hence
\[
2^{2 I(V;A|B)} = \frac{2^{2 h(A|B)}}{2^{2 h(A|VB)}} \geq \frac1{1+\frac1{P_B}}\cdot\frac1D \ .
\]

Gathering the above equations, tuple $(D,D_E)$ verifies the following inequalities:
\begin{IEEEeqnarray*}{l}	
\frac1{1+\frac1{P_B}}\cdot\frac1D	
		\leq 1 + \frac{P}{P_Y}											\ ,\\
D_E		\leq \frac1{1 + \frac1{P_E}}									\ ,\\
D_E		\leq \frac{1+\frac1{P_B}}{1+\frac1{P_E}}\cdot D
						\cdot 2^{2\left[ I(T;Y|Q) - I(T;Z|Q) \right]}	\ .
\end{IEEEeqnarray*}

Depending on the values of $P_Y$ and $P_Z$, there are two different cases, investigated in the following paragraphs.

\subsection{$P_Y < P_Z$}

Assume in this section that $P_Y<P_Z$.
According to Remark~\ref{rem:marginal}, the Markov chain $X\mkv Y\mkv Z$ can be assumed without loss of generality.
Hence, from the long Markov chain $Q\mkv T\mkv X\mkv Y\mkv Z$, $I(T;Y|Q) - I(T;Z|Q) = I(T,Y|QZ) \leq I(X;Y|Z)$, and tuple $(D,D_E)$ verifies the following inequalities:
\begin{IEEEeqnarray*}{l}
\frac1{1+\frac1{P_B}}\cdot\frac1D	
		\leq 1 + \frac{P}{P_Y}											\ ,\\
D_E		\leq \frac1{1 + \frac1{P_E}}									\ ,\\
D_E		\leq \frac{1+\frac1{P_B}}{1+\frac1{P_E}}\cdot D
						\cdot 2^{2 I(X;Y|Z)}	\ .
\end{IEEEeqnarray*}

Then, from the Markov chain $X\mkv Y\mkv Z$, there exists a random variable $\widetilde N_Z\sim\cN(0,P_Z-P_Y)$, independent of $Y$, such that $Z = Y + \widetilde N_Z$. 
The entropy power inequality (EPI)~\cite{shannon1948mathematical,elgamal2010lecture} yields:
\[
2^{2 h(Z)} \geq 2^{2 h(Y)} + 2^{2 h(\widetilde N_Z)}	\ .
\]
From the latter equation and inequality $\Var{Y}\leq P+P_Y$,
\begin{IEEEeqnarray*}{rCl}
\label{eq:gaussian:B:Cs}
I(X;Y|Z)	&=&		h(Y) - h(Y|X) - h(Z) + h(Z|X)						\\
			&\leq&	h(Y) - h(N_Y) \\
			&-& \frac12\log\left( 2^{2 h(Y)} 
						+ 2^{2 h(\widetilde N_Z)} \right) + h(N_Z)		\\		
			&\leq&	\frac12\log\left( \frac1{ 1 + \frac{P_Z-P_Y}{P+P_Y}}\,
									  \frac{P_Z}{P_Y} \right)			\\	
			&=&		\frac12\log\left( \frac{1+\frac{P}{P_Y}}
										{1+\frac{P}{P_Z}} \right)		\ .\yesnumber
\end{IEEEeqnarray*}
Tuple $(D,D_E)$ thus verifies the following inequalities:
\begin{IEEEeqnarray*}{l}
\frac1{1+\frac1{P_B}}\cdot\frac1D	
		\leq 1 + \frac{P}{P_Y}											\ ,\\
D_E		\leq \frac1{1 + \frac1{P_E}}									\ ,\\
D_E		\leq \frac{1+\frac1{P_B}}{1+\frac1{P_E}}\cdot D
						\cdot \frac{1+\frac{P}{P_Y}}{1+\frac{P}{P_Z}}	\ .
\end{IEEEeqnarray*}

\subsection{$P_Y \geq P_Z$}

Assume in this section that $P_Y\geq P_Z$.
Then, $I(T;Y|Q) \leq I(T;Z|Q)$, and tuple $(D,D_E)$ verifies the following inequalities:
\begin{IEEEeqnarray*}{l}	
\frac1{1+\frac1{P_B}}\cdot\frac1D	
		\leq 1 + \frac{P}{P_Y}							\ ,\\
D_E		\leq \frac1{1 + \frac1{P_E}}					\ ,\\
D_E		\leq \frac{1+\frac1{P_B}}{1+\frac1{P_E}}\cdot D	\ .
\end{IEEEeqnarray*}

This concludes the proof of the converse part of Proposition~\ref{prop:gaussian:B}.

\section{Proof of the Converse Part of Proposition~\ref{prop:gaussian:EZ}}
\label{app:gaussian:EZ}

Assume that $P_B > P_E$ and $P_Y \geq P_Z$, and let $(D,D_E)$ be an achievable tuple.
From Proposition~\ref{prop:digital:special:Z}, there exist random variables $U$, $V$, $X$ with joint distribution $p(uvabexyz) = p(u|v)p(v|a)p(ae)p(b|e)\,p(x)p(yz|x)$,\footnote{%
	Since it is assumed that $P_B > P_E$, according to Remark~\ref{rem:marginal}, the Markov chain $A\mkv E\mkv B$ can be assumed here without loss of generality.}
and a function $\hat A$, verifying
\begin{IEEEeqnarray*}{rcl}
I(V;A|B)&\leq& I(X;Y) 						\ ,\\[.1em]
D 		&\geq& \bE\big[d(A,\hat A(V,B))\big] 	\ ,\\[.1em]
\Delta	&\leq&  h(A|VB) + I(A;B|U) - I(A;E|U) \ ,\\[.1em]
\Var{X}	&\leq& P 								\ ,
\end{IEEEeqnarray*}
where $\Delta=\frac12\log(2\pi e\,D_E)$.

From the long Markov chain $U\mkv V\mkv A\mkv E\mkv B$, $I(A;B|U) - I(A;E|U) = - I(A;E|UB) \leq - I(A;E|VB)$, and tuple $(D,\Delta)$ verifies the following inequalities:
\begin{IEEEeqnarray}{rcl}
I(V;A|B)&\leq& I(X;Y) 						\ ,\\[.1em]
D 		&\geq& \bE\big[d(A,\hat A(V,B))\big] 	\ ,\\[.1em]
\Delta	&\leq&  h(A|VB) - h(E|VB) + h(E|AB)	\ ,\label{eq:gaussian:EZ:Delta}\\[.1em]
\Var{X}	&\leq& P 								\ .
\end{IEEEeqnarray}
Moreover, the side informations write:
\begin{IEEEeqnarray*}{rCL}
E &=& A + N_E							\ ,\\
B &=& A + N_E + \bar N_B = E + \bar N_B	\ ,
\end{IEEEeqnarray*}
where $\bar N_B\sim\cN(0,P_B-P_E)$ is independent of $A$ and $N_E\sim\cN(0,P_E)$.

In order to find an upper bound on the r.h.s.\ of~\eqref{eq:gaussian:EZ:Delta}, we need the following expansion of $E$, for any $\gamma\in\bR$:
\begin{equation}
\label{eq:gaussian:EZ:expE}
E = \gamma B + (1-\gamma) A + C \ ,
\end{equation}
where
\[
C = (1-\gamma)N_E - \gamma \bar N_B \ .
\]
Note that $(A,B,C)$ is a Gaussian vector, and that $A$ and $C$ are independent for any $\gamma$.
The usefulness of the above expansion comes from the fact that $C$ is also independent of $B$ if $\gamma=\frac{P_E}{P_B}$:
\begin{IEEEeqnarray*}{rCL}
\bE[BC]	&=& (1-\gamma)\,\bE\left[ B N_E \right] - \gamma\,\bE\left[ B \bar N_B \right]	\\
		&=& (1-\gamma)P_E - \gamma\,(P_B-P_E)	\\
		&=& P_E - \gamma P_B					\\
		&=& 0									\ .
\end{IEEEeqnarray*}
Finally, since $V$ only depends on $A$, $C$ is independent of $(V,A,B)$.

Using expansion~\eqref{eq:gaussian:EZ:expE}, we now write 
\begin{IEEEeqnarray*}{rCl}
h(E|VB)	&=& h(\gamma B + (1-\gamma) A + C|VB)	\\
		&=& h((1-\gamma) A + C|VB)				\ .
\end{IEEEeqnarray*}
And from the above paragraph, the conditional EPI holds between $A$ and $C$ (given $(V,B)$):
\[
2^{2h((1-\gamma) A + C|VB)}
	\geq 2^{2h((1-\gamma) A|VB)} + 2^{2h(C|VB)} \ .
\]
Since $C$ is independent of $(V,A,B)$, the last entropy can be written as
\begin{IEEEeqnarray*}{rCL}
h(C|VB)	&=& 	h(C|AB)		\\
		&=&		h(E|AB)		\ ,
\end{IEEEeqnarray*}
where the last equality follows from expansion~\eqref{eq:gaussian:EZ:expE}.

Gathering the above equations, \eqref{eq:gaussian:EZ:Delta} yields
\begin{IEEEeqnarray*}{rCl}
\label{eq:gaussian:EZ:Delta2}
\Delta	&\leq&	h(A|VB) \\
                &-& \frac12\log\left( (1-\gamma)^2\,2^{2 h(A|VB)} + 2^{2 h(E|AB)} \right) + h(E|AB)	\\
		&=&		 \frac12\log\left( \frac1{\displaystyle \frac1{2^{2 h(A|VB)}} + \frac{(1-\gamma)^2}{\displaystyle2^{2 h(E|AB)}}} \right)	\\
		&=&		 \frac12\log\left( \frac{2\pi e}{\displaystyle \frac{2\pi e}{2^{2 h(A|VB)}} + \frac1{P_E} - \frac1{P_B}} \right) \yesnumber\ ,
\end{IEEEeqnarray*}
where the last equality follows from $\Var{E|AB} = P_E \left(1-\frac{P_E}{P_B}\right)$ after some manipulations.

Now, from standard properties of differential entropy, and from the fact that distortion measure~$d$ is the Euclidean distance on $\bR$, the following sequence of inequalities holds true:
\[
2^{2 h(A|VB)}/(2\pi e) \leq \Var{A|VB} \leq \bE\big[d(A,\hat A(V,B))\big] \ ,
\]
and hence
\[
2^{2 I(V;A|B)} = \frac{2^{2 h(A|B)}}{2^{2 h(A|VB)}} \geq \frac1{1+\frac1{P_B}}\cdot\frac1D \ .
\]

Gathering the above equations, tuple $(D,D_E)$ verifies the following inequalities:
\begin{IEEEeqnarray*}{l}	
\frac1{1+\frac1{P_B}}\cdot\frac1D	
		\leq 1 + \frac{P}{P_Y}								\ ,\\
D_E		\leq \frac1{\frac1D + \frac1{P_E} - \frac1{P_B}}	\ .
\end{IEEEeqnarray*}
This proves the converse part of Proposition~\ref{prop:gaussian:EZ}.

\section{Proof of Proposition~\ref{prop:gaussian:EY}}
\label{app:gaussian:EY}

Assume that $P_B > P_E$ and $P_Y < P_Z$, and let $(D,D_E)$ be an achievable tuple.
From Theorem~\ref{th:outer}, there exist random variables $U$, $V$, $Q$, $T$, $X$ with joint distribution $p(uvqtabexyz) = p(uv|a)p(abe)\,p(q|t)p(tx)p(y|x)p(z|y)$,\footnote{%
	Since it is assumed that $P_Y<P_Z$, according to Remark~\ref{rem:marginal}, the Markov chain $X\mkv Y\mkv Z$ can be assumed here without loss of generality.}
and a function $\hat A$, verifying
\begin{IEEEeqnarray*}{rcl}
I(V;A|B) &\leq& I(T;Y) 											\ ,\\
D 		&\geq& \bE\big[d(A,\hat A(V,B))\big] 						\ ,\\
\Delta	&\leq&  h(A|UE) -\Big[ I(V;A|B) - I(U;A|B)  \\
\IEEEeqnarraymulticol{3}{r}{- \Big( I(T;Y|Q) - I(T;Z|Q) \Big) \Big]_+} 	\ ,\\
\Var{X} & \leq &P 												\ ,
\end{IEEEeqnarray*}
where $\Delta=\frac12\log(2\pi e\,D_E)$.

From the Markov chain $Q\mkv T\mkv X\mkv Y\mkv Z$, tuple $(D,\Delta)$ verifies the following inequalities:
\begin{IEEEeqnarray}{rcl}
I(V;A|B) &\leq& I(X;Y) 						\ ,\label{eq:gaussian:EY1}\\
D 		&\geq& \bE\big[d(A,\hat A(V,B))\big]	\ ,\label{eq:gaussian:EY2}\\
\Delta	&\leq&  h(A|UE) \nonumber\\
&-& \Big[ I(V;A|B) - I(U;A|B) - I(X;Y|Z) \Big]_+
											\ ,\label{eq:gaussian:EY3}\\
\Var{X} &\leq& P 								\ .\label{eq:gaussian:EY4}
\end{IEEEeqnarray}
Moreover, from the proof of Theorem~\ref{th:outer} (see \eqref{eq:outer:defU}, \eqref{eq:outer:defV} in Appendix~\ref{app:outer:proof}), we can restrict our attention to auxiliary variables $U$, $V$ s.t.\ $U\mkv V\mkv A\mkv E$ form a Markov chain.

We introduce two parameters:
$\nu = 2^{2h(A|VB)}/(2\pi e)$,
$\mu = 2^{2h(A|UB)}/(2\pi e)$.
From the fact that conditioning reduces the entropy and classical properties of the differential entropy, the above parameters are bounded as follows:
\begin{equation}
\label{eq:gaussian:EY:consnumu}
\nu \leq \mu \leq \frac1{1+\frac1{P_B}} \ .
\end{equation}

We now write \eqref{eq:gaussian:EY1}--\eqref{eq:gaussian:EY4} as functions of these parameters.
First, recalling that distortion measure $d$ is the Euclidean distance on $\bR$,
\begin{equation}
\label{eq:gaussian:EY:Dnu}
\bE\big[d(A,\hat A(V,B))\big] \geq \Var{A|VB} \geq \nu \ . 
\end{equation}

Since $P_B > P_E$, and according to Remark~\ref{rem:marginal}, we can assume that $A\mkv E\mkv B$ form a Markov chain.
Then, following the argument of Appendix~\ref{app:gaussian:EZ} (based on expansion~\eqref{eq:gaussian:EZ:expE} together with the conditional EPI), we can easily prove the following equation (similar to~\eqref{eq:gaussian:EZ:Delta2}):
\begin{IEEEeqnarray*}{rCl}
\label{eq:gaussian:EY:hAUEmu}
h(A|UE)	&\leq&	h(A|UB)-\frac12\log\Big( (1-\gamma)^2\,2^{2 h(A|UB)} \nonumber\\
	\IEEEeqnarraymulticol{3}{R}{ +  2^{2 h(E|AB)} \Big) + h(E|AB)}	\\
		&=&		\frac12\log\left( \frac{2\pi e}{\frac1\mu + \frac1{P_E} - \frac1{P_B} } \right)
				\yesnumber\ .
\end{IEEEeqnarray*}

Since the Markov chain $X\mkv Y\mkv Z$ is assumed, \eqref{eq:gaussian:B:Cs} also holds  here:
\begin{equation}
\label{eq:gaussian:EY:IXYZzeta}
I(X;Y|Z) \leq \frac12\log\left( \frac{1+\frac{P}{P_Y}}
									{1+\frac{P}{P_Z}} \right)		\ .
\end{equation}

Gathering the above equations, tuple $(D,D_E)$ verifies the following inequalities:
\begin{IEEEeqnarray*}{l}	
\nu \leq \mu \leq \frac1{1+\frac1{P_B}}								\ ,\\
 \frac1{1+\frac1{P_B}}\cdot\frac1\nu	\leq 1 + \frac{P}{P_Y} 			\ ,\\
D			\geq \nu			 									\ ,\\
D_E			\leq \frac1{\frac1\mu + \frac1{P_E} - \frac1{P_B}}		\ ,\\
D_E			\leq \frac1{\frac1\mu + \frac1{P_E} - \frac1{P_B}}\cdot\frac\nu\mu\cdot
						\frac{1+\frac{P}{P_Y}}{1+\frac{P}{P_Z}}		\ .
\end{IEEEeqnarray*}

Eliminating parameter $\nu$ and $\mu$ proves Proposition~\ref{prop:gaussian:EY}.

\section{Proof of Proposition~\ref{prop:gaussian:hybrid}}
\label{app:gaussian:hybrid}

In this section, we prove a sequence of lemmas which together prove Proposition~\ref{prop:gaussian:hybrid}.
To that end, using auxiliary variables~\eqref{eq:hybrid:defU}--\eqref{eq:hybrid:defX}, we show that any tuple $(D,D_E)$ verifying conditions~\eqref{eq:GaussD}--\eqref{eq:GaussRate} in Proposition~\ref{prop:gaussian:hybrid} lies in region $\cR_\text{hybrid}^P$.

\subsection{Conditional Covariance of Gaussian Variables}

The following lemma can be found in~\cite[Appendix~A.2]{rasmussen2006gaussian}:

\begin{lemma}[Conditional covariance matrix of Gaussian vectors]
\label{lem:gaussian:varCond}
Let $P$, $Q$ be two jointly Gaussian random vectors with covariance matrix
\[
\Gamma_{PQ} = 
	\left[\begin{IEEEeqnarraybox}[\IEEEeqnarraystrutmode][c]{,c/c,}
		A & C\T	\\
		C & B
	\end{IEEEeqnarraybox}\right] \ .
\]
Then the conditional covariance matrix $\Gamma_{(P|Q)}$ of $P$ given $Q$ verifies the following equality
\begin{equation}
\Gamma_{(P|Q)} = A - C B^{-1} C\T \ .
\end{equation}
\end{lemma}

From the above lemma, we can easily derive the following corollary, which gives the conditional variance for two scalar Gaussian random variables.
\begin{corollary}[Conditional variance of Gaussian variables]
\label{coro:gaussian:varCond}
Let $P$ and $Q$ be two jointly Gaussian random variables. Then
\begin{equation}
\Var{P|Q} = \frac{\det{\Gamma_{PQ}}}{\Var{Q}} \ .
\end{equation}
\end{corollary}

\subsection{Preliminary Lemmas}

\begin{lemma}
\label{lem:gaussian:I_V_A}
With definition~\eqref{eq:hybrid:defV}, 
\begin{equation}
I(V;A) = \frac12 \log\left( 1 + \frac{\alpha^2}{\gamma^2} \right) \ .
\end{equation}
\end{lemma}

\begin{IEEEproof}
From definition~\eqref{eq:hybrid:defV}, the covariance matrix of $(A,V)$ is given by
\[
\Gamma_{AV} = 
	\left[\begin{IEEEeqnarraybox}[\IEEEeqnarraystrutmode][c]{,c/c,}
		1		& \alpha			\\
		\alpha	& \alpha^2+\gamma^2
	\end{IEEEeqnarraybox}\right]	\ .
\]
Lemma~\ref{lem:gaussian:I_V_A} then directly follows from equality
\[
I(V;A) = \frac12\log\left( \frac{\Var{A}}{\Var{A|V}} \right)	\ ,
\]
and Corollary~\ref{coro:gaussian:varCond}. 
\end{IEEEproof}

\begin{lemma}
\label{lem:gaussian:Var_VsBY}
With definitions~\eqref{eq:hybrid:defV}, \eqref{eq:hybrid:defX}, 
\begin{equation}
\Var{V|BY} = \gamma^2\, 
		\frac	{1 + \frac1{P_B} + \frac{\alpha^2}{\gamma^2} + \frac{P}{P_Y}(\alpha+\beta)^2}
				{1 + \frac1{P_B} + \frac{P}{P_Y}\left( 1 + \frac{\gamma^2}{P_B} \right)}	\ .
\end{equation}
\end{lemma}

\begin{IEEEproof}
From definitions~\eqref{eq:hybrid:defV}, \eqref{eq:hybrid:defX}, the covariance matrix of $(V,B,Y)$ is given by
\[
\Gamma_{VBY} = 
	\left[\begin{IEEEeqnarraybox}[\IEEEeqnarraystrutmode][c]{,c/c/c,}
		\alpha^2+\gamma^2				& \alpha		& (\alpha\beta-\gamma^2)\sqrt{P}	\\
		\alpha							& 1+P_B			& \beta\sqrt{P}						\\
		(\alpha\beta-\gamma^2)\sqrt{P}	& \beta\sqrt{P}	& P+P_Y
	\end{IEEEeqnarraybox}\right]	\ .
\]
This equation comes from the following sequence of equalities, using~\eqref{eq:gaussX(A,V)}:
\begin{IEEEeqnarray*}{rCl}
\bE[VY]	&=& \bE[VX]													\\
		&=& \big( (\alpha+\beta)\bE[VA] - \Var{V} \big)\sqrt{P}	\\
		&=& \left( \alpha\beta -\gamma^2 \right)\sqrt{P}			\ .
\end{IEEEeqnarray*}
Lemma~\ref{lem:gaussian:Var_VsBY} then follows from Lemma~\ref{lem:gaussian:varCond} after some straightforward manipulations.
\end{IEEEproof}

Letting $P_B$ tend to zero in the above lemma yields the following corollary (which can also been proved independently using similar argument):
\begin{corollary}
\label{coro:gaussian:Var_VsAY}
With definitions~\eqref{eq:hybrid:defV}, \eqref{eq:hybrid:defX},
\begin{equation}
\Var{V|AY} = \frac	{\gamma^2}{1 + \gamma^2 \frac{P}{P_Y}}	\ .
\end{equation}
\end{corollary}

\begin{lemma}
\label{lem:gaussian:Var_AsBY}
With definitions~\eqref{eq:hybrid:defV}, \eqref{eq:hybrid:defX}, 
\begin{equation}
\Var{A|BY} = \frac	{1 + \gamma^2\frac{P}{P_Y}}
					{1 + \frac1{P_B} + \frac{P}{P_Y}\left( 1 + \frac{\gamma^2}{P_B} \right)}	\ .
\end{equation}
\end{lemma}

\begin{IEEEproof}
From definitions~\eqref{eq:hybrid:defV}, \eqref{eq:hybrid:defX}, the covariance matrix of $(A,B,Y)$ is given by
\[
\Gamma_{ABY} = 
	\left[\begin{IEEEeqnarraybox}[\IEEEeqnarraystrutmode][c]{,c/c/c,}
		1				& 1				& \beta\sqrt{P}	\\
		1				& 1+P_B			& \beta\sqrt{P}						\\
		\beta\sqrt{P}	& \beta\sqrt{P}	& P+P_Y
	\end{IEEEeqnarraybox}\right]	\ .
\]
Lemma~\ref{lem:gaussian:Var_AsBY} then follows from Lemma~\ref{lem:gaussian:varCond} after some straightforward manipulations.
\end{IEEEproof}

\begin{lemma}
\label{lem:gaussian:I_X_ZsE}
With definitions~\eqref{eq:hybrid:defV}, \eqref{eq:hybrid:defX}, 
\begin{equation}
I(X;Z|E) = \frac12\log\left(
				\frac	{1 + \frac1{P_E} + \frac{P}{P_Z}\left( 1 + \frac{\gamma^2}{P_E} \right)}
						{1 + \frac1{P_E}} \right)	\ .
\end{equation}
\end{lemma}

\begin{IEEEproof}
Lemma~\ref{lem:gaussian:I_X_ZsE} directly follows from Corollary~\ref{lem:gaussian:varCond} together with equality $\Var{Y|X}=P_Y$, expansion
\[
I(X;Z|E) = h(Z|E) - h(Z|X) \ ,
\]
which comes from the Markov chain $Z\mkv X\mkv E$,
and the following expression of the covariance matrix of $(Z,E)$:
\[
\Gamma_{ZE} = 
	\left[\begin{IEEEeqnarraybox}[\IEEEeqnarraystrutmode][c]{,c/c,}
		P+P_Z			& \beta\sqrt{P}	\\
		\beta\sqrt{P}	& 1+P_E
	\end{IEEEeqnarraybox}\right]	\ .
\]
\end{IEEEproof}

\subsection{End of Proof}

We now combine the above lemmas to prove that the inequalities~\eqref{eq:hybrid1}--\eqref{eq:hybrid4} and \eqref{eq:power} are verified by variables~\eqref{eq:hybrid:defU}--\eqref{eq:hybrid:defX} under conditions~\eqref{eq:GaussD}--\eqref{eq:GaussRate}.

As a matter of fact, inequality~\eqref{eq:hybrid1} is verified with definition~\eqref{eq:hybrid:defU}.
From \eqref{eq:hybrid:defX} and \eqref{eq:defgamma}, $X\sim\cN(0,P)$ and the power constraint~\eqref{eq:power} is also verified.

\subsubsection{Proof of~\eqref{eq:hybrid2}}

From Lemma~\ref{lem:gaussian:Var_VsBY} and equality $\Var{V} = \alpha^2+\gamma^2$, $I(V;BY)$ can be written as
\begin{IEEEeqnarray}{rcl}
I(V;BY) &=&\nonumber \\
\IEEEeqnarraymulticol{3}{R}{ \frac12\log\left( \left(1 + \frac{\alpha^2}{\gamma^2}\right)
		\frac	{1 + \frac1{P_B} + \frac{P}{P_Y}\left( 1 + \frac{\gamma^2}{P_B} \right)}
				{1 + \frac1{P_B} + \frac{\alpha^2}{\gamma^2} + \frac{P}{P_Y}(\alpha+\beta)^2}
		\right) }\ .
\label{eq:gaussian:I_V_BY}
\end{IEEEeqnarray}
This equality together with Lemma~\ref{lem:gaussian:I_V_A} and constraint~\eqref{eq:GaussRate} proves \eqref{eq:hybrid2}.

\subsubsection{Proof of~\eqref{eq:hybrid3}}

In the quadratic Gaussian case considered in Section~\ref{sec:gaussian}, distortion measure $d$ is the Euclidean distance on $\bR$:
\[
\bE\big[d(A,\hat A(V,B,Y))\big] = \bE\left[ \big(A-\hat A(V,B,Y)\big)^2 \right] \ .
\]
Moreover, in the proposed scheme, function $\hat A$ is the MMSE estimator of $A$ from $(V,B,Y)$, therefore
\[
\bE\big[d(A,\hat A(V,B,Y))\big] = \Var{A|VBY} \ .
\]

We now use the Markov chain $V\mkv(A,Y)\mkv B$ to expand the following conditional entropy:
\[
h(A|VBY) = h(A|BY) + h(V|AY) - h(V|BY)	\ ,
\]
and since the above random variables are jointly Gaussian, this yields
\[
\bE\big[d(A,\hat A(V,B,Y))\big] = \frac{\Var{A|BY}\Var{V|AY}}{\Var{V|BY}} \ .
\]

Gathering Lemmas~\ref{lem:gaussian:Var_VsBY}, \ref{lem:gaussian:Var_AsBY} and Corollary~\ref{coro:gaussian:Var_VsAY}, the above equation can be written as
\[
\bE\big[d(A,\hat A(V,B,Y))\big] 
	= \frac1{1 + \frac1{P_B} + \frac{\alpha^2}{\gamma^2} + \frac{P}{P_Y}(\alpha+\beta)^2} \ ,
\]
and hence \eqref{eq:hybrid3} is verified under constraint~\eqref{eq:GaussD}.

\subsubsection{Proof of~\eqref{eq:hybrid4}}

From Corollary~\ref{coro:gaussian:varCond} with the covariance matrix $\Gamma_{AE}$ given below, Lemmas~\ref{lem:gaussian:I_V_A} and~\ref{lem:gaussian:I_X_ZsE}, we can easily prove the following equality:
\begin{IEEEeqnarray}{rcl}
&&h(A|E) - I(V;A) - I(X;Z|E)\nonumber \\
	&&= \frac12\log\left(	\frac{2\pi e}{
			\left(1 + \frac{\alpha^2}{\gamma^2}\right)
			\left(1 + \frac1{P_E} + \frac{P}{P_Z}\left( 1 + \frac{\gamma^2}{P_E} \right) \right) }
		\right) \ .
\label{eq:gaussian:equivPart}
\end{IEEEeqnarray}
\[
\Gamma_{AE} = 
	\left[\begin{IEEEeqnarraybox}[\IEEEeqnarraystrutmode][c]{,c/c,}
		1	& 1			\\
		1	& 1+P_E
	\end{IEEEeqnarraybox}\right]	\ .
\]

Then, letting $P_B$ tend to zero and replacing $P_Y$ by $P_Z$ in~\eqref{eq:gaussian:I_V_BY} yields the following equality (which can also been proved independently using argument similar to the one Lemma~\ref{lem:gaussian:Var_VsBY}):
\begin{equation}
\label{eq:gaussian:I_V_AZ}
I(V;AZ) = \frac12\log\left( \left(1 + \frac{\alpha^2}{\gamma^2}\right)
							\left(1 + \gamma^2 \frac{P}{P_Z} \right)
					\right) \ .
\end{equation}

Inequality \eqref{eq:hybrid4} then follows under constraint~\eqref{eq:GaussDE} from~\eqref{eq:gaussian:I_V_BY}--\eqref{eq:gaussian:I_V_AZ} and definition $D_E = 2^{2\Delta}/(2\pi e)$.

This concludes the proof of Proposition~\ref{prop:gaussian:hybrid}.
\endIEEEproof

\section{Proof of the Direct Part of Theorem~\ref{th:gaussian:noB}}
\label{app:gaussian:noB:direct}

Letting $P_B$ tend to infinity, \eqref{eq:GaussD}--\eqref{eq:GaussRate} write
\begin{IEEEeqnarray}{l}
D	\geq \frac1{1 + \frac{\alpha^2}{\gamma^2}
				+ \frac{P}{P_Y}(\alpha+\beta)^2}		\ ,\label{eq:Gauss0DAch}\\
D_E	\leq \frac1{1 + \frac1{P_E} + \frac{P}{P_Z}\left( 1 + \frac{\gamma^2}{P_E} \right)} \nonumber\\
		\cdot \min\left\{ 
			\frac	{1 + \frac{P}{P_Y}}
					{1 + \frac{\alpha^2}{\gamma^2} + \frac{P}{P_Y}(\alpha+\beta)^2}
			\,;
			1 + \gamma^2\,\frac{P}{P_Z}
		\right\} 										\ ,\label{eq:Gauss0DeltaAch}\\
\frac{\alpha^2}{\gamma^2} + \frac{P}{P_Y} (\alpha+\beta)^2
	\leq \frac{P}{P_Y} 									\ .\label{eq:Gauss0RateAch}
\end{IEEEeqnarray}
We then check that these equations are verified with definitions~\eqref{eq:defalpha}, \eqref{eq:defbeta} under constraint~\eqref{eq:Gauss0DE}.
Recall that we consider here any distortion level: 
$$
D\in\left[	\frac1{1+\frac{P}{P_Y}} , \frac{1+\frac{P}{P_Z}}{1+\frac{P}{P_Y}} \right].
$$

\subsection{Proof of \eqref{eq:Gauss0DAch}}

From definitions~\eqref{eq:defalpha}, \eqref{eq:defbeta}, on one hand:
\begin{IEEEeqnarray*}{rl}
&\alpha^2
	= \left( \frac
			{ \gamma^2 \sqrt{\frac1{D}\left( \frac{P}{P_Y}-\frac{P}{P_Z} \right)} - \beta \gamma^2 \frac{P}{P_Y} }
			{ 1 + \gamma^2 \frac{P}{P_Y} } \right)^2			\\
	&= \gamma^4\,\frac
			{ \frac1{D}\left( \frac{P}{P_Y}-\frac{P}{P_Z} \right) + \beta^2\left(\frac{P}{P_Y} \right)^2 
				- 2 \beta\frac{P}{P_Y}\sqrt{\frac1{D}\left( \frac{P}{P_Y}-\frac{P}{P_Z} \right)} }
			{ \left( 1 + \gamma^2 \frac{P}{P_Y} \right)^2 } 	\ ,
\end{IEEEeqnarray*}
on the other hand:
\[
(\alpha+\beta)^2
	= \frac	
			{ \beta^2	+ \frac{\gamma^4}{D}\left( \frac{P}{P_Y}-\frac{P}{P_Z} \right)
				+ 2 \beta \gamma^2 \sqrt{\frac1{D}\left( \frac{P}{P_Y}-\frac{P}{P_Z} \right)} }
			{ \left( 1 + \gamma^2 \frac{P}{P_Y} \right)^2} 		\ .
\]
The denominator in~\eqref{eq:Gauss0DAch} thus can be written as
\begin{IEEEeqnarray*}{rl}
\label{eq:denom}
1 + \frac{\alpha^2}{\gamma^2}	&+ \frac{P}{P_Y}(\alpha+\beta)^2\\ 
	&= 1 + \frac
			{ \frac{\gamma^2}{D}\left( \frac{P}{P_Y}-\frac{P}{P_Z} \right) + \gamma^2\beta^2\left(\frac{P}{P_Y} \right)^2}{ \left( 1 + \gamma^2 \frac{P}{P_Y} \right)^2}	\\ 
			\IEEEeqnarraymulticol{2}{r}{+ \frac{\beta^2\,\frac{P}{P_Y}	+ \frac{\gamma^4}{D}\frac{P}{P_Y}\left( \frac{P}{P_Y}-\frac{P}{P_Z} \right) }
			{ \left( 1 + \gamma^2 \frac{P}{P_Y} \right)^2}}	\\
	&= 1 + \frac
			{\frac{\gamma^2}{D}\left( \frac{P}{P_Y}-\frac{P}{P_Z} \right) + \beta^2\,\frac{P}{P_Y} }
			{ 1 + \gamma^2 \frac{P}{P_Y}}	\\
	&=  \frac
			{ 1 + \frac{P}{P_Y} + \frac{\gamma^2}{D} \frac{P}{P_Y} 
			- \frac{\gamma^2}{D} \frac{P}{P_Z} }
			{ 1 + \gamma^2 \frac{P}{P_Y}}					\ ,\yesnumber
\end{IEEEeqnarray*}
where the last equality follows from~\eqref{eq:defgamma}.

Now, from definitions~\eqref{eq:defgamma}, \eqref{eq:defbeta}:
\begin{equation}
\label{eq:relGamma}
1 + \gamma^2\frac{P}{P_Z} = D \left(1+\frac{P}{P_Y}\right) \ ,
\end{equation}
and hence \eqref{eq:denom} can be written as
\begin{IEEEeqnarray*}{rCl}
\label{eq:1/D}
1 + \frac{\alpha^2}{\gamma^2}	+ \frac{P}{P_Y}(\alpha+\beta)^2
	&=&  \frac
			{ \frac1D + \frac{\gamma^2}{D}\frac{P}{P_Y} }
			{ 1 + \gamma^2 \frac{P}{P_Y}}						\\
	&=&  \frac1D												\ .\yesnumber
\end{IEEEeqnarray*}
This proves \eqref{eq:Gauss0DAch}.

\subsection{Proof of \eqref{eq:Gauss0DeltaAch}}

First, from~\eqref{eq:relGamma} and \eqref{eq:1/D}, the two arguments of the $\min\{\cdot\,;\cdot\}$ in~\eqref{eq:Gauss0DeltaAch} are equal:
\begin{IEEEeqnarray*}{rCl}
\frac{1 + \frac{P}{P_Y}}{1 + \frac{\alpha^2}{\gamma^2} + \frac{P}{P_Y}(\alpha+\beta)^2}
		&=& D \left(1+\frac{P}{P_Y}\right) 		\\
		&=&	1 + \gamma^2\,\frac{P}{P_Z}
\end{IEEEeqnarray*}
Then, from \eqref{eq:relGamma} once again, the first term in the r.h.s.\ of \eqref{eq:Gauss0DeltaAch} can be written as
\[
\frac1{1 + \frac1{P_E} + \frac{P}{P_Z}\left( 1 + \frac{\gamma^2}{P_E} \right)}
	= \frac1{1 + \frac{P}{P_Z} + \frac{D}{P_E} \left(1+\frac{P}{P_Y}\right)} \ ,
\]
and since \eqref{eq:Gauss0DE} can be written as, for $D \leq \frac{1+\frac{P}{P_Z}}{1+\frac{P}{P_Y}}$,
\[
D_E	\leq \frac1{ \frac1D \cdot \frac{1+\frac{P}{P_Z}}{1+\frac{P}{P_Y}} + \frac1{P_E}} \ ,
\]
this proves \eqref{eq:Gauss0DeltaAch}.

\subsection{Proof of \eqref{eq:Gauss0RateAch}}

Inequality~\eqref{eq:Gauss0RateAch} directly follows from \eqref{eq:1/D} and $D \geq \frac1{1 + \frac{P}{P_Y}}$.

This concludes the proof of the direct part of Theorem~\ref{th:gaussian:noB}.
\endIEEEproof

\section{Proof of the Converse Part of Proposition~\ref{prop:gaussian:noB:digital}}
\label{app:gaussian:noB:digital}

Let $(D,D_E)\in\cR_\text{digital}^P$ i.e., such that there exist random variables $U$, $V$, $Q$, $T$, $X$, and a function $\hat A$, with joint distribution $p(uvqtaexyz) = p(u|v)p(v|a)p(ae)p(q|t)p(tx)p(y|x)p(z|y)$,\footnote{%
	Since it is assumed that $P_Y<P_Z$, according to Remark~\ref{rem:marginal}, the Markov chain $X\mkv Y\mkv Z$ can be assumed here without loss of generality.}
and verifying
\begin{IEEEeqnarray*}{rcl}
I(U;A) &\leq& I(Q;Y) 																	\ ,\\[.1em]
I(V;A) &\leq& I(T;Y) 																	\ ,\\[.1em]
D 		 &\geq& \bE\big[d(A,\hat A(V))\big] 											\ ,\\[.1em]
\Delta	 &\leq&  h(A|UE) \,\\[.1em]
&-& \Big[ I(V;A|U) - \Big( I(T;Y|Q) - I(T;Z|Q) \Big) \Big]_+ 	\ ,\\[.1em]
\Var{X}  &\leq& P 																	\ ,
\end{IEEEeqnarray*}
where $\Delta=\frac12\log(2\pi e\,D_E)$.

From the Markov chain $Q\mkv T\mkv X\mkv Y\mkv Z$, tuple $(D,\Delta)$ verifies the following inequalities:
\begin{IEEEeqnarray}{rcl}
I(U;A) &\leq& I(Q;Y) 							\ ,\label{eq:gaussian:noB:digital:0}\\
I(V;A) &\leq& I(X;Y) 							\ ,\label{eq:gaussian:noB:digital:1}\\
D 		 &\geq& \bE\big[d(A,\hat A(V))\big] 	\ ,\label{eq:gaussian:noB:digital:2}\\
\Delta	 &\leq&  h(A|UE) \nonumber\\
&-& \Big[ I(V;A|U) - I(X;Y|QZ) \Big]_+ 
											\ ,\label{eq:gaussian:noB:digital:3}\\
\Var{X} &\leq& P 								\ .\label{eq:gaussian:noB:digital:4}
\end{IEEEeqnarray}

We now introduce three parameters:
$\nu	= 2^{2h(A|V)}/(2\pi e)$,
$\mu	= 2^{2h(A|U)}/(2\pi e)$,
$\zeta	=\linebreak 2^{2h(Y|Q)}/(2\pi e)$.
Since $U\mkv V\mkv A\mkv E$ and $Q\mkv X\mkv Y$ form Markov chains, from the fact that conditioning reduces the entropy and inequality $\Var{Y}\leq P+P_Y$, the above parameters are bounded as follows:
\begin{IEEEeqnarray}{l}
\nu	\leq \mu   \leq 1		\label{eq:gaussian:noB:digital:numu}\ ,\\
P_Y	\leq \zeta \leq P+P_Y	\label{eq:gaussian:noB:digital:zeta}\ .
\end{IEEEeqnarray}

We now write \eqref{eq:gaussian:noB:digital:0}--\eqref{eq:gaussian:noB:digital:4} as functions of these parameters.
First, recalling that distortion measure $d$ is the Euclidean distance on $\bR$,
\begin{equation}
\label{eq:gaussian:noB:Dnu}
\bE\big[d(A,\hat A(V))\big] \geq \Var{A|V} \geq \nu \ . 
\end{equation}

Then, from the Markov chain $U\mkv A\mkv E$, we write 
\begin{equation}
\label{eq:gaussian:noB:hAUE}
h(A|UE) = h(A|U) - h(E|U) + h(E|A) \ .
\end{equation}
Now, since $E = A + N_E$ with $N_E$ independent of $A$ (and $U$), the conditional EPI~\cite{elgamal2010lecture} yields:
\begin{equation}
\label{eq:gaussian:noB:condEPI}
2^{2 h(E|U)} \geq 2^{2 h(A|U)} + 2^{2 h(N_E)}	\ .
\end{equation}
Gathering \eqref{eq:gaussian:noB:hAUE} and \eqref{eq:gaussian:noB:condEPI}, we obtain:
\begin{IEEEeqnarray*}{rCl}
\label{eq:gaussian:noB:hAUEmu}
h(A|UE)	&\leq&	h(A|U) - \frac12\log\left( 2^{2 h(A|U)} + 2^{2 h(N_E)} \right) + h(E|A)	\\
		&=&		\frac12\log\left( \frac1{\frac1{2^{2 h(A|U)}} + \frac1{2^{2 h(N_E)}}} \right) \\
		&=&		\frac12\log\left( \frac{2\pi e}{\frac1\mu + \frac1{P_E}} \right)
				\yesnumber\ .
\end{IEEEeqnarray*}

From the Markov chain $Q\mkv X\mkv Y\mkv Z$, there exists a random variable $\widetilde N_Z\sim\cN(0,P_Z-P_Y)$, independent of $(Q,X,Y)$ such that $Z = Y + \widetilde N_Z$. 
Then the conditional EPI~\cite{elgamal2010lecture} yields:
\[
2^{2 h(Z|Q)} \geq 2^{2 h(Y|Q)} + 2^{2 h(\widetilde N_Z)}	\ .
\]
From the latter equation,
\begin{IEEEeqnarray*}{rCl}
I(X;Y|QZ)	&=&		h(Y|Q) - h(Y|X) - h(Z|Q) + h(Z|X)				\\
			&\leq&	h(Y|Q) - h(N_Y) \nonumber\\
			\IEEEeqnarraymulticol{3}{r}{ -\frac12\log\left( 2^{2 h(Y|Q)} 
						+ 2^{2 h(\widetilde N_Z)} \right) + h(N_Z)	}\\		
			&\leq&	\frac12\log\left( \frac{\zeta}{\zeta + P_Z - P_Y}\,
									  \frac{P_Z}{P_Y} \right)		\ .
\end{IEEEeqnarray*}

Gathering the above equations, tuple $(D,D_E)$ verifies \eqref{eq:gaussian:noB:digital:numu}, \eqref{eq:gaussian:noB:digital:zeta}, and 
\begin{IEEEeqnarray*}{l}	
\frac1\mu	\leq \frac{P+P_Y}{\zeta}										\ ,\\
\frac1\nu	\leq 1+\frac{P}{P_Y} 											\ ,\\
D			\geq \nu			 											\ ,\\
D_E			\leq \frac1{\frac1\mu + \frac1{P_E}} 							\ ,\\
D_E			\leq \frac1{\frac1\mu + \frac1{P_E}}\cdot\frac\nu\mu\cdot
					\frac1{1 + \frac{P_Z-P_Y}{\zeta}}\cdot\frac{P_Z}{P_Y}	\ .
\end{IEEEeqnarray*}
Eliminating parameters $\zeta$, $\mu$ and removing redundant inequalities in the above system prove the converse part of Proposition~\ref{prop:gaussian:noB:digital}.
\endIEEEproof

\section{Proof of Proposition~\ref{prop:gaussian:noB:analog}}
\label{app:gaussian:noB:analog}

Consider any distortion level $D\in\left[	\frac1{1+\frac{P}{P_Y}} , 1 \right]$.
The analog scheme of Proposition~\ref{prop:gaussian:noB:analog} then consists in sending a scaled version of the source over the channel:
\begin{equation}
\label{eq:analog:defX}
X = \sqrt{\tau}\,A \ ,
\end{equation}
where $\tau = P_Y \left(\frac1D - 1\right)$.
Note that, since $D\geq \frac1{1+\frac{P}{P_Y}}$, $\Var{X} = \tau \leq P$ and the power constraint~\eqref{eq:power} is verified.
Bob then simply computes the MMSE estimate $\hat A$ of $A$ from $Y$.

In such an analog scheme, the mean distortion at Bob can be written as
\begin{IEEEeqnarray*}{rCL}
\bE\big[d(A,\hat A(Y))\big] 
 	&=&	\Var{A|Y}				\\
	&=&	\frac{P_Y}{\tau + P_Y}	\\
	&=& D						\ ,
\end{IEEEeqnarray*}
where the next-to-last equation follows after Corollary~\ref{coro:gaussian:varCond} together with the covariance matrix of $(A,Y)$:
\[
\Gamma_{AY} = 
	\left[\begin{IEEEeqnarraybox}[\IEEEeqnarraystrutmode][c]{,c/c,}
		1 			& \sqrt{\tau}	\\
		\sqrt{\tau}	& \tau + P_Y	
	\end{IEEEeqnarraybox}\right] \ .
\]

The equivocation rate at Eve is $h(A|EZ)$ and quantity $D_E$ then can be written as
\begin{IEEEeqnarray*}{rCL}
D_E &=& \Var{A|EZ}															\\
	&=& \frac1{1 + \frac1{P_E} + \frac{\tau}{P_Z}}							\\
	&=& \frac1{1 + \frac1{P_E} + \left(\frac1D - 1\right)\frac{P_Y}{P_Z}}	\ ,
\end{IEEEeqnarray*}
where the next-to-last equation follows after some straightforward manipulations from Lemma~\ref{lem:gaussian:varCond} and the covariance matrix of $(A,E,Z)$:
\[
\Gamma_{AEZ} = 
	\left[\begin{IEEEeqnarraybox}[\IEEEeqnarraystrutmode][c]{,c/c/c,}
		1 			& 	1				&	\sqrt{\tau}	\\
		1 			& 	1+P_E			&	\sqrt{\tau}	\\
		\sqrt{\tau}	& 	\sqrt{\tau}		&	\tau + P_Y	
	\end{IEEEeqnarraybox}\right] \ .
\]

This proves Proposition~\ref{prop:gaussian:noB:analog}.
\endIEEEproof

\begin{IEEEbiographynophoto}{Joffrey Villard} 
(S'09-M'12) was born in Saint-\'{E}tienne, France, in 1985. He received the Dipl.Ing. degree in digital communication and electronics in 2008, the M.Sc. degree in wireless communication systems in 2008, and the Ph.D. degree in 2011, all from SUPELEC, Gif-sur-Yvette, France. From 2008 to 2011, he was with the Department of Telecommunications of SUPELEC. He is currently a Platform R\&D Engineer at WITHINGS, Issy-les-Moulineaux, France. His research interests include information theory, source coding, statistical inference, and signal processing for wireless sensor networks.
\end{IEEEbiographynophoto}

\begin{IEEEbiographynophoto}{Pablo Piantanida}
(S'04-M'08) received the B.Sc. and M.Sc degrees (with honors) in Electrical Engineering from the University of Buenos Aires (Argentina), in 2003, and the Ph.D. from the Paris-Sud University (France) in 2007. In 2006, he has been with the Department of Communications and Radio-Frequency Engineering at Vienna University of Technology (Austria). Since October 2007 he has joined in 2007 the Department of Telecommunications, SUPELEC, as an Assistant Professor in network information theory. His research interests include multi-terminal information theory, Shannon theory, cooperative communications, physical-layer security and coding theory for wireless applications. 
 \end{IEEEbiographynophoto}

\begin{IEEEbiographynophoto}{Shlomo Shamai (Shitz)}
 received the B.Sc., M.Sc., and Ph.D. degrees in
electrical engineering from the Technion---Israel Institute of Technology,
in 1975, 1981 and 1986 respectively.

During 1975-1985 he was with the Communications Research Labs,
in the capacity of a Senior Research Engineer. Since 1986 he is with
the Department of Electrical Engineering, Technion---Israel Institute of
Technology, where he is now a Technion Distinguished Professor,
and holds the William Fondiller Chair of Telecommunications.
His research interests encompasses a wide spectrum of topics in information
theory and statistical communications.

Dr. Shamai (Shitz) is an IEEE Fellow, a member of the Israeli Academy of
Sciences and Humanities and a Foreign Associate of the
US National Academy of Engineering. He is the recipient of the 2011
Claude E. Shannon Award.
He has been awarded the 1999 van der Pol Gold Medal of the Union Radio
Scientifique Internationale (URSI), and is a co-recipient of the 2000 IEEE
Donald G. Fink Prize Paper Award, the 2003, and
the 2004 joint IT/COM societies paper award, the 2007 IEEE Information
Theory Society Paper Award, the 2009 European Commission FP7, Network of
Excellence in Wireless COMmunications (NEWCOM++) Best Paper Award,
and the 2010 Thomson Reuters Award for International Excellence
in Scientific Research.  He is also the recipient of
1985 Alon Grant for distinguished young scientists and the 2000 Technion Henry
Taub Prize for Excellence in Research.
He has served as Associate Editor for the Shannon Theory of the IEEE
Transactions on Information Theory, and has also served twice on the
Board of Governors of the Information Theory Society.
He is a member of the Executive Editorial Board of the IEEE Transactions
on Information Theory.
\end{IEEEbiographynophoto}

\bibliographystyle{IEEEtran}
\bibliography{secure_transIT}

\end{document}